\input harvmac


\noblackbox


 \font\cmsss=cmss10 at 8pt \font\affil=cmti10 at 9pt \font\titlerm=cmr10 scaled\magstep3
\font\titlerms=cmr7 scaled\magstep3 \font\titlermss=cmr5 scaled\magstep3 \font\titlei=cmmi10 scaled\magstep3
\font\titleis=cmmi7 scaled\magstep3 \font\titleiss=cmmi5 scaled\magstep3 \font\titlesy=cmsy10 scaled\magstep3
\font\titlesys=cmsy7 scaled\magstep3 \font\titlesyss=cmsy5 scaled\magstep3 \font\titleit=cmti10 scaled\magstep3
\font\ticp=cmcsc10 \font\ttsmall=cmtt10 at 8pt

\input epsf
%


%


\def\itm{\noindent $\bullet$ \ }

\def\[{\left [}
\def\]{\right ]}
\def\({\left (}
\def\){\right )}
\def\<{\langle}
\def\>{\rangle}

\def\eg{{\it e.g.}}
\def\ie{{\it i.e.}}

\def\etc{{\it etc.}}


\def\a{\alpha}
\def\b{\beta}
\def\g{\kappa}

\def\Om{\Omega}
\def\lam{\lambda}

\def\s{\sigma}


\def\CH{{\cal H}}

\def\CO{{\cal O}}

\def\CS{{\cal S}}
\def\CU{{\cal U}}


\def\S{{\bf S}}
\def\A5S5{{\rm AdS}_5 \times \S^5}

\def\l{\ell}

\def\Tr{{\rm Tr}}


\def\Scri{{\cal I}}
\def\scri{{\cal I}}
\def\scrip{{\cal I}^+}
\def\io{\alpha}

\def\dta{\left({\partial \over\partial t}\right)^a}
\def\pd{{\dot \phi}}
\def\pp{\phi'}
\def\rh{{\hat r}}
\def\rb{{\bar r}}

\def\VR{V_{\rm eff}}
\def\Vmax{V_{\rm max}}
\def\td{\dot{t}}
\def\rd{\dot{r}}

\def\rmin{r_{\rm min}}
\def\rmax{r_{\rm max}}
\def\len{{\cal L}}
\def\dS{de Sitter}
\def\SAdS{Schwarzschild-AdS}

\def\blocktext#1{\vskip6pt\noindent{\it #1}\vskip6pt\noindent}

\input inflation_paper.defs


\lref\MarolfFY{
  D.~Marolf, {\it States and boundary terms: Subtleties of Lorentzian AdS/CFT},
  JHEP {\bf 0505}, 042 (2005)
  [arXiv:hep-th/0412032].
}
\lref\LoukoHC{
  J.~Louko and D.~Marolf,
  {\it Single-exterior black holes and the AdS-CFT conjecture},
  Phys.\ Rev.\ D {\bf 59}, 066002 (1999)
  [arXiv:hep-th/9808081].
}

\lref\KKLT{ S. Kachru, R. Kallosh, A. Linde and S. Trivedi, {\it de Sitter vacua in string theory}, Phys. Rev. {\bf
D68} (2003) 046005 [arXiv:hep-th/0301240].}

\lref\KKLTexample{ F. Denef, M. Douglas, B. Florea, A. Grassi and S. Kachru, {\it Fixing all moduli in a simple F-theory
compactification}, [arXiv:hep-th/0503124]; F. Denef, M. Douglas and B. Florea, {\it Building a better racetrack}, JHEP
{\bf 0406} (2004) 034 [arXiv:hep-th/0404257]; P.~S.~Aspinwall and R.~Kallosh,
  {\it Fixing all moduli for M-theory on K3 x K3},
  [arXiv:hep-th/0506014].}

\lref\nonsusyKKLT{ J. Conlon, F. Quevedo and K. Suruliz, {\it Large-volume flux compactifications: Moduli spectrum and
D3/D7 soft supersymmetry breaking}, [arXiv:hep-th/0505076]; P. Berglund and P. Mayr, {\it Non-perturbative
superpotentials in F-theory and string duality}, [arXiv:hep-th/0504058]; V. Balasubramanian, P. Berglund, J. Conlon
and F. Quevedo, {\it Systematics of moduli stabilisation in Calabi-Yau flux compactifications}, JHEP {\bf 0503} (2005)
007 [arXiv:hep-th/0502058].}

\lref\eva{ A. Saltman and E. Silverstein, {\it A new handle on de Sitter compactifications}, [arXiv:hep-th/0411271]; A.
Maloney, E. Silverstein and A. Strominger, {\it de Sitter space in noncritical string theory}, [arXiv:hep-th/0205316].
 B. Acharya, {\it A moduli fixing mechanism in M-theory}, [arXiv:hep-th/0212294].
 B. de Carlos, A. Lukas and S. Morris, {\it Non-perturbative vacua for M-theory on G2 manifolds}, JHEP {\bf 0412}
(2004) 018 [arXiv:hep-th/0409255.]}

\lref\Heterotic{ G. Curio, A. Krause and D. L\"ust, {\it Moduli stabilization in the heterotic/IIB discretuum},
[arXiv:hep-th/0502168]; S. Gurrieri, A. Lukas and A. Micu, {\it Heterotic on Half-flat}, Phys.Rev. {\bf D70} (2004)
126009 [arXiv:hep-th/0408121]; K. Becker, M. Becker, K. Dasgupta, P. Green and E. Sharpe, {\it Compactifications of
Heterotic Strings on Non-Kahler complex manifolds II}, Nucl.Phys. {\bf B678} (2004) 19 [arXiv:hep-th/0310058]; M.
Becker, G. Curio and A. Krause, {\it De Sitter Vacua from Heterotic M Theory}, Nucl.Phys. {\bf B693} (2004) 223
[arXiv:hep-th/0403027]; R. Brustein and S.P. de Alwis, {\it Moduli Potentials in String Compactifications with Fluxes:
Mapping the Discretuum}, Phys.Rev. {\bf D69} (2004) 126006 [arXiv:hep-th/0402088]; S. Gukov, S. Kachru, X. Liu and
L. McAllister, {\it Heterotic Moduli Stabilization with Fractional Chern-Simons Invariants}, Phys.Rev. {\bf D69} (2004)
086008 [arXiv:hep-th/0310159]; E. Buchbinder and B. Ovrut, {\it Vacuum Stability in Heterotic M-theory}, Phys.Rev. {\bf
D69} (2004) 086010 [arXiv:hep-th/0310112]; G. Cardoso, G. Curio, G. Dall'Agata and D. L\"ust, {\it Heterotic string
theory on non-Kahler manifolds with H-flux and gaugino condensate}, Fortsch. Phys. {\bf 52} (2004) 483
[arXiv:hep-th/0310021]; G. Cardoso, G. Curio, G. Dall'Agata and D. L\"ust, {\it BPS action and superpotential for
heterotic string compactifications with fluxes}, JHEP {\bf 0310} (2003) 004 [arXiv:hep-th/0306088]
  G.~Curio and A.~Krause, {\it G-fluxes and non-perturbative stabilisation of heterotic M-theory}, Nucl.Phys.{\bf B643} (2002) 131
  [hep-th/0108220]
.}

\lref\Antoniadis{
  I.~Antoniadis and T.~Maillard,
{\it Moduli stabilization from magnetic fluxes in type I string theory},
  Nucl.\ Phys.\ B {\bf 716}, 3 (2005)
  [arXiv:hep-th/0412008];
  I.~Antoniadis, A.~Kumar and T.~Maillard,
{\it Moduli stabilization with open and closed string fluxes}, [arXiv:hep-th/0505260].}

\lref\lenny{ L. ~Susskind, {\it private communication}. }

\lref\SusskindKW{
  L.~Susskind,
  {\it The anthropic landscape of string theory},
  arXiv:hep-th/0302219.
}

\lref\BoussoXA{
  R.~Bousso and J.~Polchinski
  {\it Quantization of four-form fluxes and dynamical neutralization
  of the cosmological constant},
  JHEP {\bf 0006}, 006 (2000)
  [arXiv:hep-th/0004134].
}

\lref\vil{
  J.~Garriga and A.~Vilenkin,
  {\it Perturbations on domain walls and strings: A Covariant theory},
  Phys.\ Rev.\ D {\bf 44}, 1007 (1991).
}

\lref\vilg{
  J.~Garriga and A.~Vilenkin,
  {\it Quantum fluctuations on domain walls, strings and vacuum bubbles},
  Phys.\ Rev.\ D {\bf 45}, 3469 (1992).
}

\lref\BanksNM{
  T.~Banks,
  {\it Heretics of the false vacuum: Gravitational effects on and of vacuum decay.
  II},
  arXiv:hep-th/0211160.
}

\lref\FischlerYJ{
  W.~Fischler, A.~Kashani-Poor, R.~McNees and S.~Paban,
  {\it The acceleration of the universe, a challenge for string theory},
  JHEP {\bf 0107}, 003 (2001)
  [arXiv:hep-th/0104181].
}

\lref\BoussoNF{
  R.~Bousso,
  {\it Positive vacuum energy and the N-bound},
  JHEP {\bf 0011}, 038 (2000)
  [arXiv:hep-th/0010252].
}

\lref\alex{ A. Maloney, {\it AdS$_3$ Cosmology}, to appear. }

\lref\BanksYP{
  T.~Banks and W.~Fischler,
 {\it M-theory observables for cosmological space-times},
  arXiv:hep-th/0102077.
}

\lref\BanksDD{
  T.~Banks, M.~R.~Douglas, G.~T.~Horowitz and E.~J.~Martinec,
  {\it AdS dynamics from conformal field theory},
  arXiv:hep-th/9808016.
}

\lref\HellermanYI{
  S.~Hellerman, N.~Kaloper and L.~Susskind,
  {\it String theory and quintessence},
  JHEP {\bf 0106}, 003 (2001)
  [arXiv:hep-th/0104180].
}
\lref\WittenKN{
  E.~Witten,
  {\it Quantum gravity in de Sitter space},
  arXiv:hep-th/0106109.
}

\lref\HorowitzXK{
  G.~T.~Horowitz and D.~Marolf,
 {\it A new approach to string cosmology},
  JHEP {\bf 9807}, 014 (1998)
  [arXiv:hep-th/9805207].
}

\lref\MaldacenaRF{
  J.~Maldacena and L.~Maoz,
 {\it Wormholes in AdS},
  JHEP {\bf 0402}, 053 (2004)
  [arXiv:hep-th/0401024].
}

\lref\KrasnovZQ{
  K.~Krasnov,
 {\it Holography and Riemann surfaces},
  Adv.\ Theor.\ Math.\ Phys.\  {\bf 4}, 929 (2000)
  [arXiv:hep-th/0005106].
}

\lref\HamiltonJU{
  A.~Hamilton, D.~Kabat, G.~Lifschytz and D.~A.~Lowe,
 {\it Local bulk operators in AdS/CFT: A boundary view of horizons and
  locality},
  arXiv:hep-th/0506118.
}

\lref\GiddingsQU{
  S.~B.~Giddings,
 {\it The boundary S-matrix and the AdS to CFT dictionary},
  Phys.\ Rev.\ Lett.\  {\bf 83}, 2707 (1999)
  [arXiv:hep-th/9903048].
}

\lref\BalasubramanianDE{
  V.~Balasubramanian, P.~Kraus, A.~E.~Lawrence and S.~P.~Trivedi,
 {\it Holographic probes of anti-de Sitter space-times},
  Phys.\ Rev.\ D {\bf 59}, 104021 (1999)
  [arXiv:hep-th/9808017].
}

\lref\IsraelUR{
  W.~Israel,
 {\it Thermo Field Dynamics Of Black Holes},
  Phys.\ Lett.\ A {\bf 57}, 107 (1976).
}

\lref\MaldacenaBW{
  J.~M.~Maldacena and A.~Strominger,
{\it AdS(3) black holes and a stringy exclusion principle},
  JHEP {\bf 9812}, 005 (1998)
  [arXiv:hep-th/9804085].
}

\lref\GibbonsMU{
  G.~W.~Gibbons and S.~W.~Hawking,
{\it Cosmological Event Horizons, Thermodynamics, And Particle Creation},
  Phys.\ Rev.\ D {\bf 15}, 2738 (1977).
}

\lref\BlauCW{
  S.~K.~Blau, E.~I.~Guendelman and A.~H.~Guth,
{\it The Dynamics Of False Vacuum Bubbles},
  Phys.\ Rev.\ D {\bf 35}, 1747 (1987).
}

\lref\AminneborgPZ{
  S.~Aminneborg, I.~Bengtsson, D.~Brill, S.~Holst and P.~Peldan,
{\it Black holes and wormholes in 2+1 dimensions},
  Class.\ Quant.\ Grav.\  {\bf 15}, 627 (1998)
  [arXiv:gr-qc/9707036].
}

\lref\AguirreXS{
  A.~Aguirre and M.~C.~Johnson,
{\it Dynamics and instability of false vacuum bubbles},
  arXiv:gr-qc/0508093.
}

\lref\JohnsonPC{
   M.~C.~Johnson, {\it private communication.}
}

\lref\JacobsonMI{
  T.~Jacobson,
{\it On the nature of black hole entropy},
  arXiv:gr-qc/9908031.
}

\lref\GubserVJ{
  S.~S.~Gubser,
   {\it AdS/CFT and gravity},
  Phys.\ Rev.\ D {\bf 63}, 084017 (2001)
  [arXiv:hep-th/9912001].
}

\lref\MaldacenaRE{
  J.~M.~Maldacena,
{\it The large N limit of superconformal field theories and supergravity},
  Adv.\ Theor.\ Math.\ Phys.\  {\bf 2}, 231 (1998)
  [Int.\ J.\ Theor.\ Phys.\  {\bf 38}, 1113 (1999)]
  [arXiv:hep-th/9711200].
}

\lref\WittenQJ{
  E.~Witten,
   {\it Anti-de Sitter space and holography},
  Adv.\ Theor.\ Math.\ Phys.\  {\bf 2}, 253 (1998)
  [arXiv:hep-th/9802150].
}

\lref\GubserBC{
  S.~S.~Gubser, I.~R.~Klebanov and A.~M.~Polyakov,
{\it Gauge theory correlators from non-critical string theory},
  Phys.\ Lett.\ B {\bf 428}, 105 (1998)
  [arXiv:hep-th/9802109].
}

\lref\MaldacenaKR{
  J.~M.~Maldacena,
{\it Eternal black holes in Anti-de-Sitter},
  JHEP {\bf 0304}, 021 (2003)
  [arXiv:hep-th/0106112].
}

\lref\AlberghiKD{
  G.~L.~Alberghi, D.~A.~Lowe and M.~Trodden,
 {\it Charged false vacuum bubbles and the AdS/CFT correspondence},
  JHEP {\bf 9907}, 020 (1999)
  [arXiv:hep-th/9906047].
}

\lref\BakJK{
  D.~Bak, M.~Gutperle and S.~Hirano,
{\it A dilatonic deformation of AdS(5) and its field theory dual},
  JHEP {\bf 0305}, 072 (2003)
  [arXiv:hep-th/0304129].
}

\lref\ClarkSB{
  A.~B.~Clark, D.~Z.~Freedman, A.~Karch and M.~Schnabl,
{\it The dual of Janus $((<:)\leftrightarrow (:>))$ an interface CFT},
  Phys.\ Rev.\ D {\bf 71}, 066003 (2005)
  [arXiv:hep-th/0407073].
}

\lref\KrausIV{
  P.~Kraus, H.~Ooguri and S.~Shenker,
{\it Inside the horizon with AdS/CFT},
  Phys.\ Rev.\ D {\bf 67}, 124022 (2003)
  [arXiv:hep-th/0212277].
}

\lref\FidkowskiNF{
  L.~Fidkowski, V.~Hubeny, M.~Kleban and S.~Shenker,
 {\it The black hole singularity in AdS/CFT},
  JHEP {\bf 0402}, 014 (2004)
  [arXiv:hep-th/0306170].
}

\lref\StromingerPN{
  A.~Strominger,
{\it The dS/CFT correspondence},
  JHEP {\bf 0110}, 034 (2001)
  [arXiv:hep-th/0106113].
}

\lref\RandallEE{
  L.~Randall and R.~Sundrum,
{\it A large mass hierarchy from a small extra dimension},
  Phys.\ Rev.\ Lett.\  {\bf 83}, 3370 (1999)
  [arXiv:hep-ph/9905221].
}

\lref\Birrell{ N. Birrell and P. Davies,  {\it Quantum Fields in Curved Space}, Cambridge University Press, Cambridge
(1982). }

\lref\BoussoCB{
  R.~Bousso,
{\it Holography in general space-times},
  JHEP {\bf 9906}, 028 (1999)
  [arXiv:hep-th/9906022].
}

\lref\BoussoXA{
  R.~Bousso and J.~Polchinski,
{\it Quantization of four-form fluxes and dynamical neutralization of the
  cosmological constant},
  JHEP {\bf 0006}, 006 (2000)
  [arXiv:hep-th/0004134].
}

\lref\ColemanAW{
  S.~R.~Coleman and F.~De Luccia,
{\it Gravitational Effects On And Of Vacuum Decay},
  Phys.\ Rev.\ D {\bf 21}, 3305 (1980).
}

\lref\FarhiYR{
  E.~Farhi, A.~H.~Guth and J.~Guven,
{\it Is It Possible To Create A Universe In The Laboratory By Quantum
  Tunneling?},
  Nucl.\ Phys.\ B {\bf 339}, 417 (1990).
}

\lref\FarhiTY{
  E.~Farhi and A.~H.~Guth,
{\it An Obstacle To Creating A Universe In The Laboratory},
  Phys.\ Lett.\ B {\bf 183}, 149 (1987).
}

\lref\FischlerSE{
  W.~Fischler, D.~Morgan and J.~Polchinski,
 {\it Quantum Nucleation Of False Vacuum Bubbles},
  Phys.\ Rev.\ D {\bf 41}, 2638 (1990).
}

\lref\FischlerPK{
  W.~Fischler, D.~Morgan and J.~Polchinski,
{\it Quantization Of False Vacuum Bubbles: A Hamiltonian Treatment Of
  Gravitational Tunneling},
  Phys.\ Rev.\ D {\bf 42}, 4042 (1990).
}

\lref\LindeSK{
  A.~D.~Linde,
{\it Hard art of the universe creation (stochastic approach to tunneling and
  baby universe formation)},
  Nucl.\ Phys.\ B {\bf 372}, 421 (1992)
  [arXiv:hep-th/9110037].
}

\lref\GallowayBP{
  G.~J.~Galloway, K.~Schleich, D.~M.~Witt and E.~Woolgar,
{\it Topological censorship and higher genus black holes},
  Phys.\ Rev.\ D {\bf 60}, 104039 (1999)
  [arXiv:gr-qc/9902061].
}

\lref\GallowayBR{
  G.~J.~Galloway, K.~Schleich, D.~Witt and E.~Woolgar,
{\it The AdS/CFT correspondence conjecture and topological censorship},
  Phys.\ Lett.\ B {\bf 505}, 255 (2001)
  [arXiv:hep-th/9912119].
}

\lref\WittenXP{
  E.~Witten and S.~T.~Yau,
{\it Connectedness of the boundary in the AdS/CFT correspondence},
  Adv.\ Theor.\ Math.\ Phys.\  {\bf 3}, 1635 (1999)
  [arXiv:hep-th/9910245].
}

\lref\KachruAW{
  S.~Kachru, R.~Kallosh, A.~Linde and S.~P.~Trivedi,
{\it De Sitter vacua in string theory},
  Phys.\ Rev.\ D {\bf 68}, 046005 (2003)
  [arXiv:hep-th/0301240].
}

\lref\GuthZM{
  A.~H.~Guth,
{\it The Inflationary Universe: A Possible Solution To The Horizon And Flatness
  Problems},
  Phys.\ Rev.\ D {\bf 23}, 347 (1981).
}

\lref\LindeMU{
  A.~D.~Linde,
{\it A New Inflationary Universe Scenario: A Possible Solution Of The Horizon,
  Flatness, Homogeneity, Isotropy And Primordial Monopole Problems},
  Phys.\ Lett.\ B {\bf 108}, 389 (1982).
}

\lref\AharonyTI{
  O.~Aharony, S.~S.~Gubser, J.~M.~Maldacena, H.~Ooguri and Y.~Oz,
{\it Large N field theories, string theory and gravity},
  Phys.\ Rept.\  {\bf 323}, 183 (2000)
  [arXiv:hep-th/9905111].
}

\lref\IsraelRT{
  W.~Israel,
{\it Singular Hypersurfaces And Thin Shells In General Relativity},
  Nuovo Cim.\ B {\bf 44S10}, 1 (1966)
  [Erratum-ibid.\ B {\bf 48}, 463 (1967\ NUCIA,B44,1.1966)].
}

\lref\BoussoTV{
  R.~Bousso,
   {\it Cosmology and the S-matrix},
  Phys.\ Rev.\ D {\bf 71}, 064024 (2005)
  [arXiv:hep-th/0412197].
}

\lref\BanksXH{
  T.~Banks,
{\it Landskepticism or why effective potentials don't count string models},
  arXiv:hep-th/0412129.
}

\lref\DeWolfeUU{
  O.~DeWolfe, A.~Giryavets, S.~Kachru and W.~Taylor,
   {\it Type IIA moduli stabilization},
  JHEP {\bf 0507}, 066 (2005)
  [arXiv:hep-th/0505160].
}

\lref\linde{A.D. Linde, {\it Particle Physics and Inflationary Cosmology}, Harwood, Chur, Switzerland (1990).}

\lref\liddle{A.R. Liddle and D.H. Lyth, {\it Cosmological Inflation and Large-Scale Structure}, Cambridge University
Press, Cambridge, England (2000).}

\lref\kolb{E. W. Kolb and M. S. Turner, {\it The Early Universe}, Addison-Wesley, Redwood City (1990).}

\lref\DeWolfeUU{
  O.~DeWolfe, A.~Giryavets, S.~Kachru and W.~Taylor,
  {\it Type IIA moduli stabilization},
  JHEP {\bf 0507}, 066 (2005)
  [arXiv:hep-th/0505160].
}

\lref\BanksFE{
  T.~Banks,
{\it Cosmological breaking of supersymmetry or little Lambda goes back to  the future. II},
  arXiv:hep-th/0007146.
}

\lref\LoukoTP{
  J.~Louko, D.~Marolf and S.~F.~Ross,
 {\it On geodesic propagators and black hole holography},
  Phys.\ Rev.\ D {\bf 62}, 044041 (2000)
  [arXiv:hep-th/0002111].
}

\lref\GuthPN{
  A.~H.~Guth and E.~J.~Weinberg,

  {\it Could The Universe Have Recovered From A Slow First Order Phase
  Transition?,}
  Nucl.\ Phys.\ B {\bf 212}, 321 (1983).
}

\lref\EmparanGF{
  R.~Emparan,
  {\it AdS/CFT duals of topological black holes and the
  entropy of  zero-energy states},
  JHEP {\bf 9906}, 036 (1999)
  [arXiv:hep-th/9906040].
}

\lref\BalasubramanianKK{
  V.~Balasubramanian, V.~Jejjala and J.~Simon,
  {\it The library of Babel,}
  arXiv:hep-th/0505123.
}

\lref\BalasubramanianMG{
  V.~Balasubramanian, J.~de Boer, V.~Jejjala and J.~Simon,
  {\it The library of Babel: On the origin of gravitational
  thermodynamics,}
  arXiv:hep-th/0508023.
}


\def\Title#1{\nopagenumbers\vskip .3in\centerline{\titlefont #1}\abstractfont\vskip .4in\pageno=0}

%

\Title{\vbox{ {\centerline{Inflation in AdS/CFT} }}}
\centerline{\ticp Ben Freivogel$^{a,b}$, Veronika E. Hubeny$^{b,c}$, Alexander Maloney$^{a,d}$, } \centerline{\ticp
Robert C. Myers$^{e,f}$, Mukund Rangamani$^{b,c}$, and Stephen Shenker$^a$ \footnote{}{\ttsmall freivogel@berkeley.edu,
 veronika.hubeny@durham.ac.uk, maloney@slac.stanford.edu,}
\footnote{}{\ttsmall rmyers@perimeterinstitute.ca, mukund.rangamani@durham.ac.uk, sshenker@stanford.edu}}
\bigskip

\baselineskip 14pt \centerline {\affil $^a$ Department of Physics, Stanford University, Stanford, CA 94305, USA}
\centerline{\affil $^b$ Department of Physics \& Theoretical Physics Group, LBNL, Berkeley, CA 94720, USA}
\centerline{\affil $^c$ Department of Mathematical Sciences,  University of Durham, Durham DH1 3LE, UK}
\centerline{\affil $^d$ Stanford Linear Accelerator Center, Menlo Park, CA 94025, USA} \centerline{\it $^{e}$
Perimeter Institute for Theoretical Physics, Waterloo, Ontario, N2L 2Y5, Canada} \centerline{\affil $^{f}$ Department
of Physics, University of Waterloo, Waterloo, Ontario, N2L 3G1, Canada}
\bigskip
\bigskip
\centerline{\bf Abstract}\footnote{}{\smallskip {\cmsss SU-ITP-05-27  ~~UCB-PTH-05/30 ~~ LBNL-58913 ~~ DCPT-05/45 ~~
SLAC-PUB-11505}}

\smallskip
\baselineskip 16pt We study the AdS/CFT correspondence as a probe of inflation. We assume the existence
of a string landscape containing at least one stable AdS vacuum and a (nearby) metastable de Sitter state. Standard
arguments imply that the bulk physics in the vicinity of the AdS minimum is described by a boundary CFT.  We argue
that large enough bubbles of the dS phase, including those able to inflate, are described by mixed states in the CFT.
Inflating degrees of freedom are traced over and do not appear explicitly in the boundary description. They
nevertheless leave a distinct imprint on the mixed state.  In the supergravity approximation, analytic continuation connects AdS/CFT correlators to dS/CFT correlators.   This provides a framework for extracting further information as well.   Our work also shows that no
scattering process can create an inflating region, even by quantum tunneling, since a pure state can never evolve
into a mixed state under unitary evolution.

\bigskip
\smallskip

\Date{October 2005}

\baselineskip 14pt \listtoc \writetoc

\newsec{Introduction}

Our current understanding of the cosmological evolution of the universe relies on the existence of an early period of
inflation\foot{For reviews see \refs{\kolb, \linde, \liddle} . }. Recent data suggest that the universe is now
undergoing another period of inflation.  It is of central importance to understand this remarkable phenomenon as
deeply as possible.

\ifig\Vphi{ A typical scalar potential appearing in string theory,
with \dS\ and Anti-\dS\ minima.} {\epsfxsize=5cm \epsfysize=3.8cm
\epsfbox{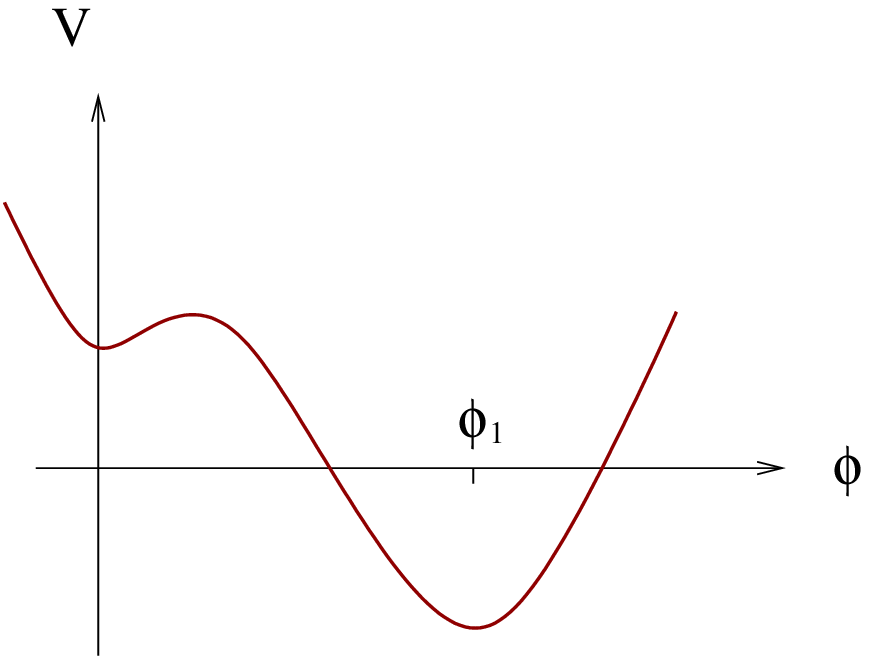}}

String theory, which currently is the only viable candidate for a
theory of quantum gravity, has had partial success in describing
inflationary physics.  In recent years there has been dramatic
progress in constructing string vacua with stabilized moduli and
positive and negative cosmological constants
\refs{\BoussoXA,\KachruAW, \KKLTexample, \nonsusyKKLT, \eva,
\Heterotic , \Antoniadis, \DeWolfeUU }. These lead to \dS\ and
Anti-de Sitter (AdS) cosmologies, respectively.  There are an
enormous number of such vacua, populating what is now called the
``string landscape'' \SusskindKW . A small piece of the landscape,
containing an AdS and a neighboring \dS\ vacuum, is sketched in
\Vphi. The richness of the landscape allows us to view the
parameters of such a potential as essentially free parameters.

These constructions do not answer many of the deep questions
raised by inflation:  How can inflation begin?  What measure
should be used on the multiverse of eternal inflation? What is the
holographic description of inflation? More generally, what degrees
of freedom are appropriate for a complete description of quantum
gravity in this domain?    A substantial amount of work has been
done on these topics \refs{\BanksFE, \BanksYP, \WittenKN,
\FischlerYJ,\HellermanYI,\BoussoNF, \StromingerPN}. We will not be
able to answer these questions in this paper, but we will try to
make some progress by embedding{  inflation in our best understood
and most powerful framework for understanding quantum gravity, the
AdS/CFT correspondence\foot{The first work on this connection is
\AlberghiKD.} \refs{\MaldacenaRE,\WittenQJ,\GubserBC,\AharonyTI}.

Consider a stable supersymmetric AdS ground state, say the one indicated in \Vphi.   The bulk correlators taken to
the AdS boundary define a conformal field theory (CFT), which encodes the bulk dynamics precisely. Given that small
fluctuations around the AdS minimum are captured by the CFT, it seems plausible that classical configurations
corresponding to the excursions to the neighboring \dS\ minimum should be encoded in the CFT somehow. For instance,
correlators of the scalar field $\phi$ describing the horizontal axis of \Vphi\ should enable one to reconstruct the
effective potential\foot{Of course, a CFT that captures aspects of the landscape as a whole must be a complicated
object indeed.}.  If we can construct a region of space where $\phi$ is displaced from the AdS minimum to the dS
minimum\foot{ The finite energy excitations described by the AdS/CFT correspondence require that $\phi$ approach the
AdS minimum at the AdS boundary.}, the behavior of such a bubble of false vacuum might probe some inflationary
physics.

The fate of such bubbles has been investigated extensively while exploring the possibility of ``creating a universe
in a laboratory'' \refs{\BlauCW,\FarhiTY,\FarhiYR}. We will discuss these results in more detail later, but for now
we will just summarize the main points.  Observed from the boundary of AdS (where the dual CFT is located),  all such
bubbles collapse into a black hole.  If the bubble was large enough initially, an inflating region forms, but it is
behind the black hole horizon.   At first glance this seems discouraging.  The standard AdS/CFT observables are only
sensitive to physics outside the horizon. But in recent years tools have been developed, based largely on
analyticity, to examine physics behind the horizon in AdS/CFT \refs{\MaldacenaRF, \LoukoTP, \KrausIV,\FidkowskiNF}.
We will discuss here how these tools enable us to observe the inflating region.

A basic obstruction to making a universe in a laboratory classically was noted by the authors of \FarhiTY. They
argued using singularity theorems in general relativity that an inflating region must classically always begin in a
singularity. But AdS/CFT quite comfortably describes geometries with both future and past singularities, like the eternal \SAdS\ black hole \refs{\MaldacenaKR,\BalasubramanianDE}. So the observations of \FarhiTY\ should not prevent us from studying inflation in AdS/CFT.

A more general worry about representing inflation in AdS/CFT is that the boundary CFT must encode a very large number
of degrees of freedom describing the inflating region. Specifically, the authors of \refs{\BanksXH, \BoussoTV} have
pointed out that in certain situations the dS entropy of the inflating region is larger than the entropy of the
\SAdS\ black hole. Notions of black hole complementarity and holography suggest that this would be hard to
accommodate.

In our picture this puzzle is resolved in a simple way. We present arguments that the geometries created by large
bubbles of false vacuum must be represented as {\it mixed} states in the boundary CFT. The large number of degrees of
freedom in the region behind the horizon are entangled with the degrees of freedom outside the horizon, as in the
Hartle-Hawking state representation of the eternal \SAdS\ black hole\foot{As we review later, the eternal \SAdS\
black hole is described as a pure entangled state in the Hilbert space of two copies of the boundary CFT, each living
on a separate AdS boundary. Tracing over one of these Hilbert spaces leads to a thermal density matrix in the other.}
\refs{\IsraelUR, \MaldacenaKR, \BalasubramanianDE}. The degrees of freedom behind the horizon are not explicitly represented;
they are traced over. They can be weakly entangled, though, so that even tracing over a large number of them can
yield a density matrix with entropy compatible with the black hole entropy.

We do not expect the degrees of freedom behind the horizon to be fully represented by a CFT, and do  not know how to
calculate the full density matrix beyond the supergravity approximation.  If we do know the density matrix,   a large amount of
information about the degrees of freedom that have been traced over can be extracted, again by using analyticity. As an example, in the eternal \SAdS\ black hole,
boundary operators on the right hand boundary can be moved to the left hand boundary by continuing in complex time.
In our situation, boundary operators on the right hand AdS boundary can be moved by a suitable continuation in
complex time to the \dS\ boundary at future (or past) infinity. The resulting correlators living on the boundary of
\dS\ have the form that one would expect from the dS/CFT correspondence \StromingerPN. This conclusion holds
classically. Quantum mechanically the \dS\ regions decay and the behavior is richer, and more mysterious.   But if these correlators can be defined nonperturbatively it should be possible to extract nonperturbative information about inflation from them.

The paper is organized as follows. In Section 2 we review the construction of false vacuum bubble spacetimes in the
thin wall approximation and discuss various aspects of these geometries.  In Section 3 we  describe the realization of
these spacetimes in AdS/CFT.  We consider the entropy puzzle and resolve it by  presenting arguments showing that the
CFT describing inflation must be in a mixed state.  We also discuss the general question of which geometries are
represented by mixed states. Section 4  deals with the signatures of inflating bubbles in the CFT correlation
functions. We  demonstrate that geodesic probes can sample the inflationary universe behind the horizon and describe
the signatures that can be gleaned from this analysis.  We discuss analytic continuation from the AdS to the \dS\  boundary.   In Section 5 we
revisit the idea of ``creating'' a universe in the laboratory from our perspective. We argue that because the CFT
dual of an inflating region is a mixed state it cannot be produced in any scattering process, including quantum
tunnelling, which is described by pure state evolution.  This agrees with some previous work \refs{\BanksNM, \lenny}.
We end with a discussion in Section 6. Some calculations are collected in appendices.

\newsec{Inflation in asymptotically AdS spacetimes}

Recently, the landscape of string theory compactifications has been shown to include both Anti-de Sitter and \dS\
minima \refs{\BoussoXA,\KachruAW, \KKLTexample, \nonsusyKKLT, \eva, \Heterotic , \Antoniadis, \DeWolfeUU }.  The
theory includes domain walls interpolating between these states, so one might expect that there are asymptotically
AdS spacetimes containing an inflating \dS\ region. For many classes of compactifications, the low energy theory is
effectively described by gravity coupled to a scalar field in a potential, as in  \Vphi. This effective potential
contains both positive and negative energy minima, with a domain wall given by field configurations interpolating
between two vacua.  Hence, by choosing carefully the initial profile of the scalar field and solving the equations of
motion, one can obtain asymptotically AdS spaces with inflating regions in the low energy effective gravitational
theory.

Although this effective model is much simpler than the full string theory, it is nevertheless quite complicated: even
solving for the exact spacetime metric requires messy numerical computations. So for much of this section we will
work in the thin wall approximation, where we can write down the metric exactly.  In this approximation we simply
match two pieces of known spacetimes together across an infinitesimally thin `domain wall', which obeys the
appropriate junction conditions \IsraelRT. The simplest requirement for this approximation to be valid is that the
width of the domain wall be less than the curvature length scales in  the geometry. This is easy to arrange. More
refined requirements will be discussed below. For  spherically symmetric spacetimes joined across a spherical shell,
the full configuration is likewise spherically symmetric, and may be viewed as a bubble of one spacetime inside the
other spacetime. Although the metric is continuous across the bubble wall, the extrinsic curvature is discontinuous
because the shell carries some energy. Einstein's equations then reduce to an effective equation of motion for this
shell.  We merely have to solve this equation to determine the shell's trajectory, and patch together the spacetimes
across the shell.

Spherical symmetry enables us to draw two-dimensional Penrose diagrams which encode the full causal structure of the
entire spacetime.  In addition, knowing the metric exactly will allow us to study the behavior of the geodesics,
Green's functions, \etc, in these spacetimes, which will ultimately be of use in extracting information about this
spacetime from its holographic dual.

We will first consider thin domain wall constructions, before broadening our discussion to include the more realistic
(scalar field) set-up towards the end. We begin by discussing what types of geometries are possible, specializing
mainly to a bubble of \dS\ inside \SAdS.  After explaining the construction and categorizing the various possible
cases, we focus on time-symmetric situations. As we will see, having a piece of \dS\ infinity (denoted $\scri$)
guarantees the existence of a \dS\ horizon; time symmetry then guarantees that its area is necessarily greater than
that of the black hole horizon.  As will be discussed in Section 3, this would seem to lead to an entropy paradox. There are also time asymmetric solutions with \dS\ $\scri$ -- in this case the \dS\ horizon may be larger or smaller
than the black hole horizon.

We will then illustrate explicitly that one can achieve essentially the same desirable ingredients (namely \dS\
$\scri$ hidden behind a horizon) for a scalar field in a suitably chosen potential\foot{ However, as we will see, the
nature of some parts of singularities and boundaries may be altered by instabilities.}. The latter is chosen by hand,
but motivated by the string landscape; we discuss what are reasonable landscape parameters to expect.

\subsec{Thin domain wall constructions}

We start by reviewing the procedure of patching together geometries across a thin junction in general relativity
\IsraelRT. This will allow us to construct classical solutions with both AdS and \dS\ regions (including $\scri$).
For simplicity, we consider only spherically symmetric geometries in four dimensions -- more general solutions are
considered in Appendix A.

We have a spherical shell, inside of which the metric is
\eqn\metin{ ds_i^2 = - f_i(r) \, dt_i^2 + {dr^2 \over f_i(r)} + r^2 \, d\Om^2 \ , }
and outside of which the metric is
\eqn\metout{ ds_o^2 = - f_o(r) \, dt_o^2 + {dr^2 \over f_o(r)} + r^2 \, d\Om^2 \ .  }
Note that having written the metrics in a static, spherically symmetric form, we must allow for the `time'
coordinates $t_{\io}$ ($\io = i,o$) to be different in each region, since this coordinate need not match across the
shell. On the other hand, $r$ is a physically meaningful coordinate -- it measures the proper size of the spheres of
a spherically symmetric spacetime -- and therefore has to vary continuously across the shell. Hence we can use the
same coordinate $r$ both inside and outside the shell.

The inside and outside geometries are patched together along a domain wall, with world-volume metric
\eqn\bubb{ ds_{bubble}^2 = -d\tau^2 + R(\tau)^2 d\Om^2 \ . }
Here $R(\tau)$ denotes the proper size of the shell as a function of its proper time $\tau$; in each part of the
spacetime its trajectory is given by $r=R(\tau)$. In the thin wall approximation we take the domain wall stress
tensor to be delta function localized on the wall surface. The equation of motion of the shell, which determines
$R(\tau)$, then follows from two matching conditions (for a review, see \eg\ \BlauCW). First, the metric must be
continuous across the domain wall. Second, the jump in extrinsic curvature across the wall is related to the stress
tensor of the bubble.  This implies that
\eqn\matchcond{ \sqrt{ \dot{R}^2 + f_i(R)} - \sqrt{ \dot{R}^2 + f_o(R)} = \kappa \, R  \ ,}
(with the sign of the radical determined by the extrinsic curvature -- see below), where $\dot{R} \equiv {dR \over d
\tau}$. The parameter $\kappa = 4 \, \pi \, G_{N} \, \s$ is related to the domain wall tension $\s$. By squaring
\matchcond\ twice, we obtain the radial equation of motion of the shell,
\eqn\shelleom{ \dot{R}^2 + \VR(R) = 0 \ ,}
with the effective potential
\eqn\Veffgen{ \VR(r) = f_o(r) - { \( f_i(r) -  f_o(r) - \kappa^2 \, r^2 \)^2 \over 4 \, \kappa^2 \, r^2 } \ .}
Equation \shelleom\ describes the one-dimensional motion of a point particle of zero energy in an effective potential
\Veffgen. Many properties of the geometry can be read off directly from the form of $\VR(r)$.  For example, if
$\VR(r) \to +\infty$ (or $\VR(r) \to C > 0$) as $r \to \infty$, the shell cannot reach the boundary.

We should note that the equation of motion \shelleom\ actually does not completely determine spacetime when, as in
\SAdS\ or \dS, $r$ is not a global coordinate\foot{In  spacetimes with horizons, such as \dS\ or \SAdS,  the static
coordinates of the type used in \metin, \metout\ are not globally well defined.  So two distinct points in the
spacetime can have the same value of $r$, $t$ and $\Omega$. Typically one can distinguish such points either by
passing to a good global coordinate chart, such as the Kruskal coordinates for \SAdS, or by a further specification
of the imaginary part of the time coordinate, $\Im(t)$.}.  This is because the effective potential \Veffgen\ was
obtained by squaring the equation for junction conditions twice, so we have lost some sign information.   In
particular, \shelleom\ does not distinguish between different points with the same value of $r$. To fix this, we must
take into account the extrinsic curvatures:
\eqn\betaio{ \b_i = {f_i(R) - f_o(R) + \kappa^2 \, R^2 \over 2 \, \kappa \, R } \  ,   \qquad  \qquad \b_o = {f_i(R)
- f_o(R) - \kappa^2 \, R^2 \over 2 \, \kappa \, R }  \ . }
Note that $\beta_{\a} = \pm \sqrt{\dot{R}^2 + f_{\a}(R)}$ automatically satisfy
\eqn\betaeom{ \beta_i - \beta_o = \kappa \, R \ .}
Physically, the extrinsic curvature is positive (negative) if the outward pointed normal points toward larger
(smaller) $r$. Hence, for a given trajectory of the domain wall (as given by \shelleom\ and \Veffgen), one can find
the extrinsic curvatures $\beta_{\a}$, and thereby determine which types of bubble trajectories are compatible and
which are inconsistent.  This allows us to construct the appropriate Penrose diagram.

We are interested in geometries describing a bubble of \dS\ in \SAdS. So the inner and outer metrics can be written
in static coordinates as \metin, \metout\ with
\eqn\metdSSAdS{ f_i(r) = 1 - \lam \, r^2 \ , \qquad  f_o(r) = 1 + r^2 - {\mu \over r} \ . }
The three independent parameters $\lam > 0$, $\mu > 0$ and $\g >0$ are related to the \dS\ cosmological constant, the
mass of the black hole and the tension of the shell\foot{ More precisely, in terms of the actual cosmological
constant $\Lambda$ and the ADM mass of the black hole $M$, $\lam = \Lambda / 3$ and $\mu = 2G_N\,M$.}. We will work in
units where the AdS radius is one. Then the relevant length scales are the AdS radius, the \dS\ radius $r_d =
1/\sqrt{\lam}$, and the black hole horizon radius $r_+$. The horizon radius $r_+$ is defined by $f_o(r_+) = 0$, so
for large $\mu$ we have $r_+ \sim \mu^{1/3}$.  We will now discuss the domain wall solutions found in this case -- a
more general family of solutions is described in Appendix A.

\ifig\Veffcases{Possible types of trajectories in an effective potential \VeffdSSAdS, with $\VR(r) \to - \infty$ both
as $r \to 0$ and as $r \to \infty$. {\bf (a):} $\Vmax > 0$, {\bf (b):} $\Vmax < 0$, and {\bf (c):} $\Vmax = 0$.
Trajectories $A$, $B$, and $D$ are time-symmetric (case $D$ describes a static shell), while the others are not
time-symmetric.} {\epsfxsize=13cm \epsfysize=4.5cm \epsfbox{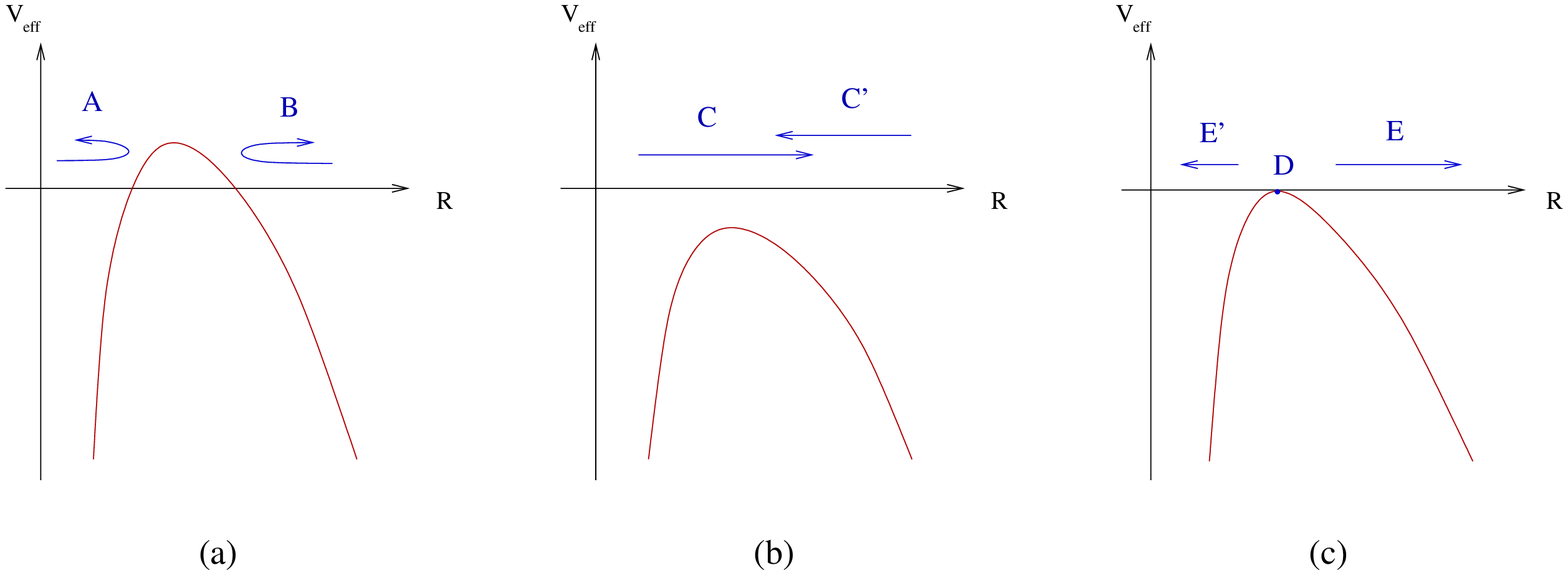}}

Evaluating \Veffgen\ for the specific case \metdSSAdS\ gives the following effective potential:
\eqn\VeffdSSAdS{ \VR(r) = - \[ { (\lam + \g^2 - 1)^2 + 4 \, \lam \over 4 \, \g^2} \] \, r^2 + 1
  + \mu \, {(1 + \lam - \g^2) \over 2 \, \g^2} \, {1 \over r}
  - {\mu^2 \over 4 \, \g^2} \, {1 \over r^4} \ . }
The behavior of the shell can be read off from this effective potential. Both the $r^2$ and the $1/r^4$ coefficients
are negative, so $\VR\to -\infty$ at $r\to 0$ and $r\to \infty$ and the potential has a maximum $\Vmax$ at some value
of $r$, say $r=r_0$. The possible domain wall trajectories $R(\tau)$ depend on the sign of $\Vmax$ (recall that the
effective `energy' is zero), as indicated in \Veffcases. If $\Vmax > 0$, as sketched in \Veffcases a, there are two
possible types of time symmetric situations: the shell can expand from zero size and recollapse (case A), or it can
contract from infinite size and re-expand (case B).  On the other hand, if $\Vmax < 0$ as in \Veffcases b, then no
time symmetric situation exists: the shell either expands (case C) or contracts (case C') on its semi-infinite
trajectory. Finally, if $\Vmax = 0$ as in \Veffcases c, then we can consider a static shell sitting at $R(\tau) =
r_0$ (case D). Such a case requires a certain fine-tuning of the parameters to obtain $\Vmax = 0$, as well as of the
initial conditions: $R(\tau_0)=r_0,\ {d \over d\tau}R(\tau_0)=0$.  If the latter is relaxed, the bubble may expand
forever (case E) or collapse (case E'). Of course, the time reverse where the bubble slowly settles to
$R(\tau\rightarrow\infty)=r_0$ is also possible.

We can now write down the Penrose diagrams for these various cases. Note that to do so we must take into account the
sign of the extrinsic curvature as mentioned above.  The details of this extrinsic curvature analysis are contained
in Appendix A -- we will simply quote the answers here. We will take the spacetime inside the bubble to be on the
left of the wall trajectory in the Penrose diagram, and the outside spacetime on the right.

\ifig\dSSAdSPDs{Sketches of Penrose diagrams for {\bf (a)} \dS,  {\bf (b)} \SAdS, and {\bf (c)} \dS/\SAdS\ domain
wall spacetimes, with constant-$r$ surfaces indicated.  The dashed vertical lines correspond to the points $r=0$ in
\dS, the dashed diagonal lines are horizons, the horizontal squiggly lines are the singularities, and the bold lines
indicate the boundaries.  The thick dotted lines indicate a possible trajectory of a shell across which the two
spacetimes (a) and (b) may be patched together to obtain (c).} {\epsfxsize=13.5cm \epsfysize=4.2cm
\epsfbox{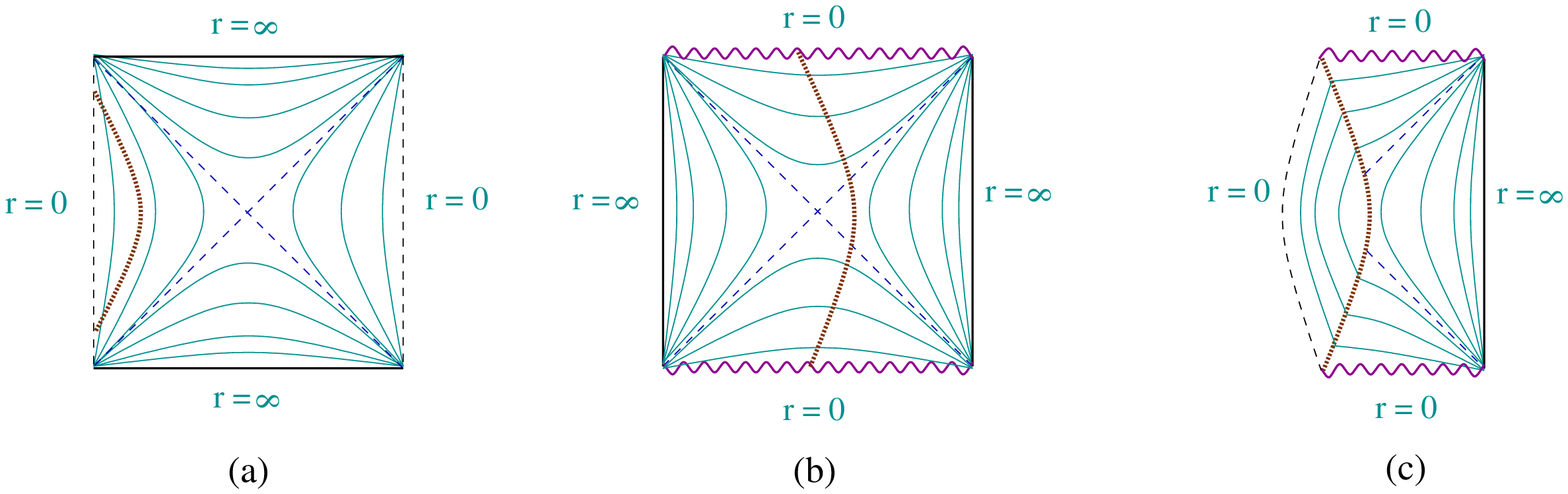}}

The Penrose diagrams\foot{These Penrose diagrams, as well as the constant-$r$ surfaces, are merely sketches; in
actuality, the singularity would be curved in, \etc, as in \FidkowskiNF\ for \SAdS. Since they nevertheless capture
many features of the causal structure, they are presented here and in the subsequent figures as sketches for ease of
visualization.} for \dS\ and \SAdS\ are given in \dSSAdSPDs a and \dSSAdSPDs b, along with constant-$r$ surfaces. In
the \dS\ geometry, $r$ increases from 0 at the `origin' (indicated by the dashed vertical lines in \dSSAdSPDs a),
through the cosmological horizon $r = r_d$ (diagonal dashed lines), to $r=\infty$ at \dS\ $\scri$ (bold horizontal
lines).  In \SAdS, on the other hand, $r=0$ at the singularities (indicated by the horizontal squiggly lines in
\dSSAdSPDs b), increases through the black hole horizon  $r= r_+$ (diagonal dashed lines), and becomes infinite at
the AdS boundary (bold vertical lines). A possible trajectory of the shell is further sketched on both spacetimes as
a thick dotted curve.  The corresponding junction spacetime is found by patching the two Penrose diagrams together
along the shell, as shown in \dSSAdSPDs c. Recall that the shell's trajectory must of course pass through the same
values of $r$ on both sides, given by $r=R(\tau)$. This means that if the shell starts out from zero size, expands,
and recontracts, its trajectory must correspondingly start and end on an origin of \dS, and on a singularity in
\SAdS, as sketched.  Note that in the resulting diagram (\dSSAdSPDs c), $r=0$ on the left, top, and bottom of the
diagram, and $r=\infty$ only on the right vertical line.

As is apparent from the Penrose diagrams, in a time symmetric set-up the shell reaches its maximum/minimum size $R_t$
at the $t=0$ slice (which passes horizontally through the middle of the diagrams and forms a symmetry axis).  In \dS,
this size $R_t$ is necessarily bounded from above by the \dS\ radius $r_d$ (\ie, $R_t \le r_d$), whereas in \SAdS,
$R_t$ is bounded from below by the black hole horizon $r_+$ (so that $r_+ \le R_t$).  We conclude that
\eqn\rrel{ r_+ \le r_d \qquad {\rm  for \ all \ time \ symmetric \ configurations.} }
It immediately follows that the black hole entropy is smaller than the \dS\ entropy for time-symmetric domain wall
configurations; the implications of this surprising fact are discussed in the next section. However, we will see that
there also exist time asymmetric solutions where the the black hole  entropy is larger than the \dS\ entropy.

\ifig\dSSAdSpossibsAB{Shell trajectory corresponding to the time-symmetric cases (A) and (B) of \Veffcases, sketched
on the \dS\ and \SAdS\ Penrose diagrams.} {\epsfxsize=12cm \epsfysize=7.1cm
\epsfbox{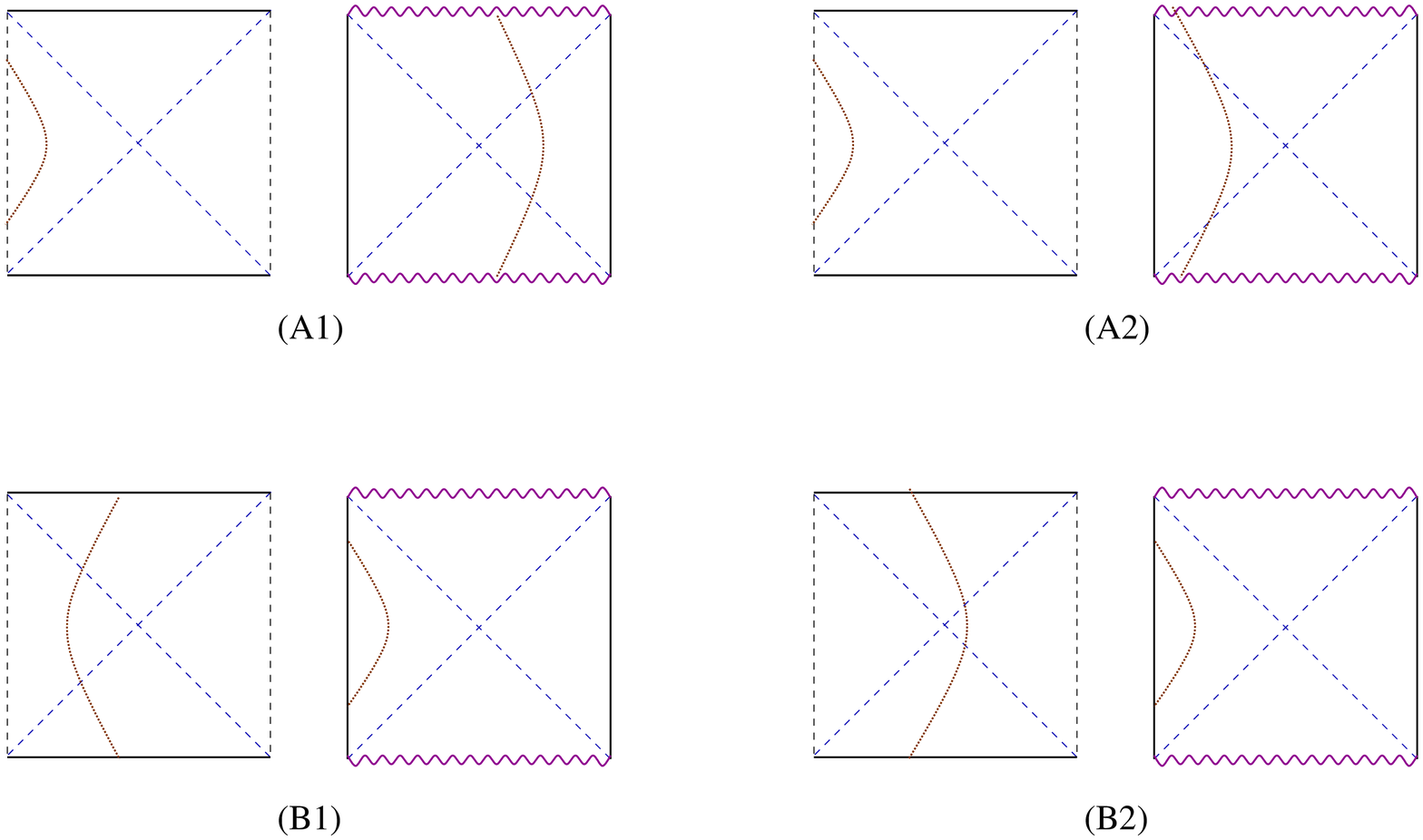}}
\ifig\dSSAdSpossibsCD{Shell trajectory corresponding to cases (C), (D) and (E) of \Veffcases, sketched on the \dS\
and \SAdS\ Penrose diagrams. In all cases, we glue together the left spacetime and the right spacetime across the
domain wall, discarding part of each diagram.} {\epsfxsize=12cm \epsfysize=7.1cm
\epsfbox{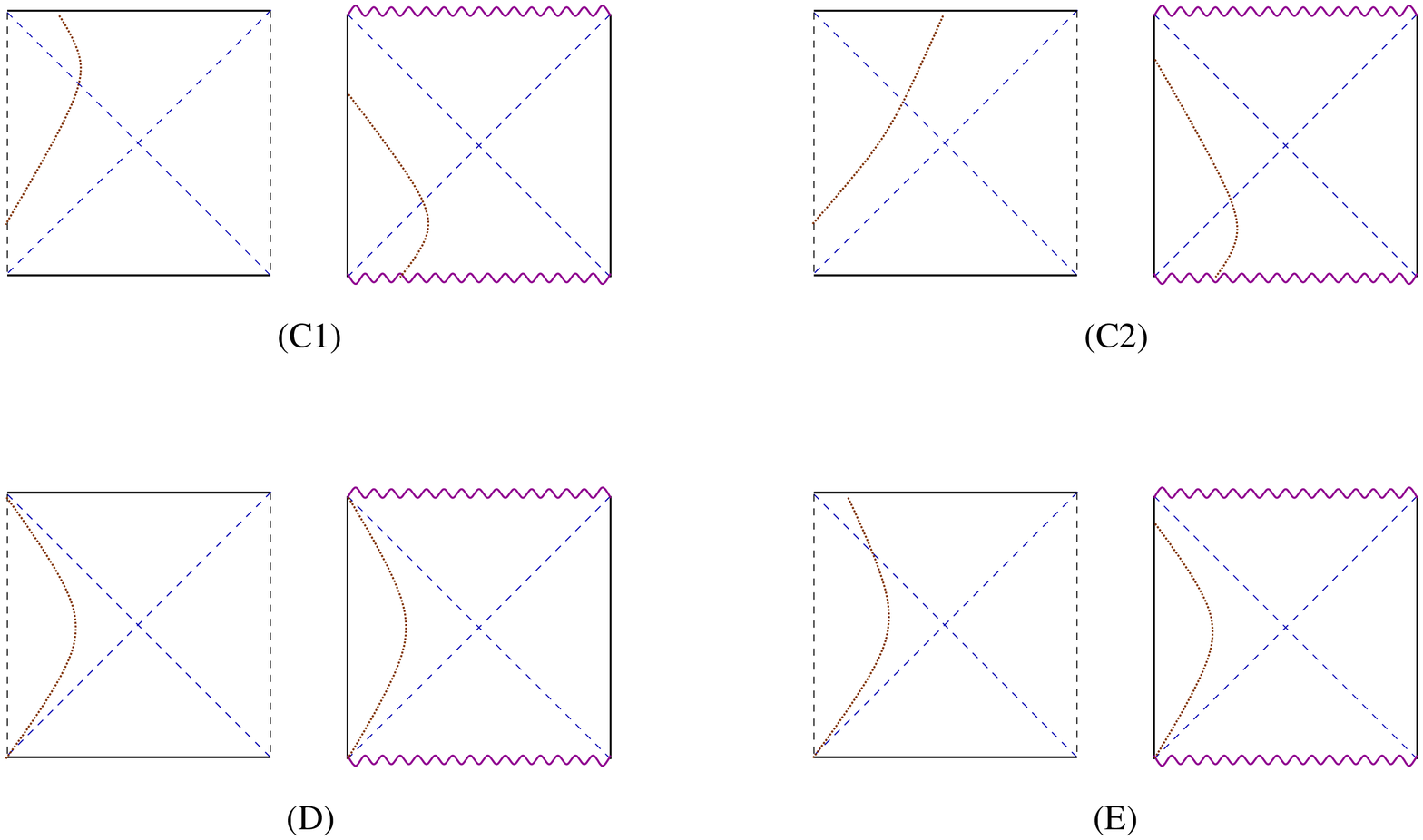}}

We will now examine in greater detail what types of trajectories (and corresponding Penrose diagrams) are admissible.
The effective potentials for the different cases indicated in \Veffcases, along with distinct possibilities for the
extrinsic curvatures, are plotted in \sampleVeff\ in Appendix A. The time symmetric trajectories, corresponding to
the top two cases (A and B) drawn in \sampleVeff, are depicted on the spacetime diagrams in \dSSAdSpossibsAB.  The
distinguishing feature between A1 (B1) and A2 (B2) is the sign of $\b_o$ ($\b_i$) at the turning point; the former
include fewer bifurcation points. Similarly, the time asymmetric trajectories, described by the cases (C) and (E) of
\sampleVeff, as well as the static case (D), are shown on the respective spacetimes in  \dSSAdSpossibsCD.  Along with
(C1) and (C2), which start from $r=0$ and expand forever, we could of course also have their time reverse, as
indicated by case (C') of \Veffcases; and similarly for case (E).  As previously, the distinction between (C1) and
(C2) comes from the sign of $\b_i$ at large $r$. Case (D) is somewhat special: it corresponds to a static shell. Here
we can have a global Killing field which is timelike everywhere outside the horizons. However, this geometry does not
contain a piece of \dS\ $\scri$.

\ifig\dSSAdSpossPDs{Sketches of the full Penrose diagrams combined from the corresponding cases of \dSSAdSpossibsAB\
and \dSSAdSpossibsCD. Metrically, the space on the left of the shell (thick dotted curve) is \dS, while the space on
the right of the shell in \SAdS.} {\epsfxsize=11cm \epsfysize=7cm \epsfbox{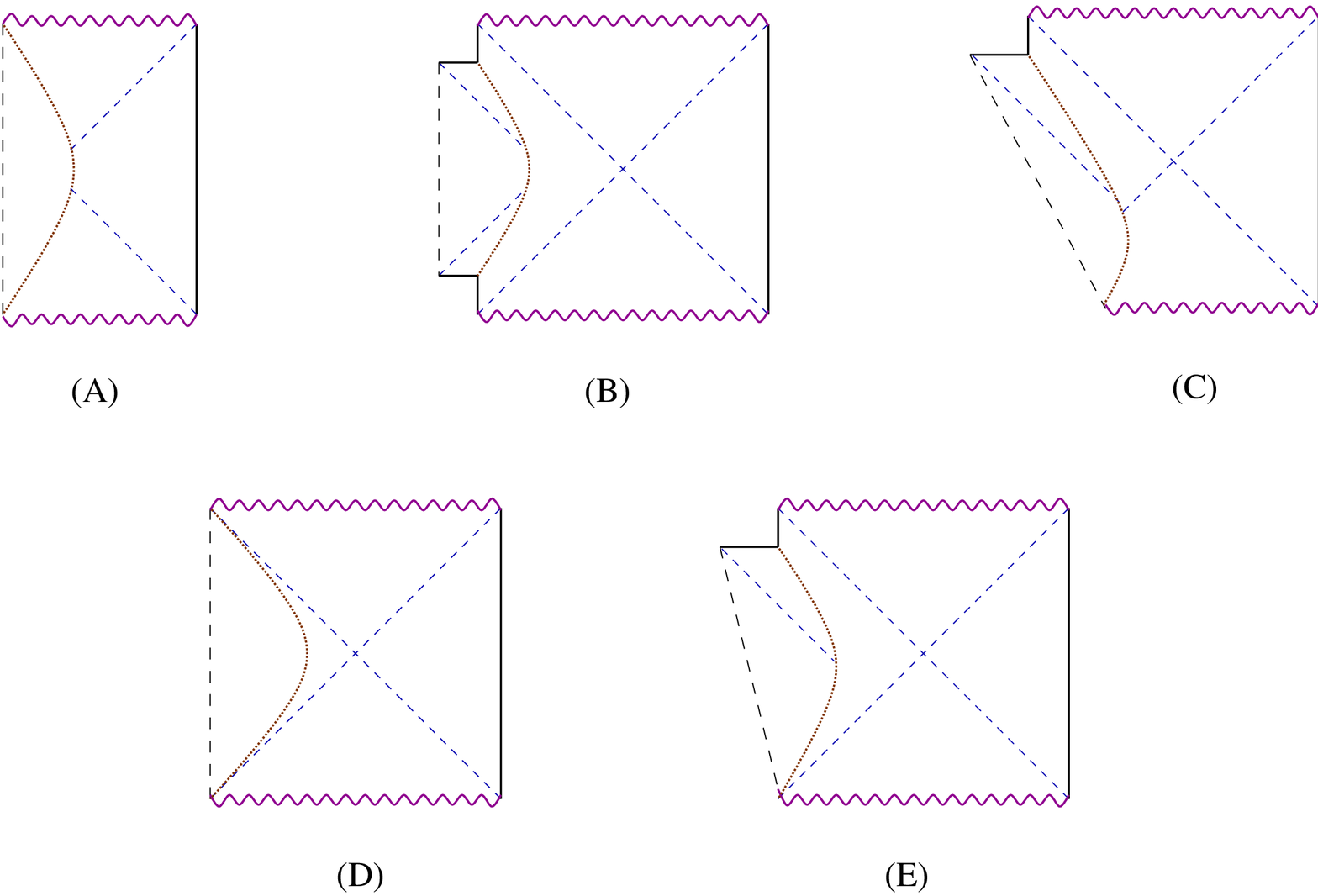}}

We can now combine Penrose diagrams for the full junction \dS/\SAdS\ spacetime, as in \dSSAdSPDs c. The result is sketched in \dSSAdSpossPDs.

Of the geometries described above, case A is an example of a false vacuum bubble that is excited in the true vacuum
which re-collapses. These geometries are very similar to time symmetric spacetimes representing black hole collapse,
except that the interior geometry is one with a different value of the cosmological constant. Cases B, C, E are the
most interesting ones from our perspective, since here we see the presence of an inflating region of spacetime with
\dS\ $\scrip$. These are the geometries we will be interested in describing holographically from the boundary field
theory living on the AdS boundary on the right. A crucial feature in these geometries is that the inflating region is
hidden behind a black hole horizon from the AdS boundary. We shall later show that this situation is generic as long
as the matter fields making up the domain wall satisfy the null energy condition.

Note that in case C, the \dS\ horizon is not necessarily larger than the \SAdS\ horizon. In particular, which area is
bigger depends on whether the \dS\ horizon (as drawn by the left diagonal dashed line in \dSSAdSpossPDs C) crosses
the shell earlier or later than the black hole horizon. Since the radial coordinate increases monotonically along the
shell, if the \dS\ horizon intersects the shell before the black hole horizon (at smaller $r$ and lower on the
Penrose diagram), then $r_d < r_+$; conversely, if it intersects later, the \dS\ is bigger.  Which of these is the
case depends on the specific values of the parameters, but both possibilities are allowed\foot{ As explained in
Appendix C, if in addition we require an initial Cauchy slice whose area increases monotonically and whose \dS\ part
has domain of dependence which contains a piece of the \dS\ $\scri$, then we necessarily obtain $r_d < r_+$.}.
\ifig\pdcont{ Sketch of Penrose diagrams obtained by continuous deformations of the initial data and parameters,
starting with pure AdS and ending with \dS\ (including the future and past boundary $\scri^{\pm}$) in \SAdS.
Note that (b) is possible only for small black holes $r_+ < r_A$,  and in (f) the \dS\ $\scri$ is generically joined
to the \SAdS\ singularities by some metric which depends sensitively on the evolution (and hence drawn by a dotted
line) but is unimportant for our discussion.} {\epsfxsize=11cm \epsfysize=7cm \epsfbox{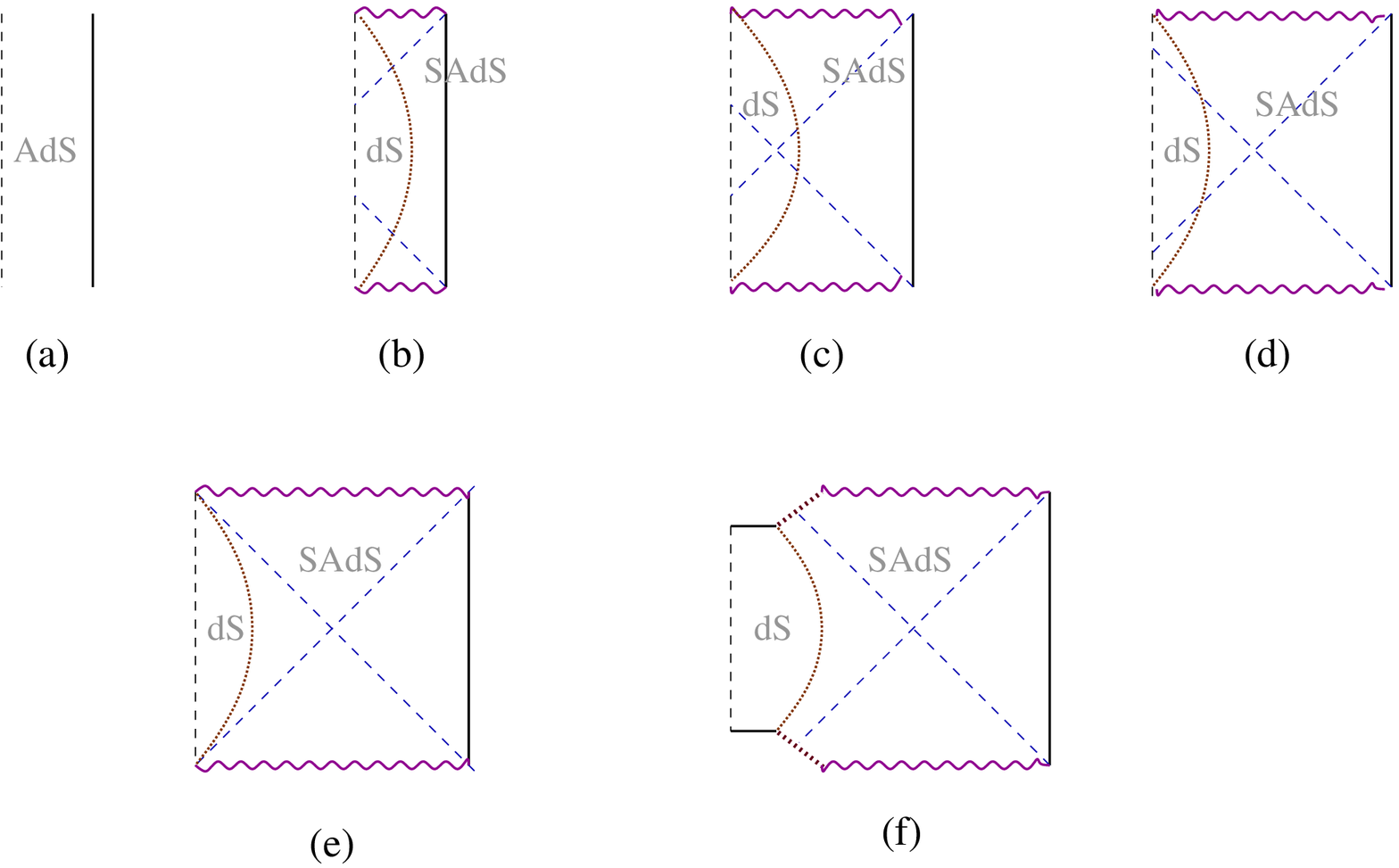}}

It is possible to start with pure AdS space and smoothly deform parameters to obtain a geometry with an inflating
region. A sequence of spacetimes which illustrates this is shown in \pdcont.  We can deform from one spacetime to the
next in \pdcont\ by smoothly adjusting the bulk initial data -- in particular, the size of the false vacuum bubble --
on the $t=0$ spacelike  slice. Note that although the local geometry does not vary much in this progression, the
global properties vary substantially from case to case.  Namely, (a) is causally trivial; in (b) the spacetime has an
event horizon; (c) acquires regions which are causally disconnected from the boundary; in (d) the shell itself passes
through such a region; in (e) the entire shell is causally disconnected from the boundary; and finally (f) acquires
additional (\dS) asymptotic regions\foot{For later discussions, we also note that running across the $t=0$ Cauchy
slice of (d), (e) or (f) the size of spheres does not vary monotonically. In fact, for (e) and (f), the same is true
for any Cauchy surface.}. Naively, one might expect that if (a) is described by a holographic dual, then so will (f),
since it seems unnatural that the holographic encoding would cease abruptly in this progression. However, as we argue
later, the nature of the state may change.

\subsec{Beyond the thin wall approximation}

We have seen that in the thin domain approximation we can find geometries that contain both \dS\ $\scri$ and AdS
$\scri$. Of course, we are really interested in the case of gravity coupled to a scalar field in a potential of the
form sketched in \Vphi. As described above, such potentials describe low energy dynamics on the string theory
landscape.  Moreover, the resulting spacetimes will be smooth.  In Appendix B we will give a detailed argument that
the basic features of the geometry do not change in this more general setup.  In particular, one can argue based on
causality that there are solutions of scalar-gravity that contain both the \dS\ and AdS $\scri$. In this section we will discuss a few characteristic features of these solutions.

Perhaps the most important feature is the fact that the \dS\ $\scri$ is causally disconnected from the AdS $\scri$ --
there is no null geodesic connecting the two.  This is a very general property of any spacetime satisfying the null
energy condition, including the scalar-gravity system under consideration.  This can be seen from Raychaudhuri's
equation for a congruence of null geodesics.  Physically, because gravity is attractive, once a set of null geodesics
start converging, they cannot diverge (unless they pass through an origin $r=0$).  This immediately rules out null
geodesics connecting the \dS\ and AdS $\scri$: a null congruence must converge to go into the bulk from an AdS
$\scri$, and diverge to reach the \dS\ $\scri$ from the bulk.

Another curious property of some of the thin wall constructions is
the presence of a part of AdS boundary on the left which
disappears and appears as the shell attains infinite size. This
second AdS boundary is an artifact of the thin wall approximation,
and does not appear in the full solutions of scalar-gravity
\BoussoTV.  Physically, any radiation emitted by the shell near
the boundary would suffer a large blue-shift as it propagates into
the bulk, and its backreaction would lead to a curvature
singularity. This is somewhat analogous to the Cauchy horizon
instability at the inner horizon of a charged or rotating black
hole. While the exact nature of this singularity is of course
difficult to determine due to its sensitivity to the initial
domain wall profile, we expect that it will meet up with the
\SAdS\ black  hole singularity in a(n almost) null fashion, as
sketched in \pdcont f. This can also be thought of as due to
cosmic censorship\foot{ We thank Gary Horowitz for pointing this
out to us. }, because of an obstacle to Cauchy evolution (though
by a boundary rather than a naked singularity). As is well known,
asymptotically AdS spacetimes are not globally hyperbolic without
specification of additional boundary conditions. Hence the
appearance of an AdS boundary implies the formation of a Cauchy
horizon for the evolution of the initial data as we cannot evolve
the spacetime in the domain of influence of this boundary\foot{
This is not an issue on the right AdS boundary, since there the
CFT tells us what boundary conditions to impose.}. Now similarly,
given generic perturbations on a Cauchy surface at finite time, we
expect that a big bang singularity will remove any such AdS
boundary in the far past. Hence it appears that these AdS
boundaries are also artifacts.

So far we have limited our discussion to spherically symmetric
geometries.  However, there is a possibility that some of these
geometries are dynamically unstable to aspherical
fluctuations\foot{We thank John McGreevy for alerting us to this
possibility.}. In particular, while the \SAdS\ and the \dS\
geometries are individually dynamically stable to fluctuations,
the shell itself might be unstable. If one considers the positions
of individual pieces of the shell as determined by the effective
potential $\VR$, the shell's deformations will grow with time.
This is because in the mechanical motion of particles away from
the extremum of the effective potential \VeffdSSAdS\ two particles
tend to accelerate away from each other. A set-up similar to ours
has been recently considered in \AguirreXS, where it was shown
that certain \dS /Schwarzschild-de Sitter domain walls are indeed
unstable. Further it appears that changing the outside geometry
there from Schwarzschild-de Sitter to \SAdS\ does not remove this
instability \JohnsonPC. It would be interesting to analyze this
potential instability in detail for our set-up. We do not expect
the instabilities to radically alter our story -- for example, in
the solutions with de Sitter $\scri$, we expect the aspherical
perturbations to remain small compared to the size of the bubble,
as in \refs{\vil, \vilg}. Finally, we should emphasize that many
of the cases considered above are fine-tuned, in the sense that
they have been chosen to be time symmetric.

\subsec{A special parameter domain}

The arguments of the following sections will be sharpest for a certain range of parameters.  This regime is given by
$r_d \ge R_t \gg r_+ \gg r_A = 1 \gg \l_s$ where, as a reminder, $\l_s$ is the string length, $r_d =
1/\sqrt{\lambda}$ is the \dS\ radius of curvature, $r_+$ is the black hole horizon radius, which for large $\mu$ is
given by $r_+ = \mu^{1/3}$, $R_t$ is the domain wall position of the time symmetric solutions (cases A, B) at the
turning point, and $r_A$ is the AdS curvature radius (which we have set to one).  In this range of parameters the
domain wall is very far away from the black hole horizon at all times, and
causally disconnected from the right AdS boundary. Solving for the
turning point of the effective potential \VeffdSSAdS, we find
\eqn\matchcondregimetwo{R_t \approx r_+/(1-\kappa^2)^{1/3} \ .}
We see that to achieve the condition $R_t \gg r_+$,  $\kappa$ must be close to one. Recall that $\kappa \sim \sigma
r_A /m_p^2$ where $\sigma$ is the tension of the domain wall.  A brief survey of known parts of the landscape reveals
regions with $\kappa \gg 1$ and regions with $\kappa \ll 1$.  There is no reason not to expect many vacua with
$\kappa \sim 1$.  Typically, if $\kappa \ll 1 $ then $R_t \rightarrow r_+$.  If $\kappa \gg 1$ then $R_t \rightarrow
r_+$ and $r_+\rightarrow 1/\kappa$. Physically $\kappa \sim 1$ in AdS units is special because then the domain wall
tension balances against the pressure which the true vacuum exerts on the domain wall. In flat space, such a balance
would only be possible for one size of the bubble because the energy due to tension is proportional to the surface
area of the domain wall, while the energy due to the pressure is proportional to the volume. In AdS, the two forces
can almost cancel for a large range of bubble sizes because at scales big compared to the AdS radius, volume is proportional to surface area. For the special case of the static domain wall, we must also set the first derivative of the effective potential to
zero, giving $\mu \sim 1/(1-\kappa^2)^2$ and $R_t \sim 1/(1-\kappa^2)$.

The curvature at the maximum of the effective potential \VeffdSSAdS\ goes to zero as $\kappa \rightarrow 1$ and
$\lambda \to 0 $.  The instability growth rate of the static domain wall is governed by this curvature and hence goes
to zero in this limit. As $\kappa \rightarrow 1$  the static domain wall becomes arbitrarily far away from the black
hole horizon and its instability becomes arbitrarily small. Note that the relevant time scale for the instability is
the Schwarzschild time of the asymptotic observer $t_o$ and not the proper time of the shell $\tau$.  The conversion
factor at large $R_t$ is $\tau \sim t_o \, (R_t/r_A)$. In a typical regime in parameter space, we have $t_{\rm
instability} \sim r_d\, r_A/R_t$, which can be taken to be large.

\newsec{Properties of the boundary CFT}

In the previous section we constructed a family of \dS/\SAdS\ domain wall spacetimes, shown in \dSSAdSpossPDs, some
of which have inflating (\dS\ $\scri$) regions. These solutions all have asymptotically AdS regions. Consider first
the pure AdS geometry in {\dSSAdSpossPDs} describing the stable ground state\foot{To ensure stability, we may take
this to be a supersymmetric minimum.} of \Vphi. If this is a vacuum of a consistent theory of quantum gravity (as we
are assuming) then this theory defines a boundary CFT which represents the bulk via the AdS/CFT
correspondence\foot{This CFT should in principle contain information about all accessible vacua in the landscape and
hence will be an extraordinarily complicated object. We expect that the wide separation of tunneling time scales will
enable us to focus on the truncated landscape of \Vphi.}. The scalar field representing the horizontal axis in \Vphi\
will be represented in the CFT and it seems plausible that one could excite a large number of its quanta to
create the initial data for the \dS/\SAdS\ spacetimes\foot{All geometrical scales can be taken much longer than
string length and the coupling can be taken weak, so these gravity solutions should approximate the behavior of the
full theory.}. However there is a puzzle about representing \dS\ degrees of freedom that at first glance casts doubt on this simple logic. This has to do with the fact that it naively appears that the boundary CFT must encode the dynamics of an enormous inflating region with far fewer active degrees of freedom accessible to it.

As we will see below, the resolution to this puzzle is that the
boundary CFT is in a mixed rather than a pure state. Some mixed
states arise by simply integrating out certain degrees of
freedom in the CFT \EmparanGF. Conversely, certain pure states are
expected to mimic mixed states to a high degree of accuracy
\refs{\BalasubramanianKK,\BalasubramanianMG}. We should emphasize that
here we are claiming that the relevant mixed states arise
because the boundary CFT is entangled with new degrees of freedom associated with \dS\ region behind the horizon.
This is analogous to the appearance of the thermal density matrix
in the standard eternal black hole \refs{\MaldacenaKR,
\BalasubramanianDE}.  We will argue that the entangled degrees of freedom
can be described as a cutoff CFT coupled to additional non-CFT modes.  
The effect of the non-CFT modes can be made parametrically small, demonstrating that the state is mixed.
\subsec{The entropy puzzle}

The arguments presented above seem to imply that the field theory dual to the geometries which incorporate an
inflating region is constructed by acting on the vacuum state with an appropriate set of local operators. This would
lead us to conclude that the spacetime geometry is dual to the field theory in a particular pure state.

This picture however cannot be right, as it leads to a paradox regarding the entropy \refs{\BanksXH, \BoussoTV}. Let
us consider a \dS\ bubble with $\scrip$. As explained at \rrel, in the time-symmetric set-up, the size of the bubble
on the $t = 0$ slice, $R_t$, is smaller than the \dS\ radius and also larger than the black hole radius $r_+$. Hence
we know that the black hole size $r_+$ is necessarily smaller than the size of the \dS\ cosmological horizon $r_d$,
implying that the black hole entropy is smaller than the entropy associated with the \dS\ false vacuum bubble.

The black hole entropy is associated with the number of active degrees of freedom in the boundary conformal field
theory. This is the picture that is naturally suggested by black hole microstate counting using D-brane
constructions. The \dS\ entropy, on the other hand, is a measure of the degrees of freedom necessary to define a
quantum gravitational theory in \dS\ space \refs{\GibbonsMU}. Given this, it is hard to imagine how a pure state that
is built out of the fewer black hole degrees of freedom can encapsulate the information  required to describe the
\dS\ space. This mismatch is what we term the {\it entropy puzzle}.

In fact, it is easy to see that this entropy mismatch can be made arbitrarily large --- after all, there is no
restriction on the allowed \dS\ size at the level of classical geometries, since we could consider arbitrarily small
positive values of cosmological constant. In a sense, we would have to use the vastly fewer degrees of freedom
accessible to the boundary observer to describe the physics of arbitrarily many degrees of freedom.   There exist
(time asymmetric) geometries where the \dS\ entropy is less than the black hole entropy.  We present the details of
these solutions in Appendix C. Although the entropy puzzle is not present for these special cases, for time symmetric
solutions the resolution lies in a different direction\foot{ For a different viewpoint on the entropy issue see
\JacobsonMI.}.

We argued above that the cosmological solutions with asymptotically AdS regions  should be described by a boundary
CFT. However, the assumption that the geometry with an inflating region is described by a pure state of the CFT leads
to an entropy puzzle. Since all of the solutions constructed in Section 2 were found by matching onto \SAdS\
solutions, the AdS boundaries are separated from the inflating region by a black-hole horizon. Furthermore, as
discussed earlier, this feature is guaranteed in any construction in classical general relativity with matter obeying
the null energy condition. It is therefore natural to expect that the boundary CFT is very similar to the thermal
field theory dual to a standard \SAdS\ black hole. A crucial feature of these spacetimes is that the conformal theory
living on a boundary is in a mixed state, with density matrix
\eqn\thermalrho{ \rho_\b=  e^{-\beta \,  H}  }
rather than a pure state\foot{Strictly speaking, this is true only for large black holes, with $r_+ > r_A$.  We will
focus on this region of parameter space.} \refs{\MaldacenaBW,\MaldacenaKR,\BalasubramanianDE}. Here $H$ is the Hamiltonian of
the CFT and $\b$ the inverse temperature. In this section we will argue that the boundary CFTs dual to inflating
geometries are also in a mixed state, whose density matrix differs from \thermalrho\ only by small corrections.

In section 3.2 we will start with a discussion of mixed states in various extensions of \SAdS, before moving on to a
more general discussion of mixed states and causal structure in section 3.3.

\subsec{Mixed states in asymptotically \SAdS\ geometries}

We will start with the most symmetric example of domain wall spacetimes, the static domain wall, shown in \pdcont e.
In this geometry the domain wall is at a fixed radial position $R(\tau) =r_0 =R_t$. Despite the fine tuning necessary
to attain this geometry, it serves as a simple example to illustrate the general principle we wish to propose.  In
fact, as discussed in section 2.3, by taking $\kappa \rightarrow 1$ we can dial $R_t \gg r_+ \gg r_A$ and make the size of the instability vanishingly small. In the thin wall approximation, the spacetime to the right of the domain
wall is identically \SAdS. If the location of the domain wall is far removed from the black hole horizon, $R_t \gg
r_+$, we have a large region of spacetime where the usual picture of a \SAdS\ black hole should hold.

We will first consider the region of the spacetime in the \SAdS\ part of the geometry with $r \le R_t$, \ie, imposing
a cutoff at a radial scale $r_c \sim R_t$. To describe the physics of just this cutoff spacetime in the dual field
theory, recall that for the usual eternal \SAdS\ black hole (see \dSSAdSPDs b) with $r_+ > r_A$ the field theory dual
is best described in the thermofield formulation. One associates a complete CFT Hilbert space to each AdS boundary of
the black hole, labeled $\CH_L$ and $\CH_R$ respectively. These Hilbert spaces are non-interacting and the geometry
is dual to a particular entangled pure state, the Hartle-Hawking state, in the tensor product Hilbert space $\CH_L
\otimes \CH_R$ \refs{\IsraelUR,\MaldacenaKR,\BalasubramanianDE}. Tracing over one of the Hilbert spaces, say $\CH_L$, leads to
a self-contained description in $\CH_R$, but in a mixed state. The density matrix is the thermal density matrix
\thermalrho\ at the black hole temperature.

Physics in a cutoff \SAdS\ background is very similar to that of
the non-cutoff geometry. Now, however, the dual CFT is replaced by
a conformal field theory cutoff at energy $E_c \sim r_c/r_A^2$. We denote the corresponding Hilbert spaces as $\CH_{L,R}^c$.
Concentrating on energy scales below $E_c$, we see that the
entangled state description in $\CH_L^c \otimes \CH_R^c$ is still
valid. So at low energies the right hand field theory will remain
in a mixed state, which is now found by entangling $\CH_R^c$ with
the cutoff theory coupled to gravity. Of course, this procedure is
ambiguous at scales near or above $E_c$, but at low energies the
density matrix is given approximately by \thermalrho, with
corrections that vanish as powers of $E/E_c$. Further, it is clear
that the mixed state description is the correct one in $\CH_R
\supset \CH_R^c$ so long as energy locality holds.  This is a
consequence of bulk locality (in $r$).

These arguments are best controlled when $R_t \gg r_+$ obtained by taking $\kappa \rightarrow 1$.  But as long as the
domain wall is some macroscopic distance from the horizon, a macroscopic fraction of the excited degrees of freedom
on the left should be entangled with those on the right, yielding a mixed state with macroscopic entropy of
entanglement.  The above arguments have been made in the $\mu >1$ ``large'' black hole regime.  But it seems  likely
that even small black holes are described by entangled states and so these considerations should also apply to the
$\mu < 1$ regime as well.

Even at scales below the cutoff, the form of the density matrix is not unambiguously defined.  In particular, the
effective action of the cutoff CFT will include some number of irrelevant operators whose presence becomes important
only at energies approaching $E_c$. Some of these effects can be calculated using bulk supergravity techniques.

We have not yet discussed the effect of the  de Sitter region to the left of the domain wall.  For all time symmetric
collapse and static geometries the de Sitter radius $r_d$ must be larger than $R_t$.  So we can treat the de Sitter
region as a piece of essentially flat space.   This results in significant modifications to the cutoff CFT because
bulk massless propagators in flat space decay like powers of the proper distance between points, while in AdS space
they decay exponentially.   So bulk massless fields can (and do) induce nonlocal terms in the effective CFT.  But we
will show that these nonlocal effects can be made arbitrarily small by taking the cutoff surface defining the CFT
much smaller than the domain wall.

To be specific take the horizon, cutoff and domain wall radii to have the following relative sizes: $ R_t \gg r_c \gg
r_+$.  Take the most extreme case, a massless bulk field dual to a marginal operator $\CO$ in the CFT.  Ignoring the de
Sitter contribution we have
\eqn\ooone{ \langle \CO \, \CO' \rangle \sim {1\over r_c^{6} \,L^6} \ .}
Here $L$ is the geodesic distance between $\CO$ and $\CO'$ on the
cutoff surface.  This formula follows from conformal invariance
for marginal operators in a $D=3$ CFT or equivalently from summing
over bulk particle paths in the AdS region to build up the
massless propagator.

The de Sitter contribution behaves differently.  Particle paths
contributing to this behavior traverse a region of AdS space to
the domain wall, then propagate in the de Sitter region before
re-entering the AdS region and returning to the cutoff surface.   We can
calculate this contribution as the product of three propagators.
Two account for the AdS propagation, each behaving like
$(r_c/R_t)^3$.  One accounts for the propagation through de Sitter
space.  This goes like $1/x^2$, the standard massless particle
propagator in four dimensions, 
where $x$ is the distance between points. One can then fold
these propagators together and integrate over joining points on
the domain wall, as described in Appendix D.  
We find that the massless propagation in the nearly flat de Sitter
region induces a correction to the correlator of order $(R_t L)^{-2}$,
which is long range compared to \ooone. Hence the \dS\ bubble introduces ``non-local'' modifications which
become appreciable in the infrared on scales longer than
$L^2 \, > \, R_t/r_c^3$. These corrections appear because, essentially, modes which are
non-normalizable in the full \SAdS\ geometry become normalizable
when the AdS boundary is "cut off" and replaced by the \dS\
bubble. This would have produced a massless graviton coupled the
cut-off CFT \refs{\RandallEE,\GubserVJ} had we been considering
the analogous constructions in higher dimensions.  
These modes are part of the non-CFT degrees of freedom necessary
to describe the region beyond the cut-off surface.  When $r_c \sim R_t$ the 
de Sitter region makes a large nonlocal modification to the CFT.  But if we 
choose $R_t \gg r_+^3 $ this correction is  small compared to \ooone.   
So in this regime we can continue to
make a controlled argument that the theory is in an entangled
state.

While we have focused the discussion above on the case of the
static domain wall, a similar situation can occur for any of the
cases where the \dS\ bubble wall passes through the region to the
left of the black hole, \eg, the cases b, c, d and e in
\dSSAdSpossPDs. In particular consider \dSSAdSpossPDs b, where the
bubble expands in the far past and future. If the minimum bubble
radius is much bigger than the radius of the black hole horizon,
$R_t \gg r_+$, there is once again a large region to the right of
the domain wall (and to the left of the horizon) where the
geometry is \SAdS, and physics may be described in terms of a
cutoff CFT. The discussion must be refined for the cases where the
bubble shrinks towards either the future or past. However, it is
clear that with some tuning, a large portion of \SAdS\ is relevant
and a cutoff CFT can describe the physics for some large interval
of time.

The arguments discussed above are our strongest evidence for the mixed state nature of the CFT description of
inflation.  

To summarize, the CFT dual to geometries with an inflating false vacuum bubble is necessarily in a mixed state. In
particular, this means that the  active degrees of freedom in the boundary field theory, whose number is given by
$\exp(S_{bh})$, are entangled in a non-trivial way with the degrees of freedom in the inflating region (which, as we
have argued before, could be much larger).  In this picture, the black hole entropy $S_{bh}$ is simply a measure of
this entanglement. The boundary observer who evaluates correlation functions in the state dual to this geometry will
conclude that the theory is in a mixed state with density matrix $\rho_{\rm bdy}$ and an entanglement entropy $S_{bh}
= S_{ent} = \Tr ( \rho_{\rm bdy} \, \log \rho_{\rm bdy}) $. In our picture, the large number of \dS\ degrees of
freedom are entangled with the black hole degrees of freedom.  However, bulk locality suggests that they are
entangled very weakly. So when these degrees of freedom are traced over, the resulting entanglement entropy $S_{ent}$
is much smaller than $S_{dS}$. Thus the mixed state picture avoids the entropy puzzle described above.

One striking aspect of this picture is the absence, say in the static domain wall case of \pdcont e, of a second
asymptotic boundary where the traced over degrees of freedom can be localized.  This is in contrast to the eternal
\SAdS\ black hole, where one traces over the degrees of freedom associated to one of the conformal boundaries. So it
is natural to ask how one should describe the degrees of freedom that are entangled with the boundary CFT.  One clue
comes from the eternal \SAdS\ black hole, where bulk fields $\phi(r,t)$ in the right hand quadrant (see \imtconv b)
can be moved to the left hand quadrant by shifting $t$ by half a Euclidean period, $- i \beta/2$. Usually this
transformation is used in the $r \rightarrow \infty$ limit where $\phi$ becomes an operator in the boundary CFT. This
shift then relates the two boundary CFTs. But we can consider finite $r$ bulk fields as well. These can be
constructed from the CFT fields by suitable coarse graining\foot{ See \refs{\BanksDD,\HamiltonJU,\GiddingsQU} for
examples of such coarse graining.}. For values of $r < R_t$, bulk fields in the left region, which can be described
by the cutoff CFT, are related by this imaginary shift in $t$ to fields in the right region. So these degrees of
freedom are accessible via analytic continuation.  We will extend these considerations in Section 4.

The fact that a large number of degrees of freedom in the inflating region are entangled, albeit weakly, with CFT
degrees of freedom allows us to use the latter to infer some properties of the former. We clearly cannot reconstruct
all the information pertaining to inflation, but by virtue of the entangled state construction we have access to some
of the information. Before proceeding to discuss how this information may be encoded in the boundary field theory, we
turn to an interesting question: what are the situations in which the boundary theory is in a mixed state?

\subsec{Conditions for the appearance of mixed states}

We have argued that a broad class of \dS\ bubble solutions must
correspond to mixed states in the CFT. However, we now wish to
consider to what extent these arguments can be applied to other
solutions, like the rapidly collapsing shells. {\it A prori} it is
not clear whether the boundary CFT is in a mixed or a pure state.
We will now proceed to discuss conditions which may delineate when
a boundary CFT will be in a mixed state. We will consider more
general situations than the cosmological solutions described
above, and describe several possible scenarios under which mixed
states arise. We will try to formulate certain criteria for the
appearance of mixed states; while some of these appear to be
sufficient to guarantee a mixed state description, we are unable
to determine which of these is necessary.

We will start by reviewing the bulk\foot{We will distinguish bulk quantum field theory data by an explicit $b$
superscript.} explanation for the appearance of mixed states. According to the AdS/CFT correspondence, correlators of
local operators in boundary field theories are found by taking bulk correlation functions to the boundary and
stripping off the appropriate powers of radial coordinate. When there are regions in the spacetime that are causally
disconnected from the boundary -- \ie, regions that are outside both the past and future light cones of the boundary
-- then typically these bulk correlation functions are evaluated in a mixed state.

To see this, consider a quantum field $\phi$ in an asymptotically AdS spacetime with a moment of time symmetry $t\to
-t$.  The Hilbert space of this scalar field can be written in a position space basis at time $t=0$.  When the
spacetime contains a region causally disconnected from the boundary, the bulk Hilbert space can be factorized into
two pieces $\CH^b=\CH_R^b\otimes \CH_L^b$, where $\CH_R^b$ is spanned by operators located inside the causal wedge of
the boundary and $\CH_L^b$ is spanned by operators in the causally disconnected region. To calculate the expectation
value of local boundary operators, we only need to calculate bulk operators inside the light cone of the boundary.
These operators act trivially on $\CH_L^b$, so they are evaluated in the mixed state found by tracing over $\CH_L^b$:
\eqn\rhor{ \rho_R^b  = \Tr_{\CH_L^b} \, |\psi^b \rangle \, \langle \psi^b| .}
Here $|\psi^b \rangle$ denotes the (pure) state of the quantum field $\phi$.  {\it A priori}, $|\psi^b\rangle$ might
be of the form $|\psi^b_L\rangle \otimes |\psi^b_R\rangle \in \CH_L^b \otimes \CH_R^b$, in which case $\rho_R^b$ has
zero entropy and  describes a pure state.  However, one can show that if this is the  case then quantum backreaction
will be large near the horizon, destroying the spacetime. This is a familiar fact for black hole or Rindler horizons
(see \eg, \Birrell). In black hole geometries the Boulware vacuum factorizes, and leads to a divergent stress tensor
at the horizon.  The same is true for the Rindler vacuum of an accelerating observer. More generally, if the state
$|\psi^b \rangle$ factorizes then the expectation value of a set of local operators jumps discontinuously as one of
the operators moves across the bifurcation point.  In particular, such a correlation function vanishes unless all of
the operators are located on the same side of the bifurcation point.  So the value of the stress tensor, which can be
found by differentiating a two point function, will typically diverge.  We conclude that the density matrix \rhor\
describes a genuine mixed state with non-zero entropy. In general for any CFT observable which is supported only in some region B of the boundary, one can show that it is fully determined by the part of the bulk spacetime which corresponds to the causal wedge of the region B \MarolfFY.

To summarize, if the standard AdS/CFT bulk to boundary dictionary is assumed in spacetimes with causally disconnected
regions, we arrive at the following criterion\foot{See Appendix E for a critique of this criterion.}
\blocktext{Criterion 1: Correlators of local operators in a
boundary CFT are evaluated in a mixed state if there exist regions
of spacetime that are causally disconnected from the boundary.}
We should note that although the description in terms of a single
boundary CFT is as mixed state, there may be additional
descriptions of the geometry in terms of a pure state.  For
example, in the eternal \SAdS\ geometry discussed above the
spacetime is described by two boundary CFTs in a particular pure
entangled state.  It is only by tracing out degrees of freedom on
one side that one obtains the mixed state description of
correlators on the other boundary\foot{There is reason to expect
that this behavior is a general feature of spacetimes with
multiple asymptotic AdS boundaries. In general, solutions to
Einstein's equations contain multiple asymptotic AdS boundaries
only under very specific circumstances
\refs{\WittenXP,\GallowayBP, \GallowayBR}.  Typically, such
solutions have singularities in both the far past and the far
future and the conformal boundaries are causally disconnected.
Thus the boundary CFTs do not interact, but are evaluated in an
entangled state of the form described here.}. We should emphasize
that the description of the \SAdS\ geometry as a thermal state
with density matrix $\rho_\b$ on the right boundary only
determines bulk correlators in the causal wedge of the right
boundary.  It does not, for example, unambiguously fix correlators
of operators near  the left boundary or correlators between
operators in the left and right causal regions. This is because
there are many choices of pure state $|\psi\rangle$ in the
boundary Hilbert space $\CH = \CH_L \otimes \CH_R$ which lead to
the same density matrix $\rho_R$ upon tracing over $\CH_L$. To
describe the entire geometry, one needs to specify the pure state
$|\psi\rangle$; typically it is specified by the Euclidean path
integral with appropriate boundary conditions. It is only once one
specifies $|\psi\rangle$ that one can, using, \eg, analytic
properties, relate correlators in the entire spacetime to those in
a single boundary CFT.

In more general spacetimes, such as the inflating geometries
described above, it is not clear how to describe the mixed
boundary state as a pure entangled state of a larger theory (or
indeed whether such a pure state description exists).  Given the
fact that mixed state correlators unambiguously fix correlators
only in the causal region, it is necessary to make an additional
assumption in order to extract behind the horizon physics.  As we
will describe in the next section, we will typically assume
analyticity in the gravity description, which in many cases
amounts to defining a bulk state $|\psi^b\rangle$ on a complete
Cauchy surface by Euclidean continuation. This leaves as implicit
the construction of a pure entangled state in the dual holographic
theory. 

One may wish to conjecture a stronger criterion where ``if" is replaced by ``if and only if". However, \EmparanGF\ considers a mixed state described by pure AdS space but where the foliation lends itself to tracing over the CFT degrees of freedom on half of the boundary. Certainly here there are no regions causally disconnected from the full boundary. Rather it is only that degrees of freedom on different components of the same boundary are entangled. Hence a stronger conjecture might be made as
\blocktext{Criterion 1': Correlators of local operators in a
boundary CFT are evaluated in a mixed state if and only if there
exist regions of spacetime that are causally disconnected from the
corresponding boundary components.}
One drawback of either of the criteria outlined above is that the
presence of a causally disconnected region is a global property of
the spacetime. In AdS/CFT, one expects bulk Cauchy evolution to
correspond to Hamiltonian evolution in the boundary. One is
therefore tempted to conclude that the prescription of Cauchy data
on a spacelike slice suffices to determine the state of the
boundary theory.  This criterion depends only on the behavior in
the neighborhood of a spacelike slice, and only very indirectly on
global properties of the spacetime. It should not be necessary to
evolve the data, and then infer from this bulk evolution that the
spacetime has a  causally disconnected region, to conclude that
dual CFT state is mixed. Furthermore, there appear to be explicit
examples with causally disconnected regions which are nevertheless
described by pure boundary states -- these are discussed in
Appendix E.

We have been careful to focus the above discussion on local
operators, whose correlation functions are easily extracted from
bulk correlators.  However, there are other possible criteria for
the existence of mixed states which do not rely on bulk field
theory arguments. For example one can motivate a criterion based
on the scales the boundary observer can probe, which we expect by
ideas of holographic renormalization to be related to the proper
size of the spheres in geometries with spherical symmetry.
Consider a spacelike slice (say along $t =0$) with metric
\eqn\asd{ ds^2_{t=0} = dr^2 + R(r)^2 \, d\Omega^2 \ . }
Geodesics with angular momentum $L$ in the geometry \asd\ achieve
a minimum radial scale, $R_{min} = L$, before turning back. If we
consider correlation functions of high dimension operators at
fixed large $r$, the two point function is given by the geodesic
length\foot{Strictly speaking, these geodesics appear as local
extrema of a path integral, and an additional calculation is
needed to determine where they are the dominant contribution.
We'll ignore this subtlety for now.}. Since the geodesic turns
around it only samples part of the geometry \asd. Further,
however, the variation of the corresponding correlator with $L$ is
a sharp probe sampling the geometry at $R_{min}$ and implicitly
the CFT at the corresponding energy scale. If $R(r)$ were
monotonic, in the limit $L \to 0$ we can probe any radial interval
down to $R(r)=0$. In contrast, if $R(r)$ is not monotonic, there
will be intervals which this probe cannot access. This restriction
on the scales that can be probed by the boundary correlators is
evocative of the c-theorem, if we assume that the size of the
spheres is a sensible measure of the effective number of degrees
of freedom. This is suggestive then that whenever we have a Cauchy
slice with a non-monotonic proper size for the spheres (as for
example in the eternal \SAdS\ black hole) there are degrees of
freedom that are not accessible to the boundary observer. One can
therefore conjecture an alternate criterion for a mixed state
description in the boundary field theory:
\blocktext{Criterion 2: When the spacetime has spherical symmetry,
boundary correlators are evaluated in a mixed state only when the
radial sizes of the spheres are non-monotonic along the spacelike
slice at $t=0$.}
We should emphasize that this differs from the first criterion
described above.  In particular, there are spacetimes where $R(r)$
is monotonic, but nevertheless the regions of small $R$ are
causally disconnected. An example is a collapse spacetime, where a
shell of matter is sent in from asymptotic infinity to create a
black hole in the interior, as discussed in Appendix E. Another
distinction is that this criterion is local in time -- Cauchy data
alone suffices to determine whether or not a state is pure. Of
course, implicitly here we are only considering time symmetric
configurations. As discussed in Appendix C, there are more general
solutions where Cauchy slices may be chosen to have the proper
sizes of spheres varying either monotonically or non-monotonically. 
It is far from clear how to extend this criterion to such cases.

This criterion was derived from holographic considerations and relies strongly on the choice of spacelike slices. One
might therefore attempt to formulate a more covariant criterion, in terms of null slices rather than spatial slices.
Demanding that the sizes of spheres be monotonic along lightlike rather than spacelike directions is equivalent to
the condition that there be no additional holographic screens, as defined by Bousso \BoussoCB. It is straightforward
to construct asymptotically AdS spacetimes for which this criterion differs from the other two, so for completeness
we summarize this as
\blocktext{Criterion 3: The correlators of operators in a boundary CFT are evaluated in a mixed state only if the
spacetime has additional holographic screens. }
The criterion conjectured here should probably be refined as one
finds additional holographic screens (beyond the AdS boundary) for
any collapsing bubbles (even \pdcont b). Naively, at least, very
light bubbles such as this would be in a pure state. Of course,
this just underscores our problem that holography in general
spacetimes remains poorly understood.

We have not undertaken an extensive classification of the
differences between these three criteria. It is however easy to
construct examples where we obtain differing results depending on
the criterion chosen. For example, in the progression sketched in
\pdcont, Criterion 3 applies to cases (b) - (f), Criterion 1 to
cases (c) - (f) and Criterion 2 to cases (d) - (f). As should be
apparent, we are unable to offer a single criterion that is both
necessary and sufficient for the CFT to be in a mixed state. This
remains an interesting open problem.

\newsec{Probes of inflation in AdS/CFT}

In the preceding sections we have shown that it is possible to construct, within the classical approximation, domain
wall spacetimes that interpolate between \dS\ and AdS.  In some cases, one can obtain a large inflating region, which
is separated from an asymptotic AdS boundary by a black hole horizon. We have argued that in these cases the dual
boundary theory is in a mixed state. In this section we will discuss the extent to which boundary CFT observables,
calculated in this mixed state,  contain information about the inflating region\foot{Throughout this section, we consider only the effects that
arise due to field propagation in the background \dS-\SAdS\
spacetime. Since the vev of the scalar field is shifted inside
the \dS\ bubble, the masses of particles may change in this region
because of couplings to the scalar --- for large mass particles
this should be a minor effect.  There may also be mixings between
different species. So in general the phyiscs will be more
complicated than we have considered.} .

At first glance, we might worry that the mixed state results from tracing over the degrees of freedom behind the
horizon, including the ones describing inflation, and so no vestige of these degrees of freedom will be visible. This
is not the case\foot{Heuristically, this can be seen by considering an entangled state in a tensor product of two
identical Hilbert spaces, each of dimension $d$,  $d \gg 1$.   While an entangled pure state in the product space is
described by $d^2$ complex parameters, The density matrix obtained by tracing over one Hilbert space is described by
$d^2/2$ complex parameters.  Roughly speaking, the density matrix can pin down ``half" of the entangled state.
Unitary transformations of the traced over space leave the density matrix unchanged and account for hidden
information. There are only $d$ independent pure states in each individual Hilbert space. The density matrix carries
far more information than the selection of a pure state. }. In fact, the traced-over degrees of freedom leave a
substantial imprint on the mixed state which can be analyzed. One well known example of this is the eternal \SAdS\
black hole where correlators in the thermal density matrix describe the degrees of freedom on the right AdS boundary.
Analytically continuing the arguments of operators in imaginary time gives correlators describing degrees of freedom
on the left boundary\foot{ For reviews of these analyticity concepts see, for example, \refs{\MaldacenaRF,
\KrausIV,\FidkowskiNF} }.  In the following section we will use analyticity to probe degrees of freedom in the inflating region. We expect correlators to be analytic for a wide variety of physically interesting states, including those under discussion here.

In section 4.1 we will follow the strategy of \refs{\LoukoTP,  \KrausIV, \FidkowskiNF} to probe physics behind the
horizon, and examine correlation functions of high dimension operators in the boundary theory. For such operators, a
boundary two point function is given by the length of the bulk spacelike geodesic which connects the two points. We
will show that the presence of \dS\ $\Scri^{+}$ leads to a singularity in the correlation functions of the boundary
theory, after appropriate analytic continuation. In section 4.2 we will describe an even more dramatic probe of the
inflating region. In the classical approximation, the boundary CFT can be analytically continued to construct dS/CFT
correlators living on the asymptotic \dS\ $\scri^\pm$. Once quantum effects are included this provides a powerful
probe of the non-perturbative physics of the inflating region.

 \subsec{Geodesics probes of domain wall spacetimes}

We will start by considering correlation functions of boundary operators $\CO(x)$ with large dimension $\Delta$
describing bulk particles with mass $m \sim \Delta$. In the limit where $m$ is large the two point function can be
evaluated in semiclassical approximation and is given by
\eqn\corrhk{ \langle \, \CO(x) \, \CO(y) \, \rangle \sim e^{- m \, \len \(x,y\)} \ } where $\len(x,y )$ is the
proper length of the spacelike geodesic connecting the two points on the boundary\foot{ There are a few subtleties in
this argument, which we will mostly neglect in the following. In particular, one must prove that the geodesics under
consideration lie on the path of steepest descent in order to contribute to the correlator. Even if this is not the
case, however, the effect described below will still be visible upon  analytic continuation of correlation functions.
To track the ``metastable'' geodesic reliably requires $m \rightarrow \infty$. The heaviest particles available as
probes are wrapped D-branes with masses of order $1/g_s$.  So this technique requires the ability to take $g_s
\rightarrow 0$.  There are examples in the landscape \DeWolfeUU\ where this is possible.}. This length is formally
infinite, so $\len\ $ is regularized by taking $x$ and $y$ slightly away from the boundary. In this section we will
focus on radial geodesics, whose form can be found by patching together geodesics on either side of the domain wall
along with an appropriate junction condition across the wall.

For radial geodesics in spacetimes of the form \metin,  the
geodesic equations are
\eqn\dwsplike{{dt_\a \over d \tau_g} = {E_\a \over f_\a(r)} \ ,
\qquad  \({dr \over d\tau_g}\)^2  = f_\a(r) + E_\a^2\ , }
where $\a={i,o}$ and $\tau_g$ is the affine parameter along the
geodesic. $E_\a$ is a conserved quantity associated with the 
Killing vectors $\left({\partial \over\partial t_\a}\right)^a$ on
either side of the domain wall. To trace geodesics through the
junction, we assume that the domain wall is transparent \ie, an
observer on the wall measures the same energy and momentum for a
geodesic on both sides. We also use that $r$ is continuous across
the junction. On the other hand, $t$ jumps discontinuously across
the domain wall. We can determine the jump in $t$ from the
normalization of the 4-velocity $u^a$ of the shell:
\eqn\tjump{ u^a \, u_a = -1 = -f_\a(r) \, \td_\a^2 + {\rd^2 \over f_\a(r)} \ .}
Using $\rd^2 = -\VR(r)$ gives
\eqn\tofrshell{ t_\a(r) = - \int {\b_\a(r) \over f_\a(r) \, \sqrt{-\VR(r) }} \, dr \ .}
To find the explicit relation between $t_i$ and $t_o$ we would have to invert \tofrshell.  Moreover, to determine
where a particular spacelike geodesic intersects the bubble trajectory, we would have to solve the implicit equations
$R(\tau) = r(\tau_g)$ and $t_{\rm bubble}(\tau) = t_\a(\tau_g)$.  Finally, to obtain $t_\a(\tau_g)$ and $r(\tau_g)$,
we must invert the expressions for $\tau_g(t)$ and $\tau_g(r)$ from \dwsplike. Therefore, except in very special
cases, one can not write a closed form expression for the geodesic.

\ifig\dSSAdSnullgeod{ A few null geodesics in thin domain wall spacetimes. {\bf (a)} Collapsing false vacuum bubble
geometry {\bf (b)} Inflating false vacuum bubble geometry. } {\epsfxsize=10cm \epsfysize=5.1cm
\epsfbox{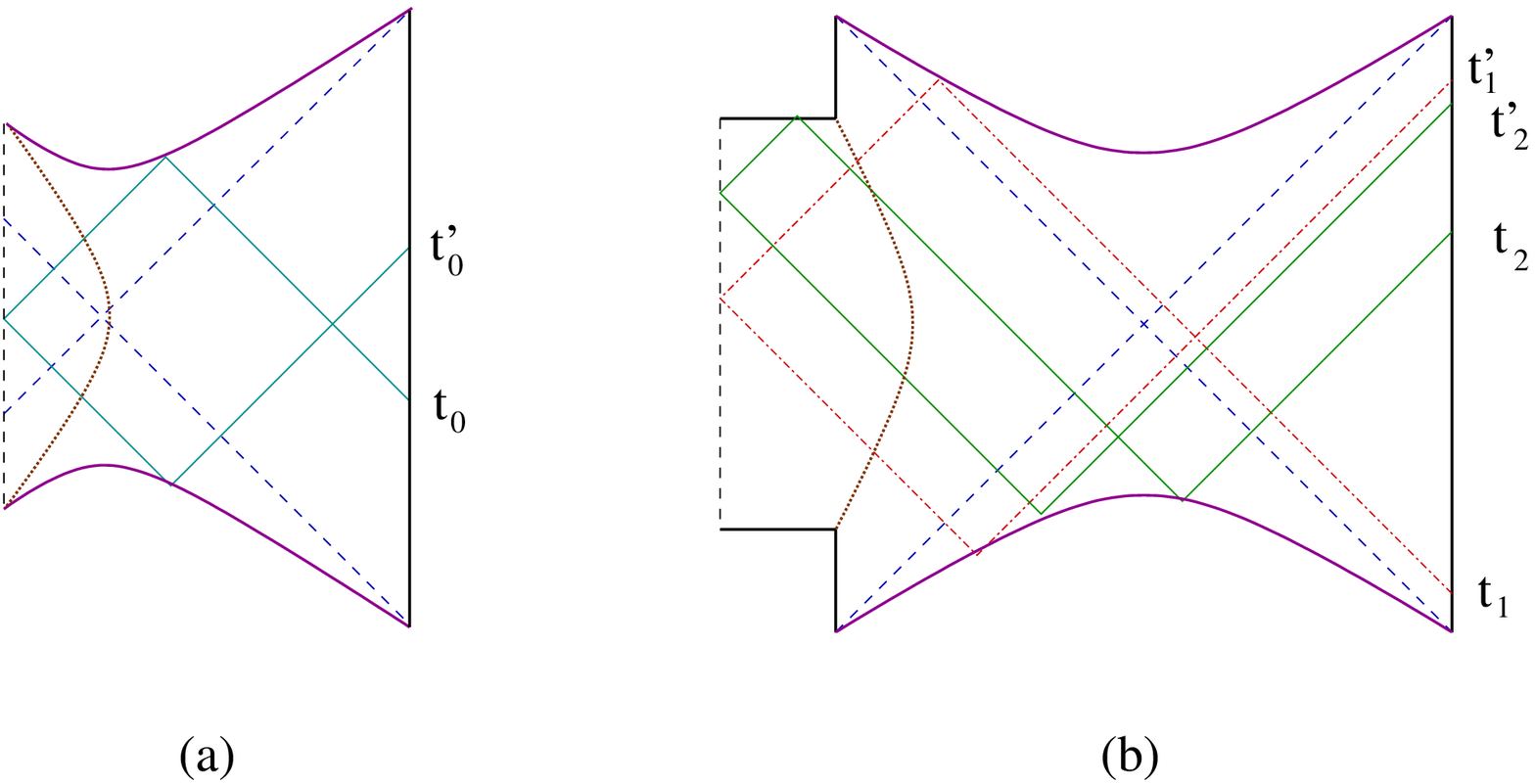}}

For certain special geodesics, $\len\ $ vanishes -- \ie, the geodesic becomes null -- and the two point function
$\langle \CO(x) \, \CO(y) \, \rangle$ has a singularity. Such null geodesics are found by taking the $E \to \infty$ limit of
the spacelike geodesics described above. In this case it is simple to solve for the geodesics explicitly. The Penrose
diagrams for two domain wall geometries, along with a few null geodesics, are sketched in \dSSAdSnullgeod. While the
black hole singularities are drawn curved \FidkowskiNF, on a true Penrose diagram, \dS\ origin and $\scri^\pm$ would
also bend.

The null geodesics \dSSAdSnullgeod\ are drawn as bouncing off the black hole singularity and \dS\ $\scri$.  This is
because our null geodesics arise as a limit of spacelike geodesics, which are repelled from both the singularity and
\dS\ $\scri$. To see this, note that near the $r=0$ singularity of \SAdS, the geodesic equations \dwsplike\ become
\eqn\bhrepulse{ \({dr \over d \tau_g}\)^2 = E^2 - {\mu \over r} \;\; \Longrightarrow \;\;  {dr \over d \tau_g}=0 \;\;
{\rm and} \;\; {d^2r \over d \tau_g^2} > 0 \;\; {\rm at} \;\; r = {\mu \over E^2} \ . }
We conclude that a spacelike geodesic is repelled by the singularity at a distance $\rmin = {\mu \over E^2}$.
Likewise, near the \dS\ $\Scri^+$ the geodesic equation becomes
\eqn\dsrepulse{ \({dr \over d \tau_g}\)^2 = E^2 - {r^2 \over r_d^2} \;\; \Longrightarrow \;\;{dr \over d \tau_g}=0
\;\; {\rm and} \;\;  {d^2r \over d \tau_g^2} < 0 \;\; {\rm at} \;\; r = E \, r_d \ . }
So a spacelike geodesic turns around at maximum radius $\rmax = E \, r_d$. This property, that spacelike geodesics
are repelled by \dS\ $\Scri^{\pm}$, is analogous to the fact that timelike geodesics are repelled by the timelike
boundary of AdS. In the limit where the geodesic becomes null, $E\to \infty$, the geodesics simply bounce off the
singularity and \dS\ $\scri$.

We conclude that CFT correlators will have additional singularities due to null geodesics behind the horizon, of the
form
\eqn\corrpred{ \langle \, \CO(t,\Om) \, \CO(s, -\Om ) \, \rangle \sim {1\over \( s - t'(t)\)^{2 \, m} }\ , }
where $t'(t)$ indicates the point where the null geodesic starting
at $t$ re-emerges on the boundary and depends on the particulars
of the geometry. The operators are at antipodal points on the
sphere because any geodesic that returns back to the AdS boundary
has to pass through the origin of \dS, where it will move to the
opposite side of the sphere.

The singularities \corrpred\ seen in the analytically continued
correlators are not time translation invariant, rather variations
in the separation $t'(t)$ and overall coefficient reveal the
interesting dynamics of the \dS\ bubble in the left causal region.
One particularly strong signal is as follows. The inflating
geometry in \dSSAdSnullgeod b gives rise to two classes of null
geodesics: those associated with the singularity at $t'_2(t_2)$,
which reflect off the \dS\ $\Scri^{+}$, and those associated with
$t'_1(t_1)$, which miss the \dS\ $\Scri^{+}$ and bounce off the
black hole singularity.  So one will find that the singular
behavior associated with $t'_1(t_1)$ is extinguished for a certain range of $t$ as the
corresponding geodesic passes the junction of the \dS\ and AdS
boundaries. Note that while we argued earlier that once
back-reaction is included this junction is replaced by a big
crunch singularity, the latter singularity does not repel
geodesics. Hence the extinction of the singularity should be even
more pronounced. This behavior will not be found in the analytically continued
correlators corresponding to non-inflating geometries like
\dSSAdSnullgeod a, so provides a distinct signature of 
inflation in the boundary CFT.

\subsec{From AdS/CFT to dS/CFT and beyond}

There are even more powerful probes of the inflating region available in this system, which build on analyticity.

\ifig\imtconv{ Conventions for imaginary part of the time coordinate. {\bf (a)} \dS\ spacetime {\bf (b)} \SAdS\
geometry.} {\epsfxsize=9cm \epsfysize=4.9cm \epsfbox{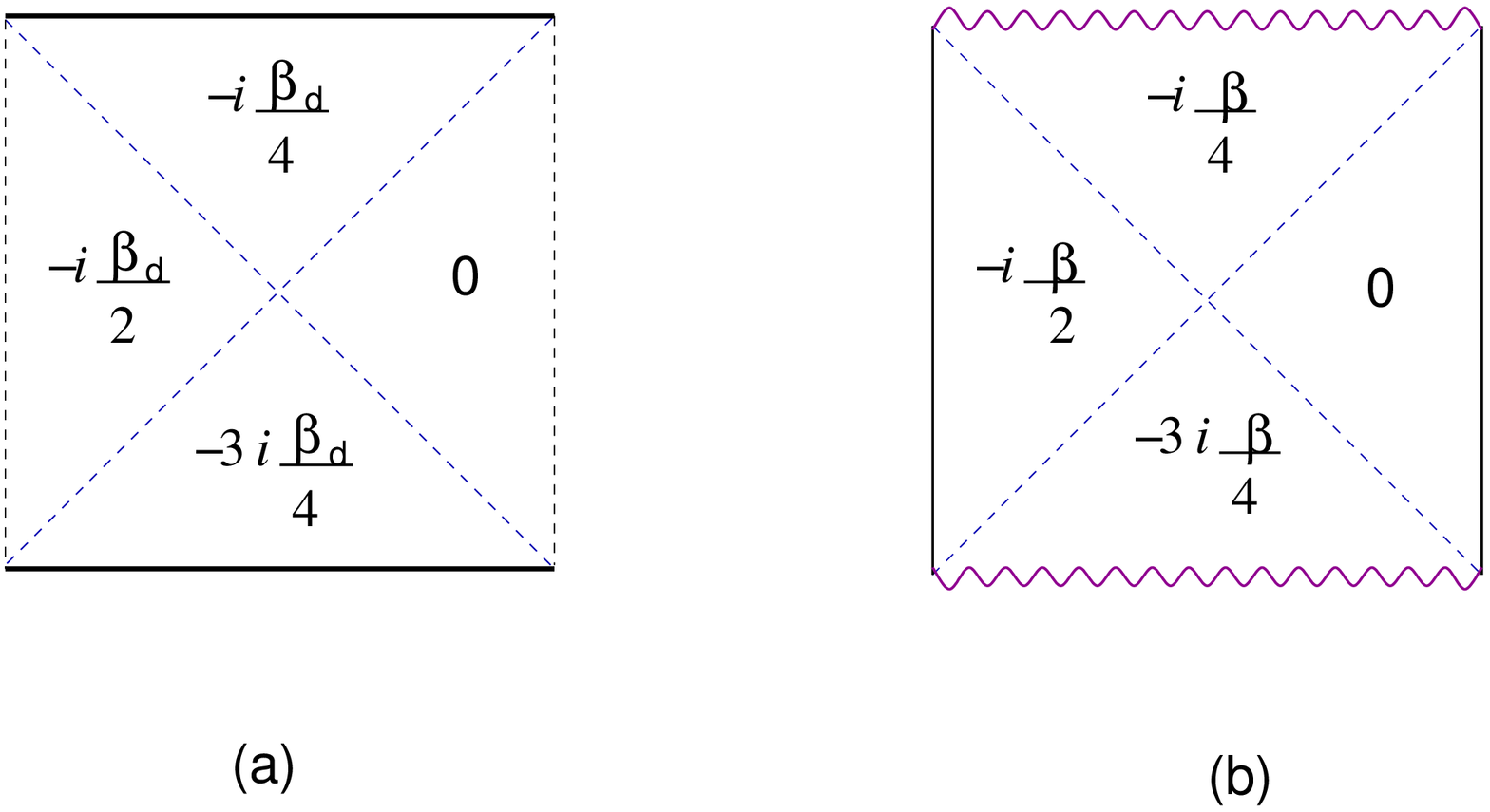}}

We start by recalling a basic manipulation in the eternal \SAdS\ black hole mentioned above. \SAdS\ can be described
by four static coordinate patches, whose time coordinates have different imaginary parts.  In our conventions, the
imaginary part of the time coordinate is shown in \imtconv b. For example, in the \SAdS\ geometry each time we cross
a horizon going counter-clockwise on the Penrose diagram, we pick up a imaginary part $- \, {i \over 4}\, \b$, where
$\b$ is the inverse Hawking temperature. So if we continue time by half a Euclidean period $\beta $, $ t \rightarrow
t-i\, \beta/2$, we take a point on the right boundary of \SAdS\ to one on the left boundary.

In the domain wall spacetimes, however, there is no left boundary. So, instead we can consider a point at large but
finite $r$. Now continue time by half a period. In the thin wall approximation the metric is just \SAdS\ (assuming
$r$ is not too large). So this continuation yields a point in the left quadrant at the same $r$ and $\Re(t)$. Now one
can move into the \dS\ region by continuing in (real) $r$. We here go beyond the thin wall approximation and assume,
as is plausible, that the domain wall is smooth. There will be a sharp signal in a correlator as one of its points
crosses the domain wall. (For concreteness imagine a two point function where both points have been continued to near
the domain wall.) This will be a first sheet effect, visible at finite $g_s$, unlike the signal due to subdominant
geodesics discussed above. It is not impossible for a two point function that is smooth on the AdS boundary to
display a sharp signal under analytic continuation. As an example imagine eternal \SAdS\ with an operator inserted on
the left boundary and the two correlated points on the right boundary. On continuation by half a period the
correlated points can collide with the left hand operator giving a large signal.  In fact, at finite $r$ the signal
will be nonsingular, but as we move to the boundary $r \rightarrow \infty$  a domain wall of fixed bulk width becomes
sharper and sharper in boundary variables. The boundary correlator will be singular, and possibly non-analytic.  This
is somewhat analogous to the Janus solution of \refs{\BakJK,\ClarkSB}.

Once we have changed $r$ enough to enter the \dS\ region, we may cross the \dS\ horizon by shifting the imaginary
part of $t$ once more. Again, this is because in static coordinates \dS\ is covered by four coordinate patches, whose
time coordinates have different imaginary parts -- see \imtconv a. So we may move through the \dS\ horizon by taking
$t \to t- i\, \beta/2 - i\, \beta_{d} /4$, where $\b_d$ is the inverse \dS\ temperature. We should mention that
smoothing out the domain wall to a small but finite thickness will change these Euclidean shifts by a small amount.
Because the metric has bounded first derivative, these changes are uniformly small.  Thus going beyond the thin wall
approximation will not alter the basic picture we are describing.

Having moved points past the \dS\ horizon we can now study behavior near \dS\ future infinity by taking $ r
\rightarrow \infty$. If we consider a bulk scalar field on this classical geometry we know that correlators near \dS\
future infinity give the conformally invariant results of dS/CFT \StromingerPN.  But correlators near the AdS
boundary give AdS/CFT results. We thus see that analytic continuation takes AdS/CFT correlators into dS/CFT
correlators. We will work out an explicit example of this in Appendix F, for the spacetimes described by Coleman and
de Luccia \ColemanAW.  In this case one can write down explicit formulas for boundary correlators that continue from
AdS/CFT to dS/CFT.  In this particular example, however, the AdS boundary will typically be destroyed by
backreaction.  So the explicit calculation in Appendix F may be thought of as a toy model for the full analytic
continuation required to go from a \SAdS\ boundary to the \dS\ boundary.

We should be able to extend these ideas beyond the classical supergravity approximation.   Assuming we have a non-perturbative definition of the boundary correlators,
continuing them   gives a precise description of certain aspects of quantum gravity  near what  was \dS\ future
infinity.   In principle, this should include
effects like bubble nucleation of other ``pocket universes'' in the far future \GuthPN\ .  
The correlators taken to the boundary are natural diffeomorphism-invariant observables.  Working at finite but large
$r$ introduces some scheme dependence probably related to the ambiguities in defining measures
in eternal inflation.  It is an interesting question to investigate which quantities have a sensible, unambiguous large $r$ limit.

\newsec{Can inflation begin by tunneling?}

After Farhi and Guth \FarhiTY\ established that beginning inflation classically required a past singularity, Farhi,
Guth and Guven \FarhiYR\ (FGG) made the interesting proposal that inflation could be initiated by quantum tunneling.
They computed a nonzero rate for this process using a Euclidean instanton.  This rate was also derived using
Hamiltonian techniques \refs{\FischlerSE,\FischlerPK} (see also \LindeSK).

Roughly speaking, FGG envisioned a process where an initial state (a ``buildable'' state) was constructed by
classical field evolution.  This state would then undergo tunneling. The initial configuration would look like the
bound trajectory on the left hand side of the effective potential\foot{In order to avoid singularity theorem constraints and hence  be buildable the initial geometry must be (a small deformation) of the lower half of
\pdcont b.} (\Veffcases a). It would tunnel ``through the effective potential'' to the unbound inflating trajectory with the same
energy on the right hand side.

A number of authors have argued that this process is not physically allowed. In particular, Banks \BanksNM\ argued
that since the de Sitter entropy of the inflating region is characteristically greater than the entropy of the black
hole surrounding it, ideas of black hole complementarity and holography prohibit the process.  Susskind \lenny\ has
given a somewhat different entropic argument that conflicts with the instanton rate.

The picture developed in this paper allows us to give a sharp argument against FGG tunneling, at least in the AdS
context.  The initial buildable state is clearly obtainable by unitary quantum time evolution, and so is a pure
state.  The final state has an inflating region, and so by the arguments in Section 3 is a mixed state. But unitary
quantum evolution cannot take a pure state to a mixed state.  So this process cannot occur.  In fact, no state
corresponding to inflation can ever result from any pure state process.  This argument is close in spirit to that of
\BanksNM, since the large entropy of the \dS\ region requires that the state be mixed.  But the argument presented
here is more general, since there exist time asymmetric situations where the \dS\ entropy is less than the black hole
entropy where the state is mixed and hence creation by any process is ruled out.

Such a simple argument demonstrates the power of embedding a physical phenomenon in a well defined non-perturbative
formalism.  But it is still important to understand the loophole in the FGG argument.

\newsec{Discussion}

We started our discussion  by assuming the existence of the string
landscape, with many \dS\ and AdS vacua. We restricted our
attention to a stable (supersymmetric) AdS vacuum and a
neighboring \dS\ minimum. Focusing on the low energy gravity
dynamics (and choosing points on the landscape, and hence
parameters, appropriately) we solved for the geometry of the
system using the thin wall approximation.  As expected from
previous work \refs{\FarhiTY,\FarhiYR}, we found parameter domains
with inflating behavior behind a black hole horizon. The stable
AdS minimum should, by general arguments, be described by a
boundary CFT.   Excitations of the CFT should probe inflationary physics. One
of our basic conclusions is that these inflating regions must be
described by a mixed state, \ie\ a density matrix in the CFT. Our
strongest argument interpreted  the static domain wall as a cutoff
version of the eternal \SAdS\ black hole, a system known to be
described by a mixed state.   The inflating geometries will
certainly be mixed if this one is.  Additional degrees of freedom besides 
the cutoff CFT are necessary to describe the region beyond the domain wall.
This mixed state description resolves an entropy puzzle because the large number of inflating
degrees of freedom need not be explicitly represented in the CFT.
This description raises several important questions.  First, as we
smoothly increase the size of the initial bubble, moving through
the progression of geometries illustrated in \pdcont, when does a
mixed state become necessary for a CFT description? We have
discussed several possible answers to this question but it still
remains open. More generally, given a rather arbitrary mixed state
in the CFT, what is its geometric interpretation? Finally, a
striking aspect of this description is the necessity of using a
mixed state to describe a geometry with one asymptotic region (as
in the static domain wall example).  Most previous examples
requiring mixed state descriptions had other asymptotic (AdS)
boundaries. They  could be given a pure state description if all
boundary degrees of freedom were kept.  Mixed states resulted when
some boundaries were traced over.   Here we do not have an
explicit representation of these extra degrees of freedom,
although we know many of their properties.  This provides a rather
well controlled system in which to search for new, non-boundary,
descriptions of non-perturbative quantum gravity.

We described techniques for probing the inflating region, even though the degrees of freedom there were not given
explicitly.  These techniques relied on analyticity.  The first used geodesics.  These geometries have nearly null
geodesics that bounce off \dS\ $\scri$, so they give subdominant singular contributions to certain correlators.  To
study them one must take $g_s$ very small, which suppresses bubble nucleation.  Chaotic eternal inflation should
still be visible with these probes.  More generally one can continue correlators to complex time.   We argued that
AdS boundary correlators should continue to \dS\ boundary correlators in the classical limit.  At finite $g_s$ the
complex pattern of bubble nucleation and other non-perturbative processes should be visible in these continued
correlators, assuming we had a way  of precisely computing them.

Even without the ability to calculate the full density matrix beyond the supergravity approximation we were able to draw some general conclusions, relying only on general features of
the picture we have developed. In particular we were able to argue that inflating regions could not be produced, even
by quantum mechanical tunneling, in a scattering process because a pure state cannot evolve into a mixed state under
Hamiltonian time evolution.   More results of this general type would certainly be welcome.

\bigskip\medskip\noindent
{\bf Acknowledgments:}
We would like to thank Ofer Aharony, Tom Banks, Raphael Bousso, Petr Ho{\v r}ava, Gary Horowitz, Shamit Kachru, Matt Lippert, Juan Maldacena, Don Marolf, Michael Peskin, Simon Ross, Yasuhiro Sekino, Eva Silverstein, Lenny Susskind, and Erik Verlinde for stimulating discussions. VH and MR would also like to thank Gary Horowitz and Sandip Trivedi for initial collaboration on related issues. VH, AM, and MR would like to acknowledge KITP and the Fields Institute
for hospitality during this work. VH and MR would also like to thank the organisers of the 2005 Amsterdam String theory workshop for hospitality.
This work was supported in part by the Stanford Institute for Theoretical Physics, NSF grant PHY-9870115, the Department of Energy under contract
DE--AC02--76SF00515, the funds from the Berkeley Center for Theoretical Physics, DoE grant DE-AC02-05CH11231, and the
NSF grant PHY-0098840. In addition, research at the Perimeter Institute is supported in part by funds from NSERC of
Canada and MEDT of Ontario. RCM is further supported by an NSERC Discovery grant.

\appendix{A}{Details of the thin wall geometries}

In Section 2 we briefly outlined the construction of  thin domain wall spacetimes and stated the main results for the
specific case of interest, the \dS/\SAdS\ junction. In this appendix we will go into more detail and derive these
results. As many of these results apply in a broader set-up than presented above, we first present the effective
potential and extrinsic curvatures for a more general junction between two spacetimes, with two free parameters (the
black hole mass and the cosmological constant) each.

\subsec{Effective potential and extrinsic curvatures}

We consider $D$-dimensional metrics of the form \metin, \metout, with
\eqn\metfD{ f_{\io}(r) = 1 - \lam_{\io} \, r^2 - {\mu_{\io} \over r^{D-3}} \ , }
where $\io$ stands for $i$  or $o$.
Here $\lam$ is related to the cosmological constant, which can have either sign,  and $\mu$ to the black hole mass.
(Thus for example, $\lam_i > 0$ and $\lam_o < 0$  with generic $\mu_{\io} > 0$ would correspond to the
Schwarzschild-\dS /\SAdS\ junction.) The effective potential is
\eqn\VeffD{\eqalign{ \VR(r) =  &- \[ \lam_o + {(\lam_o - \lam_i - \g^2)^2 \over 4 \, \g^2 }\] \, r^2  + 1
 - \[ \mu_o + {(\lam_o - \lam_i - \g^2) \, (\mu_o - \mu_i) \over 2 \, \g^2} \] \,  {1 \over r^{D-3}} \cr
&\qquad - { (\mu_o - \mu_i)^2 \over 4 \, \g^2} \,  { 1 \over r^{2D-4}}\ , }}
and the extrinsic curvatures are
\eqn\betaioD{\eqalign{ \b_i(r) &= \({\lam_o - \lam_i + \g^2 \over 2 \g}\) \, r
   + \({\mu_o - \mu_i \over 2 \g}\) \, {1 \over r^{D-2}} \cr
\b_o(r) &= \({\lam_o - \lam_i - \g^2 \over 2 \g}\) \, r
   + \({\mu_o - \mu_i \over 2 \g}\) \, {1 \over r^{D-2}}  \ , }}
for this choice of geometries.

Let us consider some generic features of the effective potential \VeffD. First, at small $r$, the last term dominates
(provided $\mu_o \ne \mu_i$ as will be the case whenever the shell carries energy), implying $\VR(r) \to -\infty$ as
$r \to 0$.  Hence it is always possible to have a shell which implodes to zero size.  Such a  shell crashes into the
black hole singularity at $r=0$ in a finite proper time.

Secondly, at large $r$, the first term in \VeffD\ dominates (again, provided its coefficient does not vanish).  The
$r^2$ coefficient may have either sign, but noting that we can re-express it as
\eqn\coeff{
 - \[ \lam_o + {(\lam_o - \lam_i - \g^2)^2 \over 4 \, \g^2 }\]  =
   - {1 \over 4 \, \g^2} \, \[ (\lam_o + \lam_i + \g^2)^2 - 4 \, \lam_o \, \lam_i \]  \ , }
we see that whenever the inside and outside cosmological constants have opposite sign, this coefficient is
necessarily negative.  This means that in such cases $\VR(r) \to -\infty$ as $r \to \infty$, so that it is possible
to have a shell which expands forever and (after infinite proper time) hits the boundary of both spacetimes. For
$\lam_{i} > 0$ this describes inflation.

On the other hand, if the $r^2$ coefficient \coeff\ of the effective potential is positive, then the boundary $r =
\infty$ is not attainable by any shell, and a time symmetric situation is always possible. We can similarly read off
the more detailed behavior of the shell by considering the extrinsic curvatures, as we do next in the 4-dimensional
\dS /\SAdS\ context. This will justify the results presented in Section 2.

\subsec{Thin wall trajectories}

Focussing now on the specific case of 4-dimensional \dS /\SAdS\ junction given by \metdSSAdS, the effective potential \VeffD\ simplifies to \VeffdSSAdS.
To see where the shell is allowed to appear on the appropriate Penrose diagrams, we will first consider the behavior
in the vicinity of small and large $r$. As mentioned in Section 2, this is determined by the sign of the extrinsic
curvatures on the two sides of the shell.  These are given by \betaioD, which  for the \dS/\SAdS\ junction simplify
to
\eqn\betadSSAdS{ \b_i(r) = {\g^2 - \lam -1 \over 2 \g} \ r   + {\mu \over 2 \g} \, {1 \over r^2} \ , \qquad \b_o(r) =
- {\g^2 +\lam + 1 \over 2 \g} \ r   + {\mu \over 2 \g} \, {1 \over r^2} \ . }
%

\ifig\alwdrs{Allowed and disallowed types of behavior of the \dS/\SAdS\ domain wall at small $r$.  The dotted curves
depict the shell's trajectory (on (a) \dS\ and (b) \SAdS\ Penrose diagrams) and the arrows the correspond to the
outward-pointing normal.  The checks indicate allowed scenario while the crosses label disallowed scenario. {\bf (a)}
In \dS, the shell starting from or ending at the left origin is allowed, whereas starting/ending on the right origin
is not allowed. {\bf (b)} In \SAdS, the shell starting towards the right from the past singularity or moving towards
the left before hitting the future singularity is allowed, whereas the opposite behavior is not allowed.}
{\epsfxsize=10cm \epsfysize=6.2cm \epsfbox{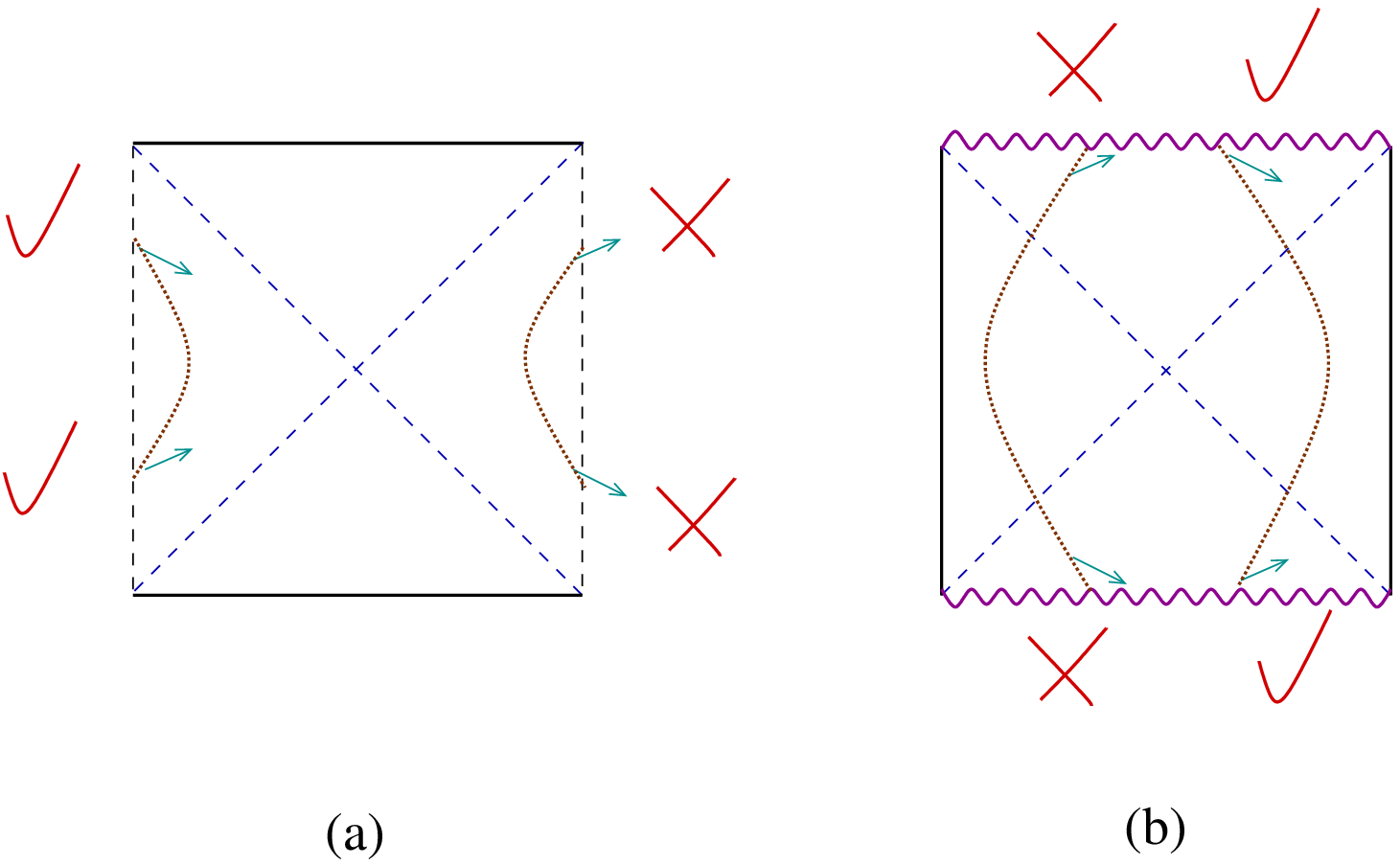}}

Clearly, as $r \to 0$, the second term dominates, and is always positive.  Hence both $\b_i(r) \to + \infty$ and
$\b_o(r) \to + \infty$ as $r \to 0$.  This means that when the shell is sufficiently small, the outward-pointed
normal has to point toward larger $r$.  \alwdrs\ summarizes the allowed and disallowed scenarios\foot{Although a
complete shell trajectory is sketched in \alwdrs, only the $r\to 0$ part is of relevance -- \ie, the full trajectory
may or may not be allowed, based on additional constraints to be discussed later.  Also, \alwdrs\ is not intended to
indicate the time of impact $t_{\io}(r \to 0)$ of the shell, but only the sign of ${dt_{\io} \over dr}(r \to 0)$.}
On the \dS\ side (where the `origin' $r=0$ are the north and south poles described by the vertical dashed lines in
\alwdrs a), the normals point toward increasing $r$ only on the left side of the Penrose diagram.  Hence the shell is
allowed to hit only the left origin but not the right origin, as indicated in \alwdrs a.  Similarly, for the \SAdS\
part of the spacetime, the shell can start out from the past singularity towards the right on the Penrose diagram,
and hit the future singularity moving towards the left, but not vice-versa, as indicated in \alwdrs b.

\ifig\alwdrL{Allowed and disallowed types of behavior of the \dS/\SAdS\ domain wall at large $r$.
{\bf (a)} In \dS, the shell starting from \dS\ $\scri^-$ towards the right or ending at $\scri^+$ veering left is
allowed for small tension (namely $\g^2 < \lam + 1$); whereas the opposite behavior is allowed for large tension
($\g^2 > \lam + 1$). {\bf (b)} In \SAdS, the shell starting from or ending at the left boundary is allowed, whereas
the the shell starting from or ending at the right boundary is not allowed.} {\epsfxsize=10cm \epsfysize=5.8cm
\epsfbox{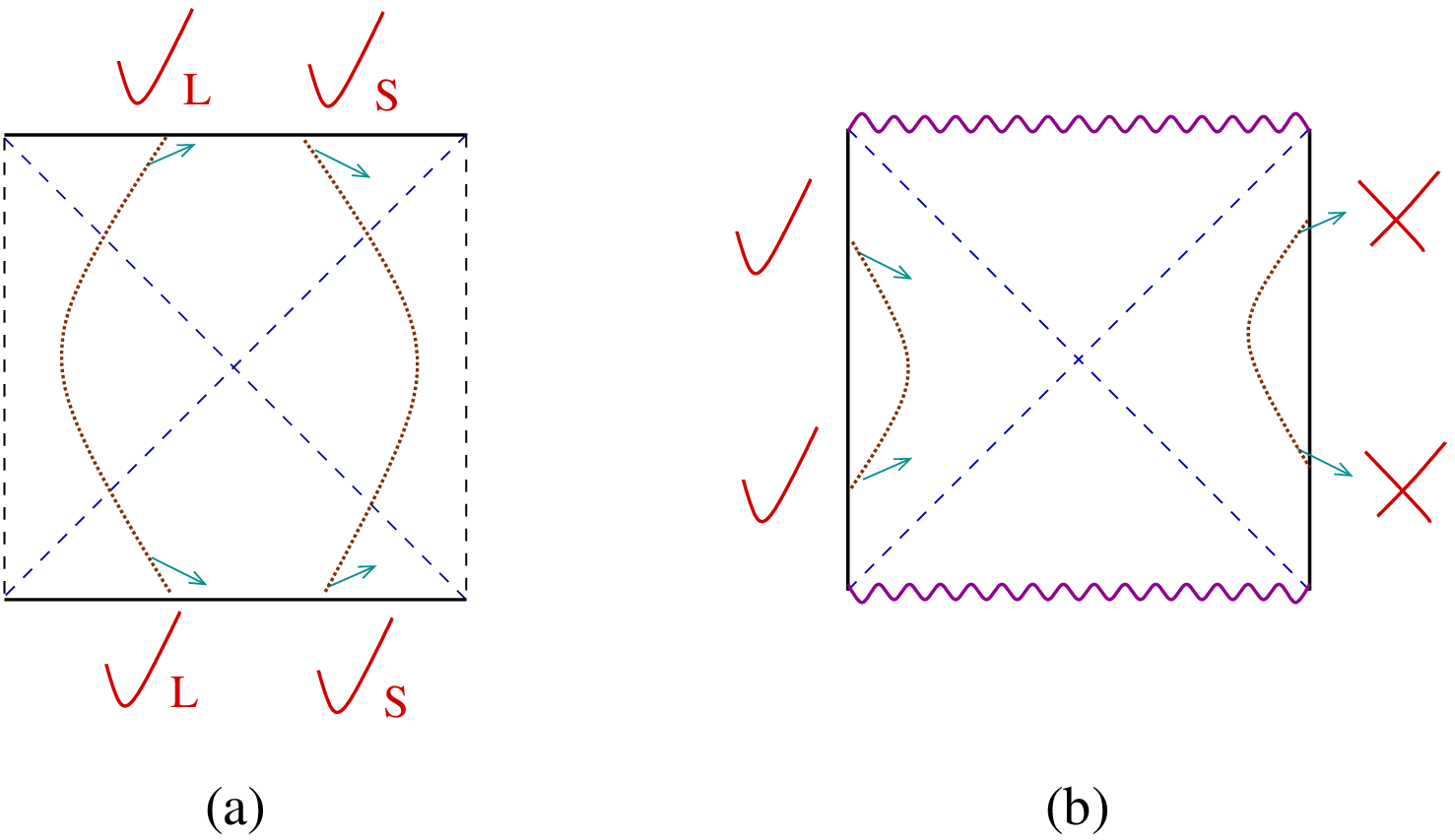}}

At large $r$, the first terms in \betadSSAdS\ dominate; here the sign of $\b_i$ depends on the shell's tension and
the \dS\ cosmological constant.  For $\g^2 < \lam + 1$, the first term is negative, so that $\b_i(r) \to -\infty$ as
$r \to \infty$, \ie, the outward normal from the \dS\ side points towards decreasing $r$.  Conversely, for large
tension $\g^2 > \lam + 1$, the extrinsic curvature remains positive, so the outward normal points toward larger $r$.
This is indicated in \alwdrL a, by the right and left trajectories, respectively. The corresponding behavior on
\SAdS\ side, sketched in  \alwdrL b, is more universal:  Here $\b_o \to -\infty$ as $r \to \infty$ for all $\g$ and
$\lam > 0$, so that the shell can hit the left boundary but not the right boundary.

\ifig\sampleVeff{Various possible effective potentials (thick, concave down curve) and extrinsic curvatures (thin,
concave up curves, where $\b_i(r) > \b_o(r)$) describing the \dS/\SAdS\ junction.  The specific parameters
$(\lam,\mu,\g)$ used were: {\bf (A1,B2)} $\lam=0.5,\mu=0.75,\g=2$;  {\bf (A2,B1)} $\lam=1,\mu=0.5,\g=1$; {\bf (C1)}
$\lam=1,\mu=2,\g=1$; {\bf (C2)} $\lam=1,\mu=1,\g=2$; and {\bf (D,E)} $\lam=2,\mu=0.89,\g=1$.} {\epsfxsize=10cm
\epsfysize=9cm \epsfbox{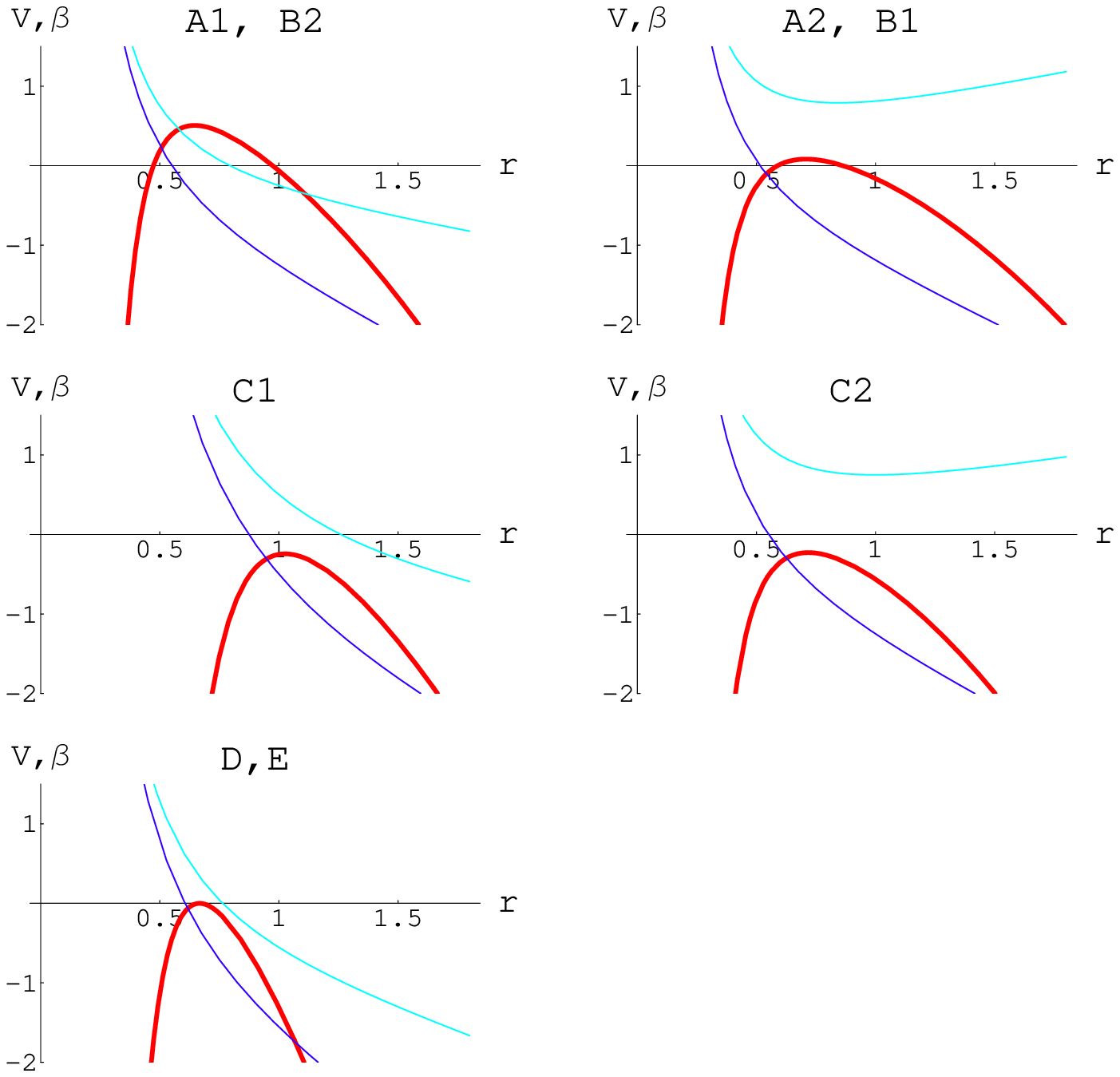}}

We can now determine the full trajectory of the shell on the Penrose diagram.  This requires us to solve for
$R(\tau)$ using the expression for $\VR$, while keeping in mind the sign of the extrinsic curvatures $\b_{\io}$.
These extrinsic curvatures are plotted along with the effective potential, for the various cases (A --- E of
\Veffcases) in \sampleVeff. The parameters $\lam$, $\mu$, and $\g$ are of order unity in all cases\foot{The
parameters chosen to plot the effective potential and extrinsic curvatures in \sampleVeff\ are chosen to clarify the
geometrical aspects and are not necessarily in the physically interesting regime.}.  The top two plots (labeled as
cases A and B, consistently with the notation employed in \Veffcases) exemplify the situations with $\Vmax >0$.  For
each potential there are two possible trajectories: (A) the shell expands from zero size and re-contracts and (B) the
shell contracts from infinite size and re-expands.  The middle two plots in \sampleVeff\ (cases C) describe a time
asymmetric situation since $\Vmax< 0$.  The two columns are distinguished by the sign of  $\b_i(r\to \infty)$.
Finally, the bottom plot depicts a fine-tuned situation with $\Vmax = 0$.

Let us now consider the corresponding extrinsic curvatures, which will enable us to construct the full Penrose
diagrams. As we can see from \sampleVeff (A1), $\b_o$ remains positive everywhere along the A trajectory, so that the
shell's turning point must lie in region I (right side) of the \SAdS\ Penrose diagram.  Conversely, in \sampleVeff
(A2), $\b_o$ becomes negative before $\VR$ becomes positive, which implies that the shell must pass through region
III rather than region I on the \SAdS\ Penrose diagram\foot{Note that for (A2), since the trajectory on the \SAdS\
diagram must still bend towards the left while passing through the left region, we would expect that this case can
occur for much smaller region in the parameter space than the more typical case (A1).   Indeed, we confirm that this
is the case by plotting the potentials as in \sampleVeff: we find that it is much harder to obtain $\Vmax >0$ with
the $\b_o$ intercept occurring at smaller $r$ than the $\VR$ intercept.}. Similarly, the sign of $\b_i$ along the B
trajectory distinguishes the cases (B1) and (B2).  In the former case $\b_i$ is positive, whereas in the latter it is
negative. Finally, as a consistency check, one can verify that we cannot have $\b_i$ becoming negative at smaller $r$
then where $\VR$ becomes positive, or $\b_o$ becoming negative at larger $r$ then where $\VR$ becomes negative
again\foot{In particular, denoting the zero-intercepts of the extrinsic curvatures by $r_{\b_i}$ and $r_{\b_o}$, we
find that $\VR'(r_{\b_i})= -2 \lam \( {\mu \over 1+ \lam - \g^2} \)^{1/3} <0$, while $\VR'(r_{\b_o}) = (3+ \lam +
\g^2) \( {\mu \over 1+ \lam + \g^2} \)^{1/3} >0$. This automatically implies that if $\Vmax > 0$, then $\VR(r_{\b_i})
> 0$ and $\VR(r_{\b_o}) < 0$. Finally, by monotonicity of the extrinsic curvatures, this implies that the $\b_i$
intercept occurs at larger $r$ than where $\VR$ first becomes positive, and similarly that the $\b_o$ intercept
occurs at smaller $r$ than where $\VR$ becomes negative again.}.

A few remarks are in order. First, where exactly the shell passes with respect to the bifurcation point of the horizons depends on the details of
the set-up, which distinguishes the cases A1 from A2, or B1 from B2, but which are not drawn separately in
\dSSAdSpossPDs; \ie, (A) of \dSSAdSpossPDs\ would strictly speaking correspond only to (A1) of \dSSAdSpossibsAB, \etc.
Second, note that the Penrose diagrams corresponding to the time reverse cases C' and E' (in the notation of
\Veffcases) would be obtained by vertically flipping the diagrams (C) and (E), respectively.

\appendix{B}{False vacuum bubbles in scalar-gravity systems }

First, we will briefly review the general formalism for obtaining a full spacetime by using the initial-value ($3+1$)
formulation of GR.  Given an initial slice, the induced 3-metric, the corresponding extrinsic curvature, and the
initial values and velocities of the matter fields, Einstein's equations split into two equations describing the
evolution and imposing constraints on the initial data. We are interested in gravity coupled to a scalar field $\phi$
with a given potential $V(\phi)$. The action for the system will be taken to be:
\eqn\scagravact{
 S = \int \sqrt{-g} \[ \half R - \half (\nabla\phi)^2 - V(\phi) \] \ . }

Consider for concreteness a 4-dimensional, spherically symmetric spacetime, with a 3-dimensional spacelike initial
slice $\Sigma_0$. Further assume the  metric on  $\Sigma_0$
\eqn\initmet{ ds^2 = {dr^2 \over 1 - {2 m(r) \over r}} + r^2 \, ( d\theta^2 + \sin^2 \theta \, d\phi^2 ) \ . }
The unit normal to the surface $\Sigma_0$ will be taken to be $\dta$. To describe how this hypersurface fits into the
full spacetime, we have to prescribe the extrinsic curvature. For simplicity, we will choose the extrinsic curvature
to be proportional to the induced metric on the Cauchy slice,
\eqn\extcurv{ K_{ab} = h(r) \, g_{ab} \ . }
Furthermore, we have a scalar field $\phi$ with a potential $V(\phi)$. We denote the radial variation of $\phi$ by
$\pp$, and the time evolution by $\pd$. Specification of initial data will involve picking appropriate functions for
$\phi(r)$, $\pd(r)$ and $h(r)$. The constraint equations are
\eqn\constrf{ \pd \, \pp = -2 h'  }
and
\eqn\constrm{ 2 \, m' + m \, r \, \pp^2 = {r^2 \over 2} \,  \( \pd^2 + \pp^2 + 2V - 6h^2 \) \ . }

Substituting \constrf\ into \constrm\ we find
\eqn\msoln{ 2 \, m(r) = \int_0^r \rh^2 \, \( V - 3 h^2 + {2 h'^2 \over \pp^2} + {1 \over 2} \pp^2 \) \, e^{- {1 \over
2} \int_{\rh}^r \rb \, \pp^2 \, d\rb } \ d\rh  \ . }
Thus specifying the initial field profile $\phi(r)$ in a given potential $V(\phi)$ determines the mass function
$m(r)$ on $\Sigma_0$ and thus the metric. Note that we can then find the total ADM mass in the usual way, by
considering $m$ at large $r$ and possibly adding an appropriate counter-term.

We will now motivate the procedure for picking the correct initial data in a scalar-gravity system, which is
guaranteed to have the desired features, \dS\ $\scri$ and AdS boundary.

We first make a trivial remark about causality. If we take any spacetime (${\cal M}, g_{a b}$) and consider some
spacelike a-chronal (and possibly compact) surface $\Sigma$ and impose the corresponding induced metric $h_{a b}$ and
extrinsic curvature $K_{a b}$ on $\Sigma$ as our initial conditions, then within the domain of dependence of
$\Sigma$, we are guaranteed to evolve to the spacetime $g_{a b}$.

\ifig\dSpd{Initial slices on a \dS\ Penrose diagram which are guaranteed to contain a piece of \dS\ $\scri$ in their
evolution.  (The vertical lines are the poles $r=0$ of \dS, the dashed diagonal lines are the cosmological horizons
$r=r_d$.  Horizontal slices represent 3-spheres, with their equator along the vertical midpoint of the diagram.)}
{\epsfxsize=5cm \epsfysize=5cm \epsfbox{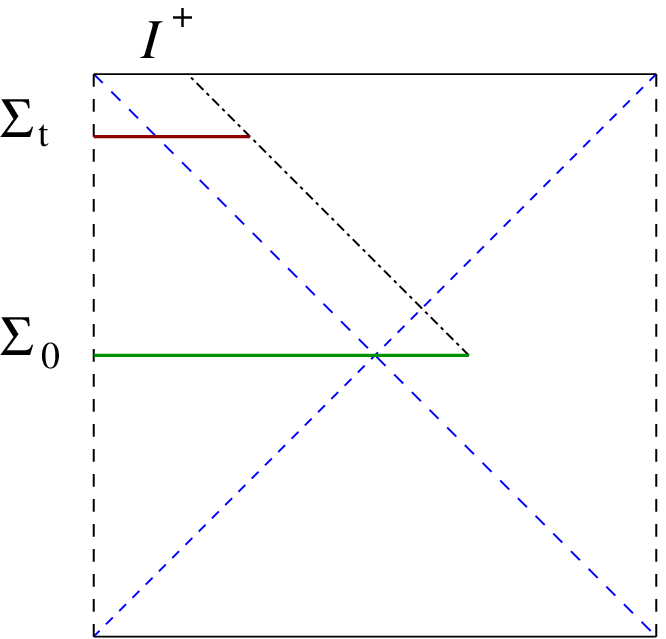}}

This implies that if we want to consider (a piece of) initial data which is guaranteed to evolve to a \dS\ $\scrip$,
we must take compatible initial conditions on a large enough initial surface (which spans the entire past domain of
influence of the desired piece of \dS\ $\scri$).  This is illustrated on the \dS\ Penrose diagram in \dSpd.  If we
wish to take the initial surface in a  time-symmetric fashion (\ie\ $\Sigma_0$ at $t=0$), then we must take
$\Sigma_0$ to cover more than half of the sphere, in which case the radial coordinate is not monotonically increasing
along the entire slice.  On the other hand, if we consider a slice at some later time $t$, say $\Sigma_t$, we can
easily contain the same piece of $\scrip$ under evolution, while $\Sigma_t$ covers only a small part of the sphere
and therefore has increasing radial coordinate.  This makes the formalism sketched above applicable; in particular,
we can use \msoln\ to find the mass.  Moreover, in a domain wall construction, if we imagine setting up a
(non-static) domain wall to pass at the right end of $\Sigma_t$ in \dSpd, we no longer have the requirement that the
\dS\ size has to be bigger than that of the black hole: all we need to satisfy is that $R(t) > r_d$ and $R(t) > r_+$
at the time $t$. Such constructions are discussed in detail in Appendix C.

\ifig\Vphir{{\bf (a)} The potential $V(\phi)$ and {\bf (b)} an initial profile of the field $\phi(r)$.}
{\epsfxsize=12cm \epsfysize=5cm \epsfbox{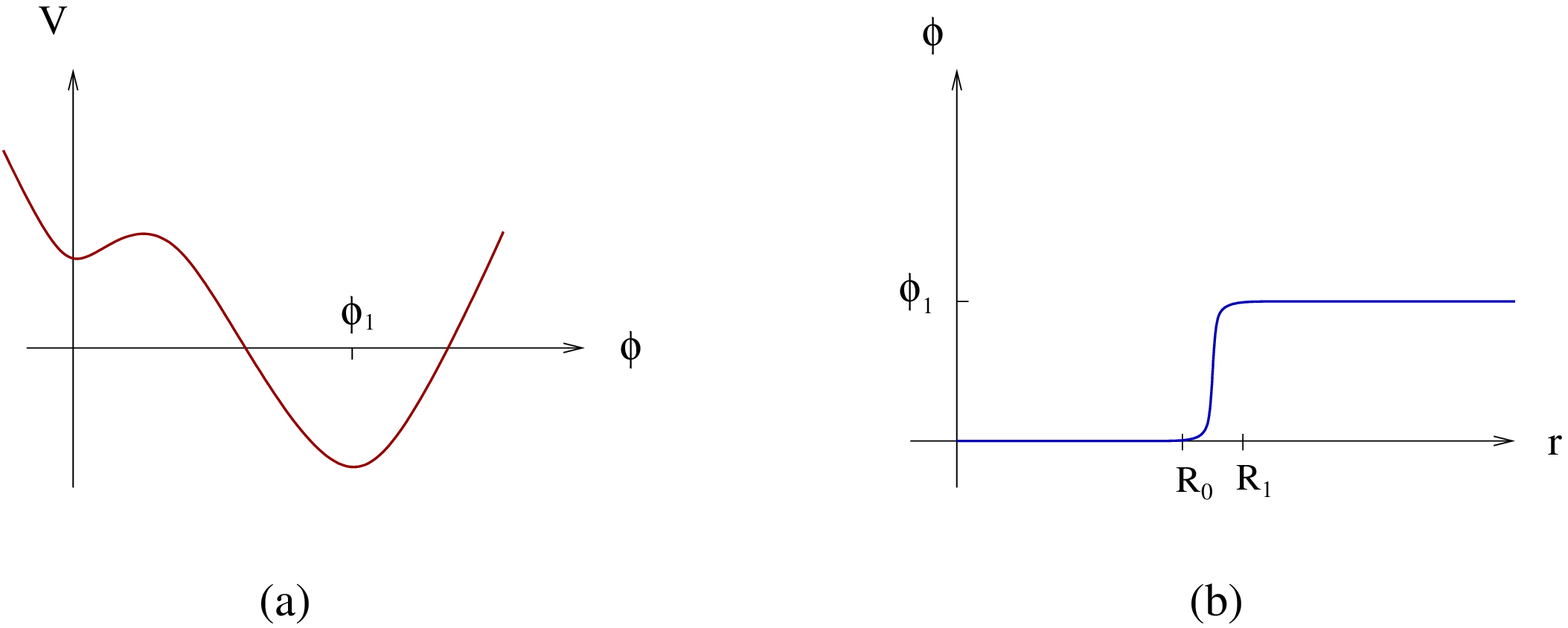}}

Let us now see more explicitly how to construct a spacetime containing both \dS\ and AdS using a scalar field in a
potential.  We use the potential $V(\phi)$ as sketched in \Vphir a.  Note that if the field $\phi(r)$ were
homogeneous and sat in the positive local minimum of the potential, namely $\phi(r)=0$, the spacetime would
correspond to a static, spherically symmetric spacetime with a positive cosmological constant, namely \dS. Likewise,
if the field sat in the negative local minimum of the potential,  $\phi(r)=\phi_1$, the spacetime would correspond to
static, spherically symmetric spacetime with a negative cosmological constant, namely AdS or \SAdS, depending on the
mass.

To interpolate between \dS\ and \SAdS, consider the field profile as sketched in \Vphir b, with $\pd=0$ in the
regions $r<R_0$ and $r>R_1$. For $r<R_0$, the field sits in the positive local minimum of the potential, so that in
the domain of dependence of $r<R_0$, the spacetime corresponds to a part of the \dS\ spacetime.  By arranging the
corresponding initial surface $\Sigma_t$ to be late enough, this will include a part of the \dS\ $\scrip$, as in
\dSpd.  Similarly, for $r>R_1$, the field sits in\foot{There is a slight subtlety: if we require the profile
$\phi(r)$ to be an analytic function of $r$, there has to be a slight deviation from $\phi$ being exactly constant in
the regions $r<R_0$ and $r>R_1$.  For \dS, this deviation does not change our arguments, since \dS\ is stable.  On
the other hand, the AdS minimum is more sensitive: if a homogeneous field is slightly off the minimum, it will big
crunch.  However, here we don't have homogeneous fields, and we can tune the deviation from the minimum to be
infinitesimal, which implies by the well-posedness of the initial value formulation in general relativity that the
resultant spacetime should be arbitrarily close to \SAdS.} the negative local minimum of the potential, corresponding
to a part of \SAdS\ spacetime.

\ifig\dSAdS{Sketch of possible causal diagram for spacetime evolved from the initial data of \Vphir.} {\epsfxsize=8cm
\epsfysize=4cm \epsfbox{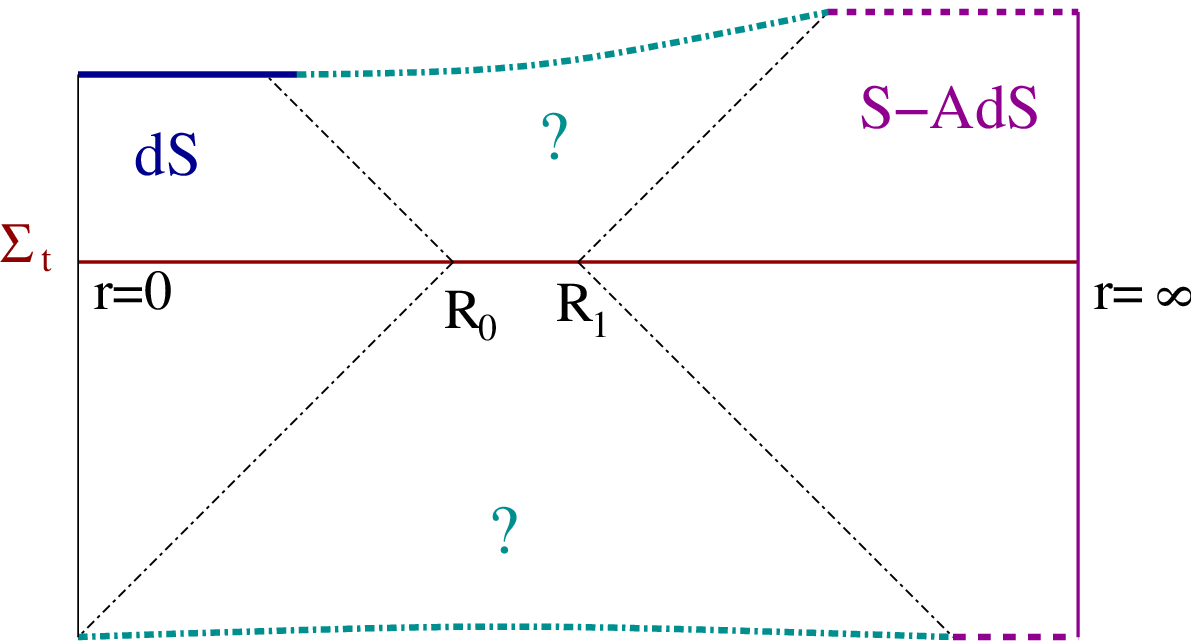}}

On a spacetime diagram, the full evolution might look like the one sketched in \dSAdS. Inside the domain of
dependence of $r<R_0$ (left wedge), we obtain \dS\ $\scrip$.  Similarly, inside the domain of dependence of $r>R_1$
(right wedge), we have \SAdS\ spacetime.  Depending on the mass (in part determined by the size of the \dS\ region),
the evolution may or may not reach a singularity.  We note in passing that if there is a singularity in this region,
as assumed in \dSAdS, then it will be of the \SAdS\ type, \ie, spacelike geodesics will bounce off the singularity.
We also note that in order to have $r$ increasing monotonically on $\Sigma_t$, we cannot have $\Sigma_t$ go through
region 3 of Schw-AdS; this implies that in \dSAdS\ there can be future or past \SAdS\ singularity, but not both. The
regions marked by `?' in \dSAdS\ refers to the domain of dependence of the interpolating region $r \in [R_0, R_1]$ in
the scalar profile \Vphir b, for which the evolution needs to be determined numerically.

The above arguments illustrate that, at least within the realm of classical general relativity coupled to a scalar
field, we can obtain \dS\ $\scri$ within an asymptotically AdS spacetime.

\appendix{C}{Construction allowing \dS\ $\scri$ and $r_d < r_+$}

In this Appendix, we build on the set-up introduced in Appendix B to give a rigorous construction of spacetimes with
\dS\ entropy smaller than the black hole entropy.
We have indicated in Section 2 that for time asymmetric configurations it is possible to obtain $r_d$ both larger and
smaller than $r_+$, depending on the parameters. In this Appendix we will demonstrate that in fact,  geometries with
$r_d < r_+$ are not only possible, but indeed required, if the initial slice $\Sigma$ has monotonically increasing
$r$ and the \dS\ $\scri$ is guaranteed by evolution from $\Sigma$.

Consider the thin domain wall spacetimes separating \dS\ and \SAdS, compatible with an initial surface $\Sigma$ on
which $r$ increases monotonically. Recall from \betadSSAdS\ that both the \dS\ and the \SAdS\ extrinsic curvatures
are positive for small $r$ and negative for large $r$ (though $\b_i(r\to \infty)$ depends on $\g$; but this will not
affect our discussion).  Moreover, since $\b_i(r) > \b_o(r)$ for all $r$ (which follows from that requirement that
the bubble have positive energy), the respective radii $r_{\b_i}$ and $r_{\b_o}$ where the \dS\ and \SAdS\ extrinsic
curvatures vanish must be related as $r_{\b_i} > r_{\b_o}$. We will now spell out the sequence of steps to determine
the Penrose diagram:

\itm To have an initial slice $\Sigma$ with monotonic $r$, the shell cannot be confined entirely to the right wedge
of \dS.  Likewise, it cannot be confined entirely to the left wedge of \SAdS. More specifically, $\Sigma$ cannot
enter into these regions. This is because we want $r$ to increase towards the right on the initial slice, whereas it
would necessarily increase towards the left in the above regions.

\itm To ensure \dS\ $\scrip$ by causality arguments, the shell must end up at $r=\infty$.  Evidently, the shell's
trajectory must intersect the boundary of the past domain of influence of the desired piece of $\scrip$; this implies
that it must pass through the upper and/or right wedge of \dS.  But since $\Sigma$ cannot enter into the right wedge
of \dS, the shell must pass through the upper wedge.  This ensures that the shell must end up at $\scrip$ since it
follows a timelike trajectory.

\itm In order for the shell to end up at $r=\infty$, it must end up in the left wedge of \SAdS\ (and in the upper
wedge of \dS, as explained in the previous point).  Since $r$ is continuous across the shell, the shell must end on
one of the AdS boundaries.  The negativity of the AdS extrinsic curvature at large $r$ implies that the shell cannot
end up at the right boundary, since this would require the extrinsic curvature to be positive (the outward normal
would necessarily have to point to larger $r$).

\itm In order for the shell to end up in the left wedge of \SAdS, it cannot pass through the right wedge of \SAdS.
This is implied simply by the fact that the shell follows a timelike trajectory.

\itm The first and the last points imply that the shell must pass through the lower wedge of \SAdS.  In fact, since
the shell follows a timelike trajectory, this further implies that it must start in the lower wedge of \SAdS, \ie, at
the singularity, $r=0$. Note that this requires that the effective potential for the domain wall motion be negative
semi-definite, $\VR(r) < 0$ for all $r$, which is easy to achieve for suitable choices of parameters.

\itm Finally, in order for the shell to start at $r=0$ and therefore the \dS\ extrinsic curvature be positive, it
must start in the left wedge of \dS.  This is again implied by the incompatibility of extrinsic curvature with
starting at the origin in the right wedge of \dS: for any such scenario, the \dS\ extrinsic curvature would have to
be negative, since the outward normal would point toward decreasing $r$.

\ifig\bubblePD{The domain wall trajectory on \dS\ and \SAdS\ Penrose diagrams which is necessitated by the type of
set-up specified above: and initial slice $\Sigma$ with monotonically increasing $r$, and a guarantee of dS $\scrip$
under evolution.}
 {\epsfxsize=7.8cm \epsfysize=4.5cm \epsfbox{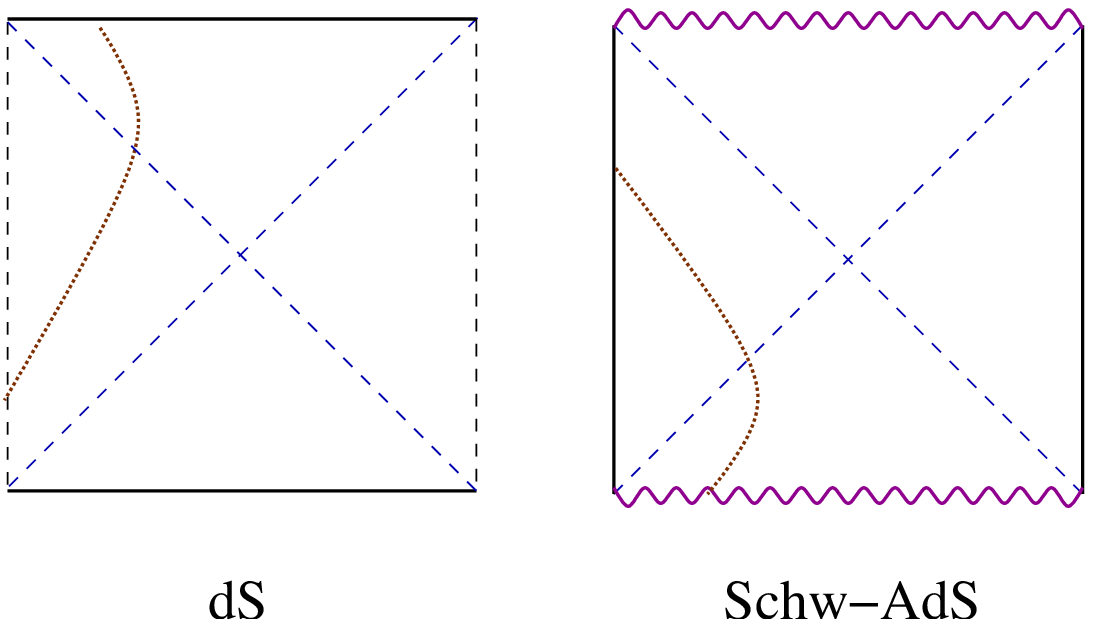}}

Thus far, we have established that from the \dS\ point of view, the shell must start on the left origin and end on
the upper $\scrip$; whereas from the \SAdS\ point of view, it must start at the past singularity and end up on the
left boundary.  To refine this picture, we can use the facts about the extrinsic curvatures further to say that the
shell must start out going to the right and end up going to the left on the Penrose diagram.  The complete trajectory
is illustrated in \bubblePD.

\ifig\bubpd{Initial slice $\Sigma$ with monotonically increasing $r$ on a combined Penrose diagram (obtained by
taking the left part of dS and right part of S-AdS in \bubblePD).  The shaded region is uncertain under a reasonable
evolution.} {\epsfxsize=5.0cm \epsfysize=4.5cm \epsfbox{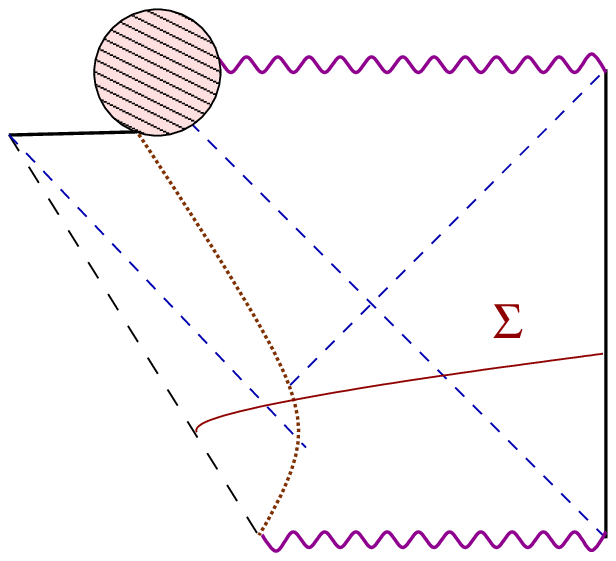}}

Finally, if we combine these on a single Penrose diagram and draw in the initial surface $\Sigma$, the resulting
diagram would look like \bubpd.  Note that the \dS\ horizon crosses the shell's trajectory lower (\ie, at smaller
$r$) than the black hole horizon, which implies that $r_d < r_+$.  Also note that on \bubblePD, the initial slice
$\Sigma$ would cut the \dS\ Penrose diagram in the upper wedge whereas it would cut the \SAdS\ Penrose diagram in the
lower wedge. While $\Sigma$ in \bubpd\ has $r$ monotonically increasing, we this property does not continue to hold
for all later spacelike slices which would cover the full Penrose diagram.  In this sense, \dSAdS\ is somewhat
misleading:  $r$ is not a good coordinate globally.

>From the above arguments we see that \bubblePD\ represents the only self-consistent set-up with thin domain wall, and
initial slice with monotonically increasing $r$ and a guaranteed piece of dS scri under evolution.  Explicitly, in
order for $r$ to increase monotonically on our initial data slice $\Sigma$, we must have $R_1 < r_+$.  On the other
hand, to guarantee that the domain of dependence of the part of $\Sigma$ with $r< R_0$ includes a part of dS scri, we
must have $R_0 > r_d$.  Since according to our setup, $R_0 < R_1$, the two conditions together guarantee that the
\dS\ radius (and therefore its entropy) is smaller than that of the black hole.

Although the above argument has been phrased in the thin wall context, it actually applies in general: assuming $r$
is monotonic on the Cauchy slice, in order to guarantee \dS\ $\scrip$ within the domain of dependence of our Cauchy
slice, we require $r_d < r_+$.

\appendix{D}{Computation of dS-SAdS Propagators}

This appendix contains the computation of correlators on the cutoff 
surface $r_c$, with a dS region behind a domain wall
at $R_t$.  For simplicity, we will take $r_c$ and $R_t$ to be much 
larger than $r_+$ and $r_A=1$ and much smaller than the de Sitter 
radius $r_d$.  We will also consider just a massless field, although
similar computations apply to massive fields.
We will restrict our attention to Euclidean space propagators.  

In this geometry, the geodesic
length between two points $(r_1,\Omega_1)$, $(r_2,\Omega_2)$ 
can be evaluated in the large $r$ limit 
\eqn\asd{\eqalign{
L 
&\sim
\ln r_1 r_2 + \ln \left(
\sin^2(\Omega_1-\Omega_2)/2 + {1\over r_1^2} + {1\over r_2^2}
\right) + \dots
.}}
We are neglecting here terms that vanish as powers of $r_+/r$, 
as well as higher powers of $1/r$.
So the propagator of a massless field ($\Delta=3$) in AdS$_4$ between a point
$(r_c,\Omega)$ and $(R_t,\Omega')$ becomes, since $R_t\gg r_c$, 
\eqn\asd{
e^{-\Delta L} \sim {1\over (r_c R_t)^3(\sin^2(\Omega - \Omega')/2 + r_c^{-2})^3}
.}
We also need the propagator of a massless field in dS.  In the limit where
the dS radius is large, this reduces to the flat space formula.  For
two points on the domain wall $(R_t,\Omega_1')$, $(R_t,\Omega_2')$, this is
\eqn\asd{
{1\over L^2} \sim {1\over R_t^2 \sin^2(\Omega_1' - \Omega_2')/2}
.} 
So the contribution to a two point function between points
$(r_c,\Omega_1)$ and $(r_c,\Omega_2)$ is found by multiplying two AdS
and one dS propagator together, 
and integrating over points $\Omega_1'$ and $\Omega_2'$ on
the $R_t$ surface.  This gives
\eqn\asd{
A(\Omega_1,\Omega_2) = 
{1\over r_c^{6} R_t^2} 
\int {d^3\Omega_1' d^3\Omega_2'\over
\sin^2(\Omega_1'-\Omega_2')/2
(\sin^2(\Omega_1-\Omega_1')/2 + r_c^{-2})^3
(\sin^2(\Omega_2-\Omega_2')/2 + r_c^{-2})^3
}
}
Note that we've pulled a power of $R_t^6$ out from the volume element on the 
two 3-spheres, so here $\Omega$ denotes a point on the unit size 3-sphere.


It is straightforward to estimate this integral.  The 
term in the denominator,
$(\sin^2(\Omega-\Omega')/2 + r_c^{-2})^{-3}$, is roughly
$r_c^{6}$ near $\Omega=\Omega'$ and order one for points where
$\Omega$ is far from $\Omega'$.  The section of the $\Omega'$ 
3-sphere where this term goes like $r_c^6$ has area $r_c^{-3}$. 
The full six dimensional integral over this domain gives a 
contribution of order $r_c^6$.
So the order one piece will be subleading when $r_c\gg 1$, and the 
integral can be approximated as
\eqn\asd{
A(\Omega_1,\Omega_2) \sim {1\over r_c^6 R_t^2}
\left({r_c^6\over \sin^2 (\Omega_1-\Omega_2)/2} + \dots 
\right) \sim {1\over R_t^2 \sin^2 (\Omega_1-\Omega_2)/2}.
}
We wish to compare this to the standard AdS massless propagator
\eqn\asd{
A_{AdS}(\Omega_1,\Omega_2) \sim 
{1\over r_c^6\sin^6(\Omega - \Omega')/2}
.}
We conclude that $A$ can be safely ignored when
\eqn\asd{
R_t \gg r_c^3 
.}

When $R_t$ is less than $r_c^3$ the new contribution to the propagator
represents a non-local contribution to the two point functions,
which scales as $L^{-2}$ rather than $L^{-6}$, where 
$L\sim \sin(\Omega_1-\Omega_2)/2$ is the proper length on the sphere.  
This term becomes important for
length scales longer than $L >\sqrt{R_t r_a^4/r_c^3}$, i.e. energy
scales below $\sqrt{r_c^2 / R_t r_a^4}$.

\appendix{E}{A pure state description of spacetimes with causally disconnected regions?}

In section 4, one of the criteria we proposed for the boundary field theory to be in a mixed state was that the
spacetime geometry have regions that are causally disconnected from the boundary. However, in some special cases we
can construct spacetimes containing a causally inaccessible region which nevertheless appear to have a pure state
description in the CFT. In this section we will describe two examples of this. In both cases, a subset of CFT
correlators are insensitive to the causally disconnected region and act as though they are in a mixed state.

\ifig\NullShAdS{Sketch of possible causal diagram for collapse spacetime created by an imploding null shell in AdS.
Here the spacetime inside the shell is pure AdS.} {\epsfxsize=3cm \epsfysize=4cm
\epsfbox{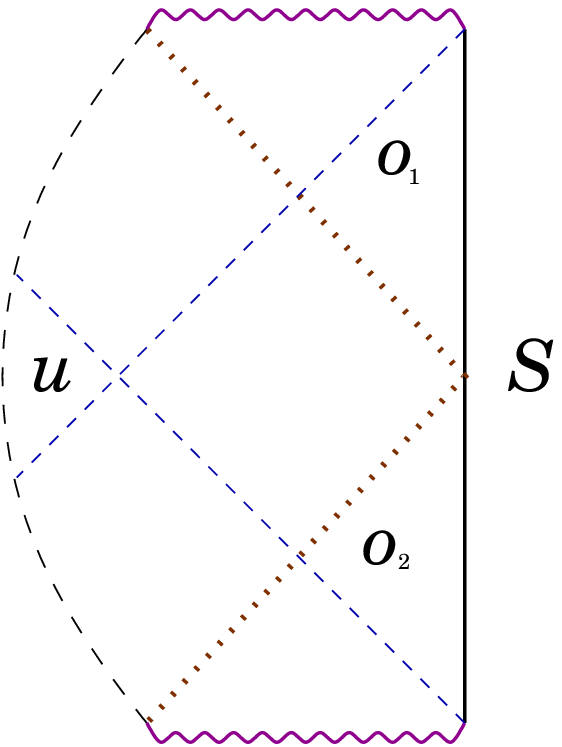}}

The simplest example is a spherically symmetric null shell of energy which is launched from the boundary at an
instant of time, say $t=0$. For simplicity we construct a time-symmetric geometry, so the shell is launched towards
the future and the past, as shown in \NullShAdS.  The overall picture, then, is of a null shell expanding out to the
boundary, reflecting off, and collapsing back in. Since the usual AdS boundary conditions are reflecting, such a
configuration can be constructed as a solution of the source-free equations of motion.

The overall structure of the spacetime depends on the mass of the shell. If the collapsing shell is massive enough to
make a stable black hole in AdS (one with $r_+ > r_A$ so as to be stable in the canonical ensemble), then the
spacetime has a region which is causally disconnected from the boundary, as shown in \NullShAdS. By making the mass
of the shell large, we can make the inaccessible region large in AdS units.

To prove the existence of an inaccessible region, note that inside the null shell the metric is pure AdS, while
outside it is \SAdS. If light rays emitted radially from the origin at $t=0$ escape out to the boundary, every other
point in the spacetime will be accessible to the boundary. By ``accessible,'' we mean that every bulk point is either
in the future lightcone or the past lightcone of at least one boundary point. On the other hand, if radial light rays
launched from $t=0, r=0$ end in the singularity, there exists a spacetime region which is causally disconnected from
the boundary. The light ray of interest propagates in pure AdS inside the shell, so we only need to know the behavior
of null geodesics in AdS to infer causal accessibility.

The question therefore is whether the radius $r_0$ at which the light ray crosses the shell is bigger or smaller than
the Schwarzschild radius of the black hole. If it is smaller, then the light ray must match onto a point inside the
horizon and end in the singularity. It turns out that the radius $r_0$ at which the light ray intersects the shell is
precisely the AdS radius $r_A$. So anytime we form a stable black hole with $r_+ > r_A$ the light rays from the
origin will be unable to escape to the boundary and we have an inaccessible region. To prove that $r_0 = r_A$, it
suffices to note that in pure AdS spacetime two radial null geodesics, one originating at the origin $r =0$ and the
other at the boundary meet precisely at $r = r_A$. It is easy to check that for pure AdS, $r_{ray}(t) = r_A  \,
\tan\(t/r_A \)$ and $r_{shell}(t) = r_A \, \cot\(t/ r_A\)$, respectively, which intersect at $r=r_A$. As an aside, it
is possible to show that time symmetric collapse geometries with $r_+ > r_A$ always contain a region causally
disconnected from the AdS boundary.

There are strong reasons to believe that this solution can correspond to a pure state in the CFT. On the time
symmetric slice, the shell solution differs from pure AdS only by the excitation of some massless fields near the
boundary. We can think of creating the state by acting on the vacuum with some bulk operators near the boundary.
Since bulk operators near the boundary are easy to map the the field theory, in the CFT we should also be able to
obtain this state by acting with an operator on the vacuum, which manifestly gives a pure state. We can imagine
exciting the massless fields in the bulk by the insertion of an operator $\CS$ in the boundary at time $t=0$; the
geometry in \NullShAdS\ corresponds to a state $\CS |0 \rangle$ of the
 field theory. For example, we can construct a null shell with the
 desired properties by a dilaton wave packet in the bulk, which in t
he boundary theory correpsonds to the  choice
\eqn\shellop{ \CS = \exp\( i \, \int d^3\vec{x} dt \, j(t) \, \Tr \(F_{\mu \nu}(x, t) \, F^{\mu \nu} (x,t)\)  \) \ .
}
By picking the support $j(t)$ we can ensure that we have a configuration with null shell that is launched
symmetrically away from the boundary and localized in time near $t=0$.

Although the field theory is in a pure state, the most obvious field theory operators are insensitive to physics in
the inaccessible region, suggesting a mixed state.  At the level of bulk field theory, the insertion of unitary
operators $\CU$ with support only in the inaccessible region will not affect spacelike separated operators. Consider
the bulk correlation functions (see \NullShAdS):
\eqn\bulkcorr{ \langle 0 | \, \CS^{\dagger} \, \CO_1 \, \CO_2 \, \cdots \CO_n \, \CS |0\rangle \qquad {\rm and }
\qquad \langle 0 | \, \CS^{\dagger} \, \CU^{\dagger} \, \CO_1 \, \CO_2 \, \cdots \CO_n \, \CU \, \CS |0\rangle \, }
If $\CU$ is unitary and spacelike separated from all the $\CO_n$, bulk locality guarantees that $[\CU, \CO_n] = 0$;
hence the two correlation functions in \bulkcorr\ are identical (using $\CU^{\dagger} \, \CU =1$). Taking the
operators $\CO_n$ out to the boundary we can recover boundary correlation functions using the usual AdS/CFT rules. So
boundary correlators which are obtained as limits of bulk correlators
 are not sensitive to excitations in the causally
inaccessible region.
 (Note that this argument is specific to insertion of unitary operators
 in the causally inaccessible region; we can
not reach the same conclusion with Hermitian operators.) Thus correlation functions of local operators can be
calculated by explicitly tracing over the inaccessible region, implying that they are evaluated in a mixed state in
the boundary description.

Nonlocal operators, such as Wilson loops, appear to be able to detect the physics of the inaccessible region. In
fact, since the $t=0$ slice is identical to pure AdS except for excitations at infinity, one expects that the
holographic mapping is unchanged and so  large Wilson loops should be able probe the ``inaccessible" region\foot{While the choice of time slicing is not unique, one expects to be able to map bulk operators to CFT operators on each time slice.}. The CFT would have to have the following strange property. In the state $\CS|0\rangle$, the bulk excitations corresponding to $\CU$ should presumably be captured by excitations of nonlocal
operators. The strange feature of these excitations is that they cannot affect any correlation function of local
gauge-invariant operators at any time. It is unclear whether the field theory needs to have this property exactly,
since our argument relies on bulk locality and thus may only be approximately true. If we were to focus only on
correlation functions of local operators in the CFT, we would see a mixed state which describes physics outside the
horizon.

A second example of a spacetime with causally disconnected regions that appears to be described by a pure state
involves the ``Swedish Geons'' of \refs{\AminneborgPZ,\KrasnovZQ,\HorowitzXK,\MaldacenaRF,\LoukoHC, \alex}. These spacetimes
are orbifolds of AdS$_3$,  which have multiple boundaries and future and past singularities, much like the BTZ black
hole.  However, unlike the BTZ black hole, these geometries have regions in the interior that are causally
disconnected from the all of the boundaries.  In fact, the casual wedge associated to each boundary is metrically
identical to the casual wedge of the BTZ black hole. There is a reasonably straightforward construction of a pure
state in these spacetimes, following the Hartle-Hawking construction of \refs{\MaldacenaRF,\LoukoHC}.  These spacetimes have a simple
Euclidean section, whose boundary is a higher genus Riemann surface.  The Hartle-Hawking state, which is a pure state
entangling degrees of freedom living on the various boundaries, is found by performing a path integral over this Riemann surface. The arguments above involving unitary operators will apply to these spacetimes as well.

\appendix{F}{Analyticity in Coleman-de Luccia spacetimes}

As an example of continuing correlators from the AdS to \dS\ boundaries, it is useful to consider the relatively simple example of Coleman and de Luccia \ColemanAW. Although once backreaction is included the AdS boundaries are removed, if one ignores backreaction this  simple geometry is an instructive toy model where many calculations can be done explicitly.  We will focus on the three dimensional case, but most of our formulas are easily be extended to higher dimensional cases. We start with the Euclidean Coleman-De Luccia geometry
\eqn\eumetric{ ds^2 = d\rho^2 + f(\rho) \,  d\Omega_2^2  \ , }
where $\rho$ runs from $0$ to $\pi r_d$. The geometry approaches Euclidean AdS (the hyperboloid) as $\rho\to 0$, and Euclidean \dS\ (the sphere) $\rho\to \pi r_d$.  So
\eqn\flimits{ f(\rho)=\left\{\eqalign{ &r_a^2 \sinh^2{ \rho \over r_a}\ , \qquad {\rm as}\; \rho \to 0\cr &r_d^2 \sin^2 {\rho\over r_d} \ , \qquad {\rm as} \; \rho \to \pi \, r_d } \right.  }
where $r_a$ and $r_d$ denote the AdS and \dS\ radii, respectively. The form of $f(\rho)$ depends on the profile of the domain wall.  In  the thin wall approximation, $f$ is found by matching the two asymptotic functions \flimits\ at an intermediate value of $\rho$.  In actuality, $f(\rho)$ is a complicated function obeying \flimits, whose exact form depends on the details of the domain wall.

If we Wick rotate one of the angles of the sphere, so that
\eqn\wickrot{ d\Omega_2^2 = dt_E^2 + \cos^2 t_E d\theta^2 \to -dt^2 + \cosh^2 t d\theta^2  \ , }
becomes the metric on a $2$ dimensional \dS\ space, then \eumetric\ describes a Lorentzian geometry interpolating between AdS and \dS.  Note that when we apply \wickrot\ to \eumetric, we obtain AdS and \dS\ coordinate patches whose constant $\rho$ slices are copies of $dS_2$. These patches cover only a portion of the full geometry, as we will describe below.

We can now evaluate the correlators of a quantum field $\phi$ in this background.  The geometry \eumetric\ is non-singular, so can be used to define a Euclidean vacuum state, in which correlators are defined by Wick rotation from Euclidean correlators on  \eumetric.  In the limit where the mass of the scalar field is large, these Euclidean two point functions may be evaluated in position space as

\eqn\twopt{ \langle \phi(x_1) \phi (x_2) \rangle \sim e^{-m \, \len(x_1,x_2)} \ , }
where $\len(x_1,x_2)$ is the proper length of the geodesic between  $x_1$ and $x_2$. In the AdS and \dS\ limits given by \flimits\ this geodesic length is

\eqn\elllimit{ \len = \left\{ \eqalign{
 &\cosh^{-1} \left( r_a^2 \( \cosh{\rho_1\over r_a} \cosh{ \rho_2\over r_a} - \sinh{\rho_1\over r_a} \sinh{\rho_2\over r_a} \cos\ell_{2} \) \right) \ , \qquad      {\rm as} \; \rho_1,\rho_2\to 0 \cr
 &\cos^{-1} \left( r_d^2 \( \cos{\rho_1\over r_d} \cos { \rho_2\over r_d} + \sin{\rho_1\over r_d} \sin{\rho_2\over r_d} \cos\ell_{2} \) \right) \ , \qquad     {\rm as} \; \rho_1,\rho_2 \to \pi \, r_d} \right.  }
where $\ell_2$ is the angular separation on the $2$ sphere.  In the interior,
\eqn\ellis{ \len = \int_{\rho_1}^{\rho_2} \, {d\rho\over \sqrt{1-L^2/f^2}} \ ,}
where $L$ is the conserved angular momentum of a geodesic,

\eqn\asd{ L^2 = f^{4} \,  \( {\dot t}^2 + \cos^2 t \, {\dot \theta}^2 \) \ , }
which is related to the $\ell_2$ by
\eqn\asd{ \ell_{2} = \int {d\rho \over f \sqrt{f^2/ L^2 -1}} \ . }

The expression \ellis\ is an analytic function as one moves from the AdS to the \dS\ regions, since the matching function $f(\rho)$ is analytic. So it allows us analytically continue correlators through the domain wall.  In fact, for particular forms of $f(\rho)$,
we can even continue correlators from the AdS boundary to the \dS\ $\scri$.

This requires an extra step, since the Wick rotation \wickrot\ of the metric \eumetric\ gives us only certain patches of AdS$_3$ and dS$_3$, respectively, whose constant $\rho$ slices are copies of dS$_2$.  Thus to go from the patch of AdS$_3$ containing the domain wall to another patch that contains the asymptotic AdS boundary we must analytically continue in both $t$ and $\rho$. To see this, recall that AdS$_3$ may be written in terms of \dS\ slices as
\eqn\adsinds{ ds^2 = d\rho^2 + r_a^2 \sinh^2 {\rho \over r_a} \, \( -dt^2 + \cosh^2 t \, d\theta^2 \) \ .}
The size of the \dS\ slices, $r_a^2 \, \sinh^2 \rho/r_a$, goes to zero as $\rho \to 0$, indicating that the dS$_2$ slices are becoming null. One may continue across this horizon to a patch foliated by hyperbolic slices
\eqn\adsinfrw{ ds^2 = -d\rho'^2 + r_a^2 \, \sin^2 {\rho' \over r_a}\, \( dt'^2 + \sinh^2 t' \, d\theta^2\) }
by taking $\rho\to \rho' = i \, \rho$ and $t\to  t'=t + i \, \pi/2$; note that the shift in $t$ is necessary to keep $d\theta^2$ spacelike\foot{ In this section we will be careful to label the various $(\rho, t)$  coordinates in  different patches of the spacetime by primes to emphasize that they describe different coordinate systems.}. The dS$_2$ slice that becomes null as $\rho \to 0$ matches onto an $H_2$ slice that becomes null as $\rho' \to0$. Then one can evolve forward in $\rho'$ until the hyperbolic slice shrinks again  to zero size as $\rho' \to \pi\,  r_a$. One then crosses this horizon by a similar analytic continuation to find a second coordinate system of the form \adsinds. This second patch is related to the original coordinate patch  by $t\to t''=t+i\, \pi$ and  $\rho \to \rho''=\rho + i \, \pi\, r_a$. The $(\rho'',t'')$  patch contains an asymptotic AdS boundary at  $\rho'' \to \infty$.

Since constant $\rho''$ slices are copies of dS$_2$, the AdS/CFT correspondence in these coordinates yields a dual boundary CFT living on dS$_2$; this is in contrast to the more familiar examples of the sphere or the plane in global and Poincar\'e coordinates, respectively.

Of course, as mentioned above, when we go beyond the thin wall approximation (as is necessary to obtain analytic  correlators) the geometry will typically develop a big crunch singularity before one can reach the AdS boundary.  This is because the patch \adsinfrw\ has a surface of infinite blueshift at $\rho'\to \pi r_a$, so any matter present will cause a strong backreaction.  However, there certainly exist choices of analytic function $f(\rho)$ where this is not the case.  In these cases the metric is completely smooth in all of the patches described above.  Although such an $f(\rho)$ will not typically solve the appropriate equations of motion, this case still presents an interesting toy model where analytic continuation can be studied explicitly.  We should emphasize that in the full \dS-\SAdS\ spacetimes the AdS boundary to the right of the black hole horizon is not removed by backreaction.  So an analytic continuation similar to the one described here should apply.

To go from the $\rho\to \pi \, r_d$ region of the \dS\ patch of dS$_3$ out to $\scri^+$  we must analytically continue in  both $t$ and $\rho$.  In particular, the patch of dS$_3$ foliated by copies of dS$_2$, with
\eqn\asd{ ds^2 = d\rho^2 + r_d^2 \sin^2 {\rho \over r_d} \, \(-dt^2 + \cosh^2 t \, d\theta^2\) \ , }
has a horizon at $\rho \to \pi \, r_d$.  By taking $t\to t'''=t + i \, \pi/2$ and $\rho \to \rho'''=i \, \rho - \pi \, R_d$ this horizon may be patched on to a region of dS$_3$ with hyperbolic slices
\eqn\asdb{ ds^2 = -d\rho'''^2 + r_d ^2 \sinh^2 {\rho''' \over r_d} \,  \(dt'''^2 + \sinh^2 t''' \, d\theta^2 \) \ .}
The \dS\ boundary is found by taking $\rho'''\to\infty$.  Note that in  this coordinate system the boundary is a copy of $H_2$; this $H_2$ covers  only part of the full de Sitter boundary $\S^2$ that one finds  in global coordinates. Correlators of  bulk fields in this coordinate system define boundary dS/CFT correlators on $H_2$, as opposed to
boundary correlators on the sphere or the plane described by \StromingerPN. Putting this together with \ellis,  we have an explicit analytic continuation of two point functions from points near an AdS boundary to points near a \dS\ boundary.

The considerations described above imply that boundary CFT correlators on the AdS boundary have very interesting analytic behavior.  Expression \elllimit\ tells us that near the AdS boundary correlators are found by  continuing $t\to t +\pi $, $\rho\to i \pi \, r_a$, and taking the large $\rho_1,\rho_2$ limit. Stripping off the factors of $e^{\rho_1}$ and $e^{\rho_2}$, this gives us the usual form for conformal two point functions on dS$_{2}$,
\eqn\twoptads{ \langle \CO_{\phi}(x_1) \, \CO_{\phi(x_2)}\rangle \sim \left({1\over\sin^2\ell_{2}/2}\right)^m + {\rm subleading} \ , }
where $\ell_{2}$ is now the geodesic length on dS$_{2}$. In the small $\ell$ limit this gives the usual short distance behavior\foot{We should emphasize that the expression \elllimit\ has branch cuts in the complex $\ell_2$ plane, so in deriving \twoptads\ we have made a specific branch choice.  This appearance of branch ambiguities is rather common when analytically continuing correlators in curved spacetime, and can be thought of as arising from the ambiguity of vacuum choice in cosmological spacetimes.  In this case, there is a clear choice of branch prescription;we simply choose the branch which matches the usual short distance behavior of the standard AdS vacuum.}
$\ell_2^{-2m}$. The more complicated form at finite distance is the standard formula for conformal two point functions on dS$_2$; it can be thought of as arising from the Weyl anomaly for conformal field theories on \dS\ backgrounds. Near the \dS\ boundary, a similar prescription can be used to define correlators of a Euclidean conformal field theory on $H_2$
\eqn\twoptds{ \langle \CO_{\phi}(x_1) \, \CO_{\phi} (x_2)\rangle \sim \left({1\over \sin^2\ell_{2}/2}\right)^{i \, m} + {\rm subleading} \ , }
where now $\ell_2$ is the geodesic length on $H_2$.  At short distances this gives the short distance behavior $\ell_2^{2\, i \, m}$ associated to a field of imaginary weight, as one usually obtains\foot{We should note that in terms of the boundary field theories, the analytic continuation from the CFT on $dS_2$ in \twoptads\ to the ECFT on $H_2$ given by \twoptds\ is $t\to t +i \, \pi/2$.} in dS/CFT.


\listrefs

\end